\documentclass[]{gCFD2e}

\usepackage{subfigure}
\usepackage{color}

\begin{document}



\title{Evaluation of explicit and implicit LES closures for Burgers turbulence}

\author{R. Maulik and O. San$^{\ast}$\thanks{$^\ast$ Email: osan@okstate.edu
\vspace{6pt}} \\ {\em{School of Mechanical and Aerospace Engineering \\ Oklahoma State University, Stillwater, Oklahoma 74078, USA}}\\\received{\today} }


\maketitle

\begin{abstract}
In this work, we perform an aposteriori error analysis on implicit and explicit large eddy simulation closure models for solving the Burgers turbulence problem. Our closure modeling efforts include both functional and structural models equipped with various low-pass filters. We introduce discrete binomial smoothing filters and an enhanced version of the Van Cittert algorithm to accelerate the convergence of approximate deconvolution processes including regularization and relaxation filtering approaches. Our implicit modeling efforts consist of various high-order schemes including compact and non-compact fifth-order upwind schemes as well as weighted essential non-oscillatory (WENO) and compact reconstructed WENO (CRWENO) schemes, and the resulting schemes are shown to effectively converge to the direct numerical simulation (DNS) for increasing resolutions. Comparing with DNS and underresolved DNS computations, our numerical assessments illustrate the ability of these methods to capture the energy content near grid cut-off scale.


\end{abstract}



\section{Introduction}
\label{sec:intro}
Turbulent flows are encountered in a variety of natural and engineering systems involving a wide range of spatial and temporal scales. All the turbulent motion is required to be resolved in a direct numerical simulation (DNS), where the full spectra of turbulence is resolved down to the Kolmogorov scale. However, the resolution requirements of small scale turbulence are computationally prohibitive to fully resolve for all associated length scales. On the other hand, large eddy simulation (LES) aims to reduce this computational expense and has proven to be a promising approach for calculations of complex turbulent flows \citep{boris1992new,lesieur1996new,piomelli1999large,meneveau2000scale}. In LES, we resolve the most energetic large scales of the turbulent motion while modeling the small scales. This allows much coarser spatial meshes for LES computations and therefore, their computational requirements are significantly lower compared to DNS. Currently, LES is considered a feasible research tool for accurate simulation of large Reynolds number flows and it is slowly augmenting Reynolds-averaged Navier-Stokes (RANS) simulations (i.e., modeling turbulent motion at all scales) in engineering design analysis.

The equations of motion for LES are derived formally by applying a low-pass spatial filter to the governing equations and solving them for the filtered quantities. Low-pass spatially filtered equations can be considered a weak form or regularized form of the governing equations to remove the resolution requirement of small-scale turbulence. Due to the nonlinearity of these governing equations, one needs to correctly treat the well-known LES closure problem \citep{sagaut2006large,berselli2006mathematics} in which interactions between the small scales and the large ones need to be modeled. In the past few decades there has been a substantial effort on developing LES closure models \citep{smagorinsky1963general,yakhot1989renormalization,yoshizawa1989subgrid,germano1991dynamic,piomelli1991subgrid,lilly1992proposed,
ghosal1995dynamic,sarghini1999scale,stolz1999approximate,hughes2000large,hughes2001multiscale,winckelmans2001explicit,
geurts2006leray}.

The simplest form of LES is just to increase the viscosity until the viscous scales are resolved by underlying computational mesh. This added viscosity is generally called the {\em eddy viscosity (EV)} and becomes the foundation of the main stream turbulent closure models. EV models are consistent with Kolmogorov's ideas about the energy spectrum of three-dimensional (3D) isotropic turbulence where energy is injected into the flow at large scales and is gradually transferred by nonlinear cascading processes to smaller and smaller scales until it is dissipated near the viscous dissipation scale \citep{frisch1995turbulence}. With these physical assumptions, the EV concept yields one of the most celebrated LES closure approaches: Smagorinsky model \citep{smagorinsky1963general}. This functional model assumes that the eddy viscosity is computed from the resolved strain rate magnitude and characteristic length scales which are assumed to be proportional to the filter width via a Smagorinsky constant. Although we determine LES equations from a low-pass filtering procedure, a filter is not specified or used in most functional model developments. Application of the Smagorinsky model to various engineering and geophysical flow problems has revealed that the constant is not single-valued and varies between 0.1 and 1.0 in literature depending on resolution and flow characteristics \citep{smagorinsky1993large,canuto1997determination,vorobev2008smagorinsky,pope2000turbulent,cushman2011introduction}. A major advance took place with the development of dynamic model proposed by \cite{germano1991dynamic} and \cite{lilly1992proposed} in which the Smagorinsky constant is self adaptively determined along with the simulation by using a low-pass spatial test filter. This has made the EV type LES closure models more widely applicable in many fields (e.g., see \cite{piomelli1999large} and \cite{meneveau2000scale}).

Alternatively, a recent mathematical closure modeling strategy, called as {\em approximate deconvolution (AD)}, has been proposed by \cite{stolz1999approximate}. It is based on the explicit filtering of the flow field without using any physical assumptions or additional phenomenological arguments, which is particularly appealing for large-scale geophysical flows with inverse energy cascade \citep{san2011approximate,san2013approximate}. Conceptually borrowed from the image processing community, this structural closure model utilizes {\em Van Cittert} iterations by employing repeated filtering operators to represent unfiltered small scale contributions \cite{germano2009new,layton2012approximate,germano2015similarity}. The AD method has been used successfully in the LES of 3D turbulent engineering flows \citep{stolz2001approximate,stolz2001approximatec,domaradzki2002direct}
and atmospheric boundary layer flows \citep{chow2005explicit,chow2009evaluation,duan2010bridging,zhou2011large}. It has also been analyzed in detail from the mathematical point of view \citep{dunca2006stolz,layton2006residual,layton2007similarity,rebholz2007conservation,stanculescu2008existence,dunca2011existence,berselli2012convergence,dunca2014error}.
In this paper, we propose an enhanced Van Cittert iteration procedure via an overrelaxation parameter $\beta$ to accelerate the convergence of AD process.

Several regularization methods have been introduced to prevent growth of energy at small scales. A regularized AD method has been proposed by \cite{stolz2001approximate} through defining a penalty term. A coupled AD and dynamic mixed scale model for LES has also been proposed by \cite{habisreutinger2007coupled} and tested for cavity flows. Another regularization process, called as {\em explicit filtering (EF)} or {\em relaxation filtering (RF)}, has also been proposed to account for the dissipative effect of residual stress in LES \citep{mathew2003explicit,mathew2006new}. It can be combined with AD process to regularize the solution or it can be used independently. The present work address both approaches with various free RF parameters. A family of selective filters has been studied by \cite{bogey2004family,bogey2006large,bogey2006computation,berland2011filter} to eliminate grid-to-grid oscillations. The performance of relaxation filtering has also been studied by \cite{fauconnier2013performance} for the Taylor-Green vortex problem. In the EF framework, flow field variables are filtered each time step to dissipate the amount of energy related to the residual stress, and thus the dissipative effects of the unresolved scales on the resolved scales are modeled. It has been noted that the free filtering parameters are crucial for the success of the method. The use of explicit filtering in LES has been investigated by \cite{bose2010grid} in order to obtain grid independent numerical solutions by using a fourth-order discrete filter. As noted by \cite{lund2003use}, by removing the high-frequency contents of the solution, explicit filtering reduces the effective resolution of the simulation compared with the dynamic range supported by the underlying mesh. Therefore, it has been concluded that a mesh should be finer than the smallest eddy that one hopes to resolve when using explicit filtering. Further details on how the explicit filtering has the effect of reducing numerical errors near the grid cut-off scale can be found in a recent work of \cite{bull2016explicit}.

Since filtering is an explicit part of the numerical simulation, the design of a low-pass filter is also crucial for success of AD and EF modeling approaches. Therefore the understanding of how the filter shape affects LES results is of paramount importance in constructing efficient and consistent LES closure models \citep{de2002sharp}. Various low-pass filtering procedures have been reported in LES literature \citep{schumann1975subgrid,jordan1996large,najjar1996study,vasilyev1998general,sagaut1999discrete,mullen1999filtering,pruett2000priori,
brandt2006priori,berland2011filter,san2015filter,bull2016explicit}. Although the filter design process is well established in digital image processing \citep{oppenheim1989discrete,jahne1997digital}, there is a limited number of studies that focuses on filter design in LES (e.g., see \cite{san2016analysis} and references therein). Examples include the differential filters \citep{germano1986differential,germano1986differential_2,najafi2015high}, Pad\'{e} (compact) filters \citep{lele1992compact,visbal2002use}, discrete selective filters \citep{bogey2004family,bogey2006large,bogey2006computation,berland2011filter}. It has been shown that the filters with complete attenuation for the smallest scales yield significantly better results to prevent energy accumulation at the grid cut-off. Our analysis here utilizes the Pad\'{e} filters which show this desired property of complete attenuation at the grid cut-off scale. We also systematically analyze the behavior of discrete binomial and binomial smoothing filters, which have been used extensively in image processing \citep{jahne1997digital}, for the AD closure modeling in LES.

Both functional EV and structural AD closure models and their variants are often considered as explicit LES closures since the underlying equations are modified and subsequently discretized. On the other hand, the term {\em implicit LES (ILES)} has been used when the truncation error of the discretization of the convective terms behaves as a sub-grid scale (SGS) term \citep{boris1992new,hickel2006adaptive,denaro2011does}. It is assumed that the numerical method provides the required properties implicitly (e.g., the correct amount dissipation of turbulent kinetic energy) and the solution procedure does not require any explicit filtering procedure to prevent energy accumulation at the grid cut-off. The interactions between SGS models and the underlying discretization schemes have also been addressed \citep{vreman1994discretization,ghosal1996analysis,majander2002evaluation,chow2003further,brandt2006priori,berland2008study}. Although the control of numerical errors caused by truncation errors is not a  trivial task in ILES, its computational efficiency makes the ILES framework (no turbulence model) attractive assuming that the numerics provide sufficient modeling of the SGS term \citep{thornber2007implicit,grinstein2007implicit,margolin2006modeling}. In the present study, we will evaluate the behavior of various fifth-order upwind biased numerical methods including compact and noncompact upwind schemes, weighted essential non-oscillatory (WENO) scheme \citep{jiang1996weno}, and recently introduced compact reconstructed WENO (CRWENO) scheme \citep{ghosh2012compact}.

This paper is devoted to the evaluation of various explicit and implicit LES closure models for solving the Burgers turbulence problem. Our analysis includes a broad range of closure modeling efforts using the EV, AD, RF, and ILES models in various forms. Particular attention is paid to the evaluation of the behaviour of the low-pass filtering and the filter shape. As discussed by \cite{falkovich2006lessons}, it has been well recognized that some important general features of turbulence can be understood by studying the Burgers equation. Although the framework of decaying Burgers turbulence is a limiting case \citep{hopf1950partial,cole1951quasi,kida1979asymptotic,frisch2001burgulence,bec2007burgers,valageas2009statistical}, it has been used to perform assessments for turbulence closure models since it retains some important properties of the Navier-Stokes equations such as a quadratic nonlinearity and a propagating shock wave as its solution \citep{love1980subgrid,de2002sharp,labryer2015framework,san2016analysis}.

The present paper is organized as follows: the governing equations and the background numerical methods are reported in Section 2. Section 3 briefly describes several variants of eddy EV models for the Burgers equation. The development of explicit filtering closure models is detailed in Section 4, with a focus on the different filtering procedures and regularization. ILES schemes are presented in Section 5. The results obtained for the ensemble of 64 sample decaying Burgers turbulence cases associated with different phases are then shown in Section 6. Our concluding remarks are finally provided in Section 7.

\section{Governing equation and numerical methods}
\label{sec:method1}

\newcommand{\partfrac}[2]{\frac{\partial #1 }{\partial #2}}
\newcommand{\partfracsec}[2]{\frac{\partial^2 #1 }{\partial #2^2}}

The viscous Burgers equation may be considered to be a test bed representing one dimensional (1D) homogeneous flows and the evolution of the velocity field $u(x,t)$ is governed by
\begin{align}
    \partfrac{u}{t} + u \partfrac{u}{x} = \nu \partfracsec{u}{x}
\end{align}
where $\nu$ is the kinematic viscosity. The nonlinear and the Laplacian terms can be seen to mimic the advective and dissipative mechanisms of complex flows and thus the Burgers equation proves useful in analyzing the characteristics of numerical schemes as well as proving useful for the development of novel subgrid models for turbulent flows\citep{kida1979asymptotic,love1980subgrid,gotoh1993statistics,bouchaud1995scaling,blaisdell1996effect,gurbatov1997decay,balkovsky1997intermittency,de2002sharp,adams2002subgrid,bec2007burgers,labryer2015framework}.

\subsection{Conservative form of the Burgers equation}
\label{sec:method2}

A general representation of the Burgers equation is
\begin{align}
    \partfrac{u}{t} + R(u) = L(u)
    \label{BurgEq}
\end{align}
where $R(u)$ and $L(u)$ are the nonlinear and linear operators given by
\begin{align}
    R(u) = \frac{1}{2} \partfrac{u^2}{x} \label{nlint} \\
    L(u) = \nu \partfracsec{u}{x} \label{lint}
\end{align}
The non-linear term here is considered in the conservative form. Further description of the various formulations for the nonlinear term may be found in \cite{blaisdell1996effect}.

\subsection{Spatial discretization}
\label{sec:method3}

Sixth-order central compact difference schemes were used for the spatial discretizations to ensure a minimal error. The first derivative can be represented compactly in the following manner \citep{lele1992compact}:
\begin{align}
    \frac{1}{3}f'_{j-1} + f'_{j} +  \frac{1}{3} f'_{j+1} = \frac{14}{9}\frac{f_{j+1}-f_{j-1}}{2h} + \frac{1}{9}\frac{f_{j+2}-f_{j-2}}{4h}\, ,
    \label{eq:com1}
\end{align}
where the superscript prime represents the first derivative and $h$ is the uniform spatial grid discretization length. Equation (\ref{eq:com1}) can be solved by the tridiagonal matrix algorithm \citep{press1992numerical} to give us a sixth order approximation of the first derivative. Similarly, a sixth order compact scheme for the second derivative is given by
\begin{align}
    \frac{2}{11} f''_{j-1} + f''_{j} +  \frac{2}{11} f''_{j+1} = \frac{12}{11}\frac{f_{j+1}-2f_{j} + f_{j-1}}{h^2} + \frac{3}{11}\frac{f_{j+2}-2f_{j} +f_{j-2}}{4h^2} \, ,
\label{eq:com2}
\end{align}
where $f^{''} = \partfracsec{f}{x}$ denotes the second derivative.

\subsection{Temporal discretization}
\label{sec:method4}
A system of semi-discrete ordinary differential equations are generated using the method of line after spatial discretizations using the compact schemes explained above. Following the spatial discretization we can represent our system by
\begin{align}
    \partfrac{u_j}{t} = \pounds (u_j)
\end{align}
where the nonlinear and linear terms are denoted in the $\pounds(u_j)$ term as follows
\begin{align}
    \pounds(u_j) = -R(u_j) + L(u_j).
\end{align}
Time integration for our system of ordinary differential equations is carried out using a third-order accurate total variation diminishing Runge-Kutta (TVDRK3) scheme \citep{gottlieb1998total}. In the following we assume that information at time level $l$ is known from which we aim to estimate our solution at time level $l+1$. The integration scheme is given as follows:
\begin{align}
    \label{eq:TVDRK}
    \begin{split}
    u^{(1)}_j & = u^{\ell}_j + \Delta t \pounds(u^{\ell}_j)  \\
    u^{(2)}_j  & = \frac{3}{4}  u^{\ell}_j + \frac{1}{4} u^{(1)}_j + \frac{1}{4}\Delta t \pounds (u^{(1)}_j)  \\
    u^{\ell+1}_j  & = \frac{1}{3}  u^{\ell}_j + \frac{2}{3} u^{(2)}_j + \frac{2}{3}\Delta t \pounds (u^{(2)}_j).
    \end{split}
\end{align}
A time step of $\Delta t = 10^{-5}$  is used for the different cases examined and it is ensured that the numerical results are devoid of errors due to time integration.

\subsection{Filtered equation for the large eddy simulation}
\label{sec:filter}
\begin{figure}[!t]
\centering
\includegraphics[width=0.45\textwidth]{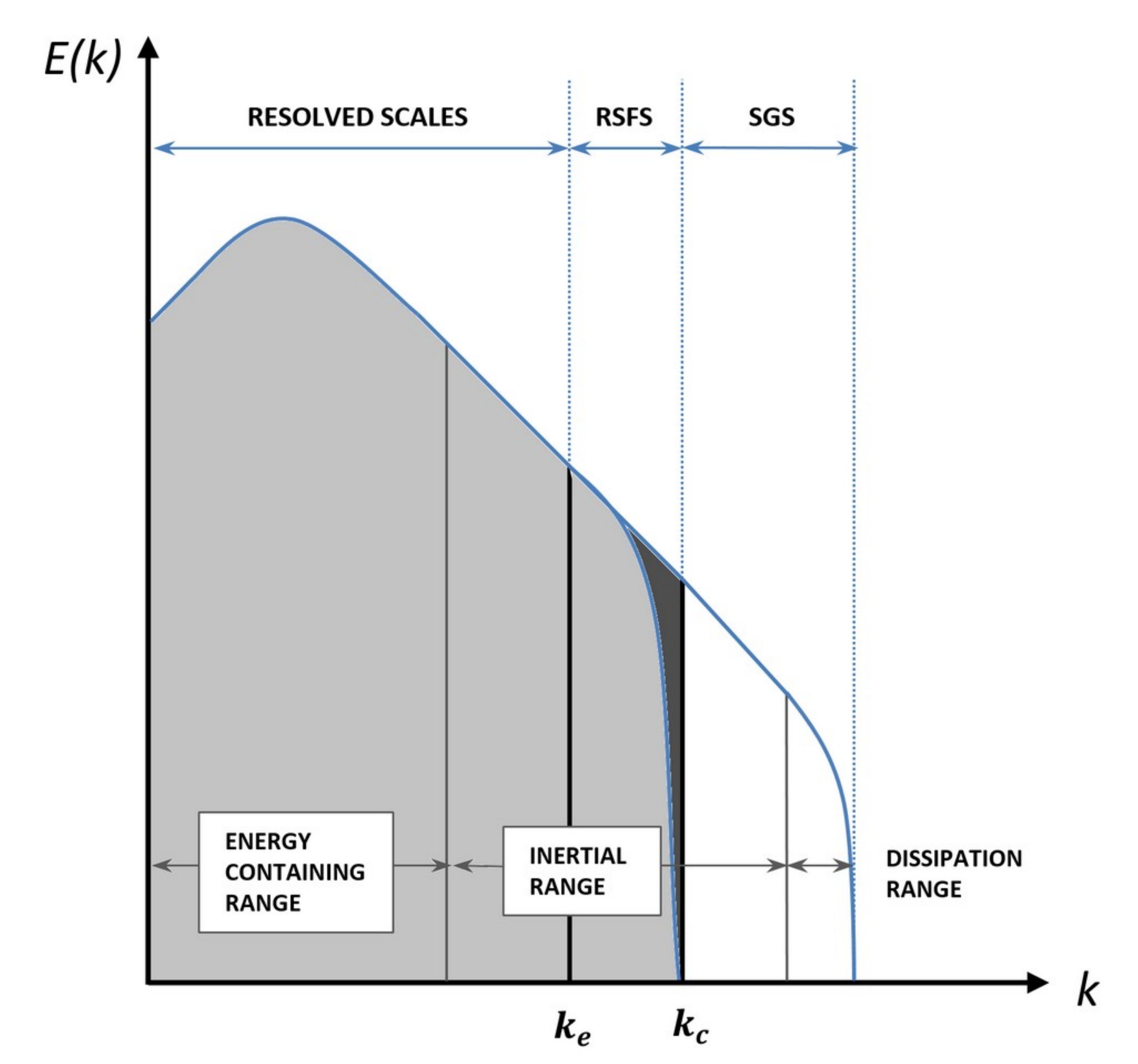}
\caption{Schematics of LES illustrating the resolved scales, resolved subfilter scales (RSFS) where $k_e$ is the effective scale for the filtering and subgrid scales (SGS) starting from the $k_c$ grid cut-off scale. Note that the light shaded area represents the scales captured by LES model and the dark shaded area refers the combination of the numerical and modeling errors in LES modeling.}
\label{fig:scm}
\end{figure}

A popular methodology to accurately capture the behavior of turbulent phenomenon at the integral scales is through the use of a low-pass filter to remove the influence of the higher wavenumbers \citep{stolz2001approximate,stolz2001approximatec,stolz2004approximatei}. These higher wavenumbers are then modeled using a subgrid scale (SGS) term. A filtering procedure may be expressed using the following convolution integral:
\begin{align}
    \bar{f}(x) = \int f(\tilde{x})G(x;\tilde{x}) d\tilde{x}
\end{align}
where $G(x;\tilde{x})$ is our low-pass filter kernel of choice. The convolution operation can also be represented using:
\begin{align}
    \bar{f} = G \ast f
\end{align}
where an overbar depicts a filtered quantity. A filtering of the 1D Burgers equation, Eq. (\ref{BurgEq}), gives us the LES equation for the time evolution of the filtered field:
\begin{align}
    \partfrac{\bar{u}}{t} + G \ast R(u) = L(\bar{u})
\end{align}
where errors of commutation have been considered negligible (i.e., we require that the filter verify the commutation property with derivatives, for example, $G \ast \partfrac{u}{t} = \partfrac{\bar{u}}{t}$ for the unsteady term and $G \ast L(u) = L(\bar{u})$ for the dissipation term). The filtering operation of the advective term leads to the formation of a source term. This source term represents the effects of the unresolved scales. Our governing equation thus becomes:
\begin{align}
    \partfrac{\bar{u}}{t} + R(\bar{u}) = L(\bar{u}) + S
    \label{eq:fge}
\end{align}
where $S = R(\bar{u}) - G \ast R(u)$ is the subfilter-scale term to model the SGS stress. Although we can compute the SGS contribution of the first term in $S$, $R(\bar{u})$, from the resolved solution $\bar{u}$, the source term also requires an LES closure model to approximate its nonlinear term, $G \ast R(u)$, as we do not know our solution field $u$. Closure of the LES governing equation requires the estimation of the nonlinear term, $G \ast R(u)$, in terms of the filtered variable $\bar{u}$. The goal of the LES, governed by Equation (\ref{eq:fge}), is to reproduce the dynamics of the filtered DNS solution by resolving largest and significant scales in the energy containing and inertial ranges, corresponding to ideally 80\% of the total kinetic energy of motion \citep{pope2000turbulent}, while modeling the low energetic small scales (e.g., see Figure (\ref{fig:scm}) for the definition of these scales). We also note that, having a little difference from the traditional LES models, various regularized models aim at reproducing the DNS solution (e.g., see \cite{geurts2003regularization,geurts2006leray,ilyin2006modified} for further discussion).

\section{Eddy viscosity models}
\label{sec:eddy}
Since the small scales in the inertial and dissipation range are not well resolved in LES, their effects are accounted for by a dissipation mechanism using a turbulent eddy viscosity, in order to avoid pile-up of energy near the cut-off wavenumber imposed by the coarse computational grid. Here, we revisit the derivation of well-known Smagorinsky model and its dynamic variants for 1D Burgers equation.
\subsection{Smagorinsky model}
\label{sec:method5}
The simplest functional SGS models are based on a turbulent viscosity and can be written for the Burgers equation as
\begin{align}
    S = \partfrac{}{x}\big(\nu_e\partfrac{\bar{u}}{x}\big)
    \label{eq:sma1}
\end{align}
where $\nu_e$ is usually referred as eddy viscosity (EV) and $\bar{u}$ is the resolved velocity. The fundamental argument of the Smagorinsky model \citep{smagorinsky1963general} is the prescription of an EV in terms of mixing length and absolute value of the strain rate tensor of the resolved flow field in order to account for the appropriate dissipation at smaller scales. For the case of 1D Burgers equation, the above prescription manifests itself in the following expression
\begin{align}
    \nu_{e} =  \ell_{0}^{2} \left| \partfrac{\bar{u}}{x} \right|
\end{align}
where $\ell_{0}$ is the mixing length and it is usually given by $\ell_{0}= C_s \delta$ in the Smagorinsky model. Coefficient $C_s$ is the Smagorinsky constant. Therefore, the well-known Smagorinsky hypothesis can be written
\begin{align}
    \nu_{e} =  (C_s \delta)^2 \left| \partfrac{\bar{u}}{x} \right|
    \label{eq:sma2}
\end{align}
where $\delta$ is formally defined by the filter width and usually set to the representative mess size. In our study we use constant grid spacing given by $\delta=h$ and refer this model as EV-LES.

\subsection{Dynamic model}
\label{sec:dyn}
Application of the Smagorinsky model to various flow problems has revealed that the $C_s$ constant is not single-valued \citep{smagorinsky1993large} and needs for an ad-hoc specification depending on the flow characteristics. On the other hand, the model by \cite{germano1991dynamic} allows a self adaptive estimate of the Smagorinsky constant from the simulation itself and has been also further improved by \cite{lilly1992proposed} who propose to calculate $C_s$ in the least square sense. For the dynamic implementation, using Equations (\ref{eq:sma1}) and (\ref{eq:sma2}), we can rewrite the filtered Burgers equation for the primary filter scale of $\delta$
\begin{align}
    \partfrac{\bar{u}}{t} + \partfrac{}{x}\big(\frac{\bar{u}^2}{2}\big) = \nu \partfracsec{\bar{u}}{x} + \partfrac{}{x}\left((C_s \delta)^2 \left| \partfrac{\bar{u}}{x} \right|\partfrac{\bar{u}}{x}\right)
    \label{eq:fil1}
\end{align}
and we can also rewrite Burgers equation with a test filter using the filter scale of $\tilde{\delta}$ (i.e., $\tilde{\delta} > \delta$)
\begin{align}
    \partfrac{\tilde{\bar{u}}}{t} + \partfrac{}{x}\big(\frac{\tilde{\bar{u}}^2}{2}\big) = \nu \partfracsec{\tilde{\bar{u}}}{x} + \partfrac{}{x}\left((C_s \tilde{\delta})^2 \left| \partfrac{\tilde{\bar{u}}}{x} \right|\partfrac{\tilde{\bar{u}}}{x}\right)
    \label{eq:fil2}
\end{align}
where superscript tilde refers to the test filter. Here we assume that the difference between test filtered fields $\tilde{u}$ and $\tilde{\bar{u}}$ becomes negligible (i.e.,$\tilde{u} \approx \tilde{\bar{u}}$) since $\tilde{\delta}$ is greater than the $\delta$. Applying the same test filter to Equation (\ref{eq:fil1}) yields
\begin{align}
    \partfrac{\tilde{\bar{u}}}{t} + \partfrac{}{x}\big(\frac{\widetilde{\bar{u}^2}}{2}\big) = \nu \partfracsec{\tilde{\bar{u}}}{x} + \partfrac{}{x}\left((C_s \delta)^2 \widetilde{\left| \partfrac{\bar{u}}{x} \right|\partfrac{\bar{u}}{x}}\right)
    \label{eq:fil1}
\end{align}
and the results is subtracted from Equation (\ref{eq:fil2}), leading to
\begin{align}
     \partfrac{}{x}\big(\frac{\tilde{\bar{u}}^2}{2}\big) - \partfrac{}{x}\big(\frac{\widetilde{\bar{u}^2}}{2}\big) = \partfrac{}{x}\left((C_s \tilde{\delta})^2 \left| \partfrac{\tilde{\bar{u}}}{x} \right|\partfrac{\tilde{\bar{u}}}{x}\right) - \partfrac{}{x}\left((C_s \delta)^2 \widetilde{\left| \partfrac{\bar{u}}{x} \right|\partfrac{\bar{u}}{x}}\right).
    \label{eq:fil3}
\end{align}
Factoring the model coefficient, Equation (\ref{eq:fil3}) can be recast in the following form
\begin{align}
     H = (C_s \delta)^2 M
    \label{eq:fil4}
\end{align}
where
\begin{align}\label{eq:hh}
     H = \partfrac{}{x}\big(\frac{\tilde{\bar{u}}^2}{2}\big) - \partfrac{}{x}\big(\frac{\widetilde{\bar{u}^2}}{2}\big)
\end{align}
\begin{align}
     M = \kappa^2 \partfrac{}{x}\left( \left| \partfrac{\tilde{\bar{u}}}{x} \right|\partfrac{\tilde{\bar{u}}}{x}\right) - \partfrac{}{x}\left(\widetilde{\left| \partfrac{\bar{u}}{x} \right|\partfrac{\bar{u}}{x}}\right)
     \label{eq:mm}
\end{align}
in which $\kappa=\tilde{\delta}/\delta$ is the filter ratio (e.g., $\kappa=2$ in the present study). The coefficient $(C_s \delta)^2$ in Equation (\ref{eq:fil4}) can be obtained by minimizing the average square error $\langle E^2 \rangle$, where the error is given by $E=H-(C_s \delta)^2 M$. In present study the averaging operator is expressed by
\begin{align}
     \langle f \rangle = \frac{1}{2\pi} \int_{0}^{2\pi} f dx
\end{align}
Differentiating the mean square error with respect to the model parameter $(C_s \delta)^2$, we get
\begin{align}
    \partfrac{\langle E^2 \rangle}{(C_s \delta)^2}  = -2\langle HM \rangle + 2(C_s \delta)^2\langle M^2 \rangle
\end{align}
where we assume that the Smagorisnky constant can be factored out of the averaging operators \citep{lilly1992proposed,mansfield1998dynamic}. Then, we can minimize the square of the error when
\begin{align}
    (C_s \delta)^2 = \frac{\langle H M \rangle}{\langle M^2 \rangle}.
\end{align}
We refer this dynamic eddy viscosity (DEV) model as DEV-LES in the present study. Instead of using the instantaneous filtered strain rate (i.e., $|\partial \bar{u} / \partial x|$ for 1D Burgers equation, e.g., see \cite{fauconnier2009family}), a variant of this model can be obtained by using the mean value of the absolute strain rate \citep{lilly1992proposed}, leading to
\begin{align}
     M = \kappa^2 \partfrac{}{x}\left( \left\langle \left| \partfrac{\tilde{\bar{u}}}{x} \right|\right\rangle \partfrac{\tilde{\bar{u}}}{x}\right) - \partfrac{}{x}\left(\widetilde{\left\langle \left| \partfrac{\bar{u}}{x} \right|\right\rangle \partfrac{\bar{u}}{x}}\right),
     \label{eq:mm2}
\end{align}
and $H$ is given by Equation (\ref{eq:hh}). We refer this averaged DEV (ADEV) model as ADEV-LES when Equation (\ref{eq:mm2}) is used.
The definition of the test filter completes the formulation of the dynamic models. The effects of various test filters will be investigated in the present study. Details of the filters will be presented in the following section.

\section{Explicit filtering models}
\label{sec:EF}
These methods allow the control of the numerical errors (e.g., due to the truncation errors and aliasing errors of high-frequencies) and model the dissipative effects of the truncated scales by computational grid (i.e., SGS contribution)  based on an explicit filtering of the flow field. Here, we focus on approximate deconvolution (AD) and relaxation filtering (RF) models as well as a coupled AD and RF approach.
\subsection{Approximate deconvolution}
\label{sec:AD1}
The AD procedure was proposed to recover the solution field $u$ from the filtered field $\bar{u}$ based on the theory of image reconstruction \citep{stolz1999approximate}. AD uses a repeated filtering procedure to obtain approximations of the unfiltered variable when the filtered variable is available. A further discussion of error estimates and convergence may be found in \cite{berselli2012convergence,layton2012approximate,dunca2014error}. AD-LES has been applied to the Burgers turbulence problem using the standard the Van Cittert method \citep{san2016analysis}. In this work, we use an enhanced version of the Van Cittert method for deconvolution based on the principle of iterative successive substitution to get an approximation for $u$ from $\bar{u}$. If $\vartheta$ is assumed to be an approximately recovered value of $u$, the LES equation becomes:
\begin{align}
    \label{eq:adm}
    \partfrac{\bar{u}}{t} + G \ast R(\vartheta) = L(\bar{u}),
\end{align}
for which an iterative procedure may be used to determine $\vartheta$ given by
\begin{eqnarray}
    \vartheta^0 &=& \bar{u} \nonumber\\
    \vartheta^i &=& \vartheta^{i-1} + \beta(\bar{u} - G \ast \vartheta^{i-1}), \quad i = 1,2,3,...,Q,
\end{eqnarray}
where increasingly accurate reconstructions are obtained with increasing $Q$. Here, the parameter $\beta$ is an over-relaxation parameter which can be used to speed up convergence of our iterative substitution process. The equation of the standard Van Cittert algorithm can be recovered when $\beta=1$. The initial guess for $\vartheta$ is the filtered value $\bar{u}$ and this guess is updated after every iteration step. Eventually, the value for $\bar{u} - G \ast \vartheta^{i-1}$ will become small and can be ignored and at that iteration the process can be assumed to have converged. It can be shown that this procedure is numerically stable for the following condition
\begin{align}
    |1 - \beta T(k)| \leq 1,
\end{align}
where $T(k)$ refers the transfer function of the filter and will be explained in forthcoming sections. The value for $\beta$ must be greater than zero to ensure convergence and less than two to prevent oscillations in the transfer function values at higher wavenumbers (i.e., $0<\beta\leq2$ when the transfer function of the filter $0\leq T(k)\leq1$). For a detailed examination of convergence conditions and limiting solution properties of the Van Cittert method, the reader may consult \cite{biemond1990iterative}. We emphasize that the Van Cittert method is popularly used with $\beta$ taken as one. The present work also reports the results of using an enhanced version of the method with a sensitivity analysis of for $\beta$. In addition, the effect of the number of the Van Cittert iterations $Q$ is also examined. Furthermore, we need to choose a consistent and computationally efficient low-pass filtering operator to completely specify the AD-LES model presented in this section.

\subsection{Low-pass filters}
A low-pass filter (considered a free modeling parameter in LES) is commonly used in explicit filtering approaches. In this work, two classes of explicit filters are considered viz. the 6th order Pad\'{e} filter and the binomial filter.

\subsubsection{Pad\'{e} filters}
\label{sec:AD2}
The discrete Pad\'{e} filter family was introduced by \citep{lele1992compact} and has been used with good results previously \citep{stolz1999approximate,stolz2001approximate,pruett2000priori,san2015filter}. The following $\alpha$ parameter Pad\'{e} filter was chosen for spatial smoothing:
\begin{align}
    \alpha \bar{f}_{j-1} + \bar{f}_{j} + \alpha \bar{f}_{j+1} = \sum_{s=0}^{3}\frac{a_{s}}{2} (f_{j-s} + f_{j+s}) ,
\label{eq:119}
\end{align}
where $\bar{f}_j$ represents the filtered value of a discrete quantity $f_j$ and the filtering coefficients are
\begin{align}
    a_0 = \frac{11}{16} + \frac{5}{8}\alpha , \quad a_1 = \frac{15}{32} + \frac{17}{16}\alpha , \quad a_2 = -\frac{3}{16} + \frac{3}{8}\alpha , \quad a_3 = \frac{1}{32} - \frac{1}{16}\alpha
\label{eq:118}
\end{align}
The parameter $\alpha$ belongs to the $[-0.5,0.5]$ range and controls the dissipation power of the smoothing filter. It is seen that higher values of $\alpha$ correspond to lower dissipation. A Fourier analysis is carried out to study the behavior of this smoothing filter in wavenumber space. Using the standard modified wavenumber analysis, a transfer function $T(k)$ can be determined that correlates the Fourier coefficients of the smoothed and unsmoothed variables as follows:
\begin{align}
    \hat{\bar{f}} = T(k) \hat{f}
\end{align}
where $\hat{\bar{f}}$ and $\hat{f}$ represent the Fourier coefficients of the filtered and unfiltered variables respectively. The classical definitions of the discrete Fourier transform have been used as follows
\begin{align}
    f_j= \sum_{m=-N/2}^{N/2-1} \hat{f}_m \exp( \mathrm{i} k x_j),
    \label{eq:traf}
\end{align}
which has been used for the discrete forward transform and for the inverse transform we have
\begin{align}
    \hat{f}_m = \frac{1}{N}\sum_{j=0}^{N-1} f_j \exp(-\mathrm{i} k x_j),
    \label{eq:traf2}
\end{align}
where $N$ is the number of spatial grid points and $i = \sqrt{-1}$. A wavenumber can be defined by $k=\frac{2\pi}{L}m$ and a transfer function can be obtained for our low pass filter defined previously
\begin{align}
    T(k) = \frac{a_0+a_1\cos(hk) +a_2\cos(2hk) + a_3\cos(3hk)}{1+2\alpha\cos(hk)} \, ,
\label{eq:tfun-1}
\end{align}
where $a_i$ has been defined above in Eq.(\ref{eq:118}) and $h$ is the spatial grid length. The stated range of the parameter $\alpha$ ensures that the transfer function remains positive definite for accurate deconvolution. The transfer function behavior for the Pad\'{e} filter has been shown in Figure (\ref{fig:tf}a). It can be seen here that the transfer function happens to attenuate the effects of the highest wavenumber completely (i.e $T_k = 0$ at $k_m = \pi/h$) while having low dissipation over the resolved inertial scales. Increasing the value of $\alpha$ from 0 to 0.5 results in lesser dissipation in the inertial zone. It must be noted here that a value of 0.5 causes no dissipation at all at any range of wavenumber and thus becomes a limiting value.

\subsubsection{Binomial filters}
\label{sec:AD3}

Binomial filters are a class of smoothing functions designed for efficient calculation. Filtered field values at a discrete point in the domain are computed using a spatial averaging operator which requires information from a stencil whose size depends on the type of binomial filter chosen. An elementary binomial filter can be considered to be (for a 1D case), a simple averaging operation given by:
\begin{align}
    B^1 = \frac{1}{2}[1\  1]
\end{align}
which may be interpreted as the computation of the filtered value at a discrete point $i$ by
\begin{align}
    \bar{f}_i = \frac{1}{2}\big(f_{i+1/2}+f_{i-1/2}\big).
\end{align}
Similarly, an example of a binomial filter for an odd numbered stencil is given by
\begin{align}
    B^2 = \frac{1}{4}[1\ 2\ 1],
\end{align}
thus, we can interpret as
\begin{align}
    \bar{f}_i = \frac{1}{4}\big(f_{i+1}+2f_{i}+f_{i-1}\big).
\end{align}

\begin{figure}[!t]
\centering
\mbox{
\subfigure[Pad\'{e} filters for AD]{\includegraphics[width=0.45\textwidth]{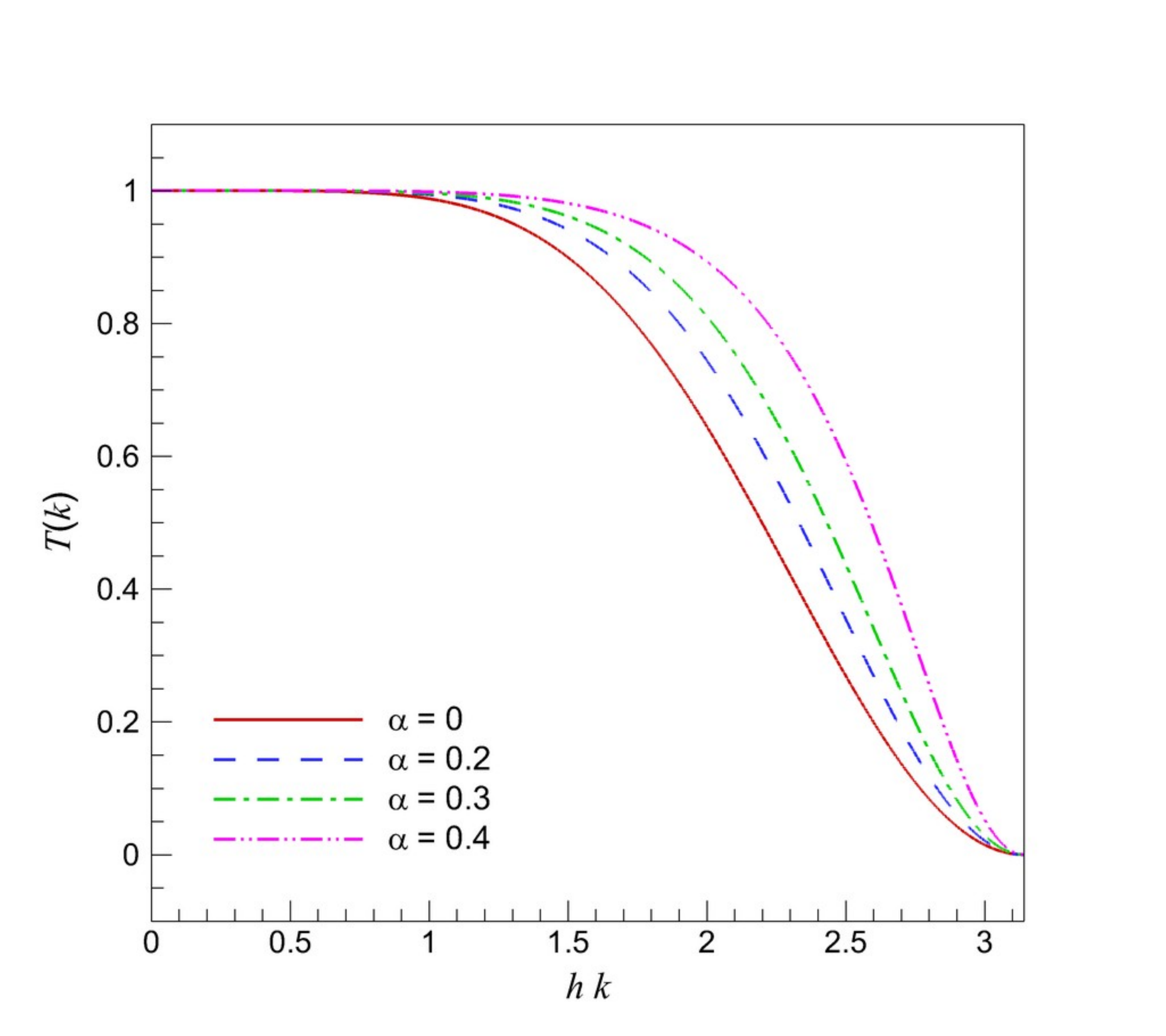}}
\subfigure[Pad\'{e} filters for EF (relaxation filters)]{\includegraphics[width=0.45\textwidth]{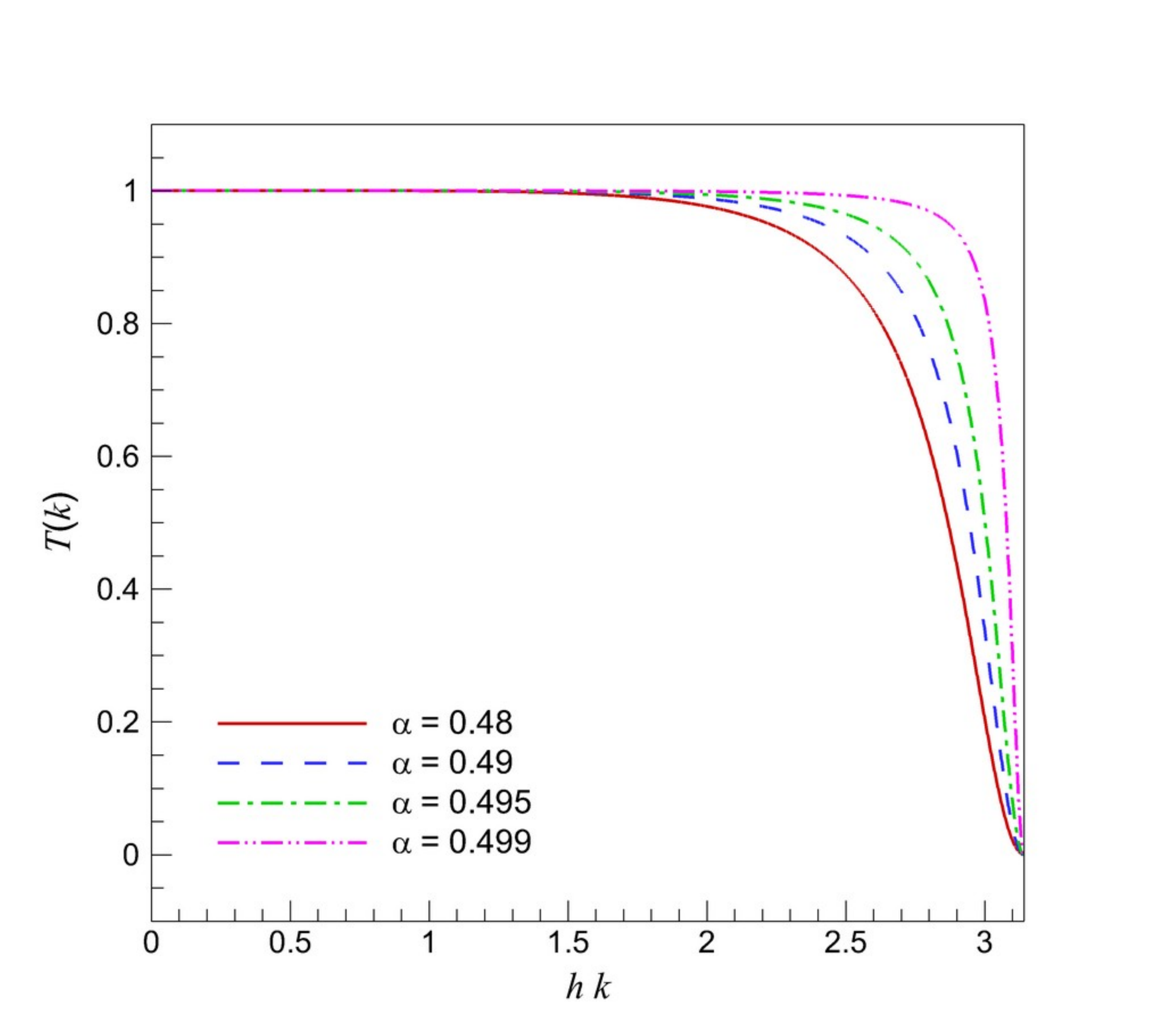}}
}
\mbox{
\subfigure[Binomial filters for AD]{\includegraphics[width=0.45\textwidth]{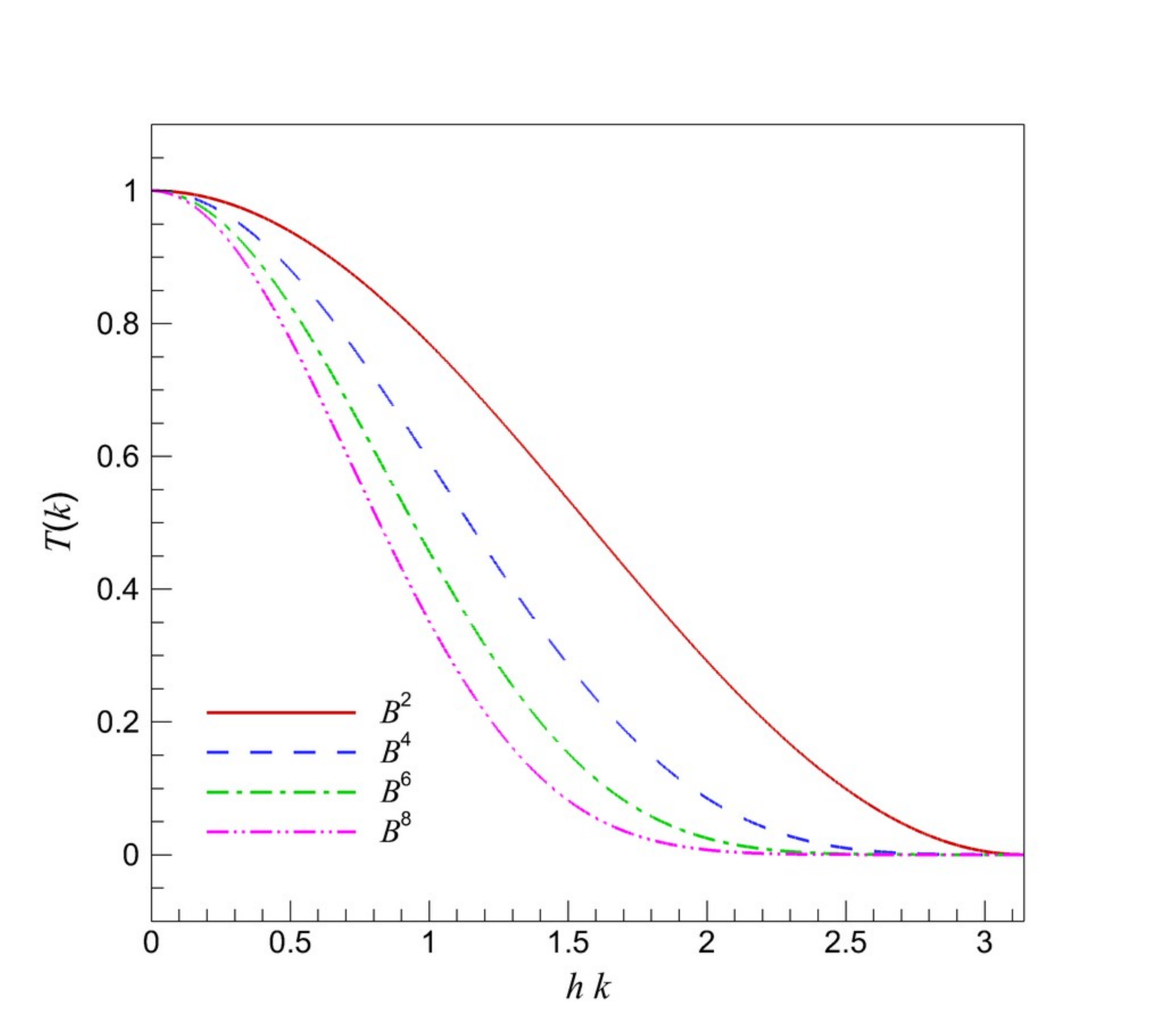}}
\subfigure[Binomial smoothing filters for AD]{\includegraphics[width=0.45\textwidth]{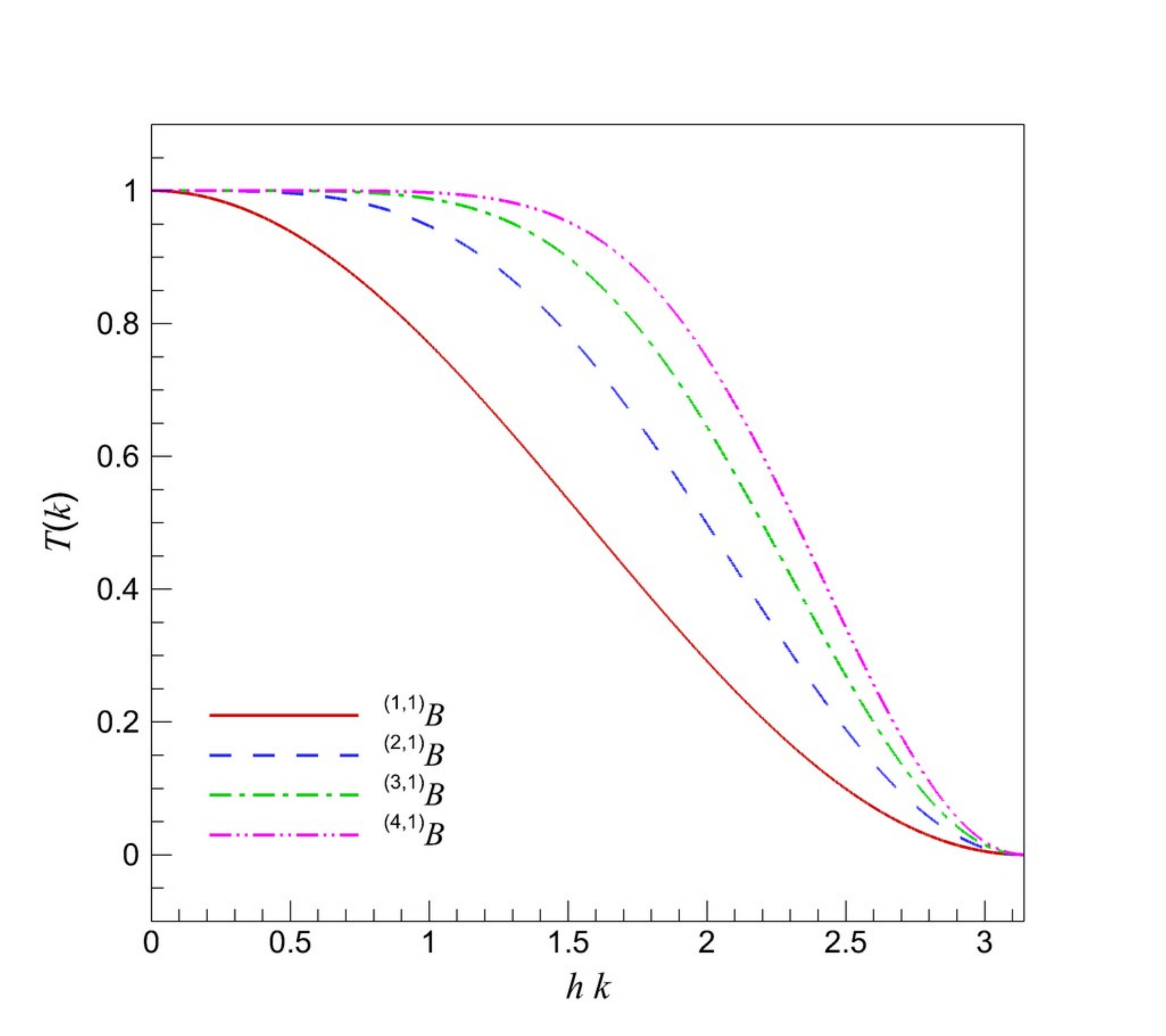}}
}\\
\caption{Transfer functions of low-pass spatial filters designed for low-pass filtering in AD and EF models.}
\label{fig:tf}
\end{figure}

Generally, a binomial filter of order $n$ is given by $\bar{f} = m[d]$ and can be obtained through the use of the Pascal triangle as shown below
\begin{center}
 \begin{tabular}{|c| c| c|}
 \hline
 n & m & d \\
 \hline
 0 & 1 & 1 \\
 \hline
 1 & 1/2 & 1 1  \\
 \hline
 2 & 1/4 & 1 2 1 \\
 \hline
 3 & 1/8 & 1 3 3 1\\
 \hline
 4 & 1/16 & 1 4 6 4 1\\
 \hline
 5 & 1/32 & 1 5 10 10 5 1\\
 \hline
 6 & 1/64 & 1 6 15 20 15 6 1\\
 \hline
 7 & 1/128 & 1 7 21 35 35 21 7 1\\
 \hline
 8 & 1/256 & 1 8 28 56 70 56 28 8 1\\
 \hline
\end{tabular}
\end{center}
where $m$ is the scaling factor of the operator. We are only interested in odd numbered stencil schemes (i.e., an even order $n$) because of underlying centred finite difference framework. This class of binomial filters can be considered a discrete approximation the Gaussian filter with various variance \citep{jahne1997digital}. A generalized transfer function of the 1D binomial filter is given by
\begin{align}
    \label{BinoNonSmoothTF}
    \hat{B}^{2R} = \frac{1}{2^R}\big(1+\cos(\pi k)\big)^R
\end{align}
The behavior of the transfer function of some 1D binomial filters with odd stencil sizes are depicted in Figure (\ref{fig:tf}c).

One of the primary advantages of the binomial class of filters is the ability to design smoothing operators by combining multiple binomial filters. This ability lets us control the amount of dissipation of the filter and it is possible to design transfer functions that minimize dissipation at lower wavenumbers and have a sharper drop off to zero at higher wavenumbers. The following relationship sets up the framework for a binomial smoothing filter design
\begin{align}
    ^{(n,l)}B = \big(I - (I-B^2)^n\big)^l.
\end{align}
This class of filters shares many important properties of the binomial filters $B^{n}$. The simplest five-point stencil operator of this type is
\begin{align}
    ^{(2,1)}B = \frac{1}{16}[-1\ 4\ 10\ 4 -1].
\end{align}
The seven point stencil filter reads as
\begin{align}
    ^{(3,1)}B = \frac{1}{64}[1\ -6\ 15\ 44\ 15\ -6\ 1],
\end{align}
and the nine-point stencil operator can be written as
\begin{align}
    ^{(4,1)}B = \frac{1}{256}[-1\ 8\ -28\ 56\ 186\ 56\ -28\ 8\ -1].
\end{align}
We note that the Pad\'{e} filter given by Equation (\ref{eq:119}) reduces to $^{(3,1)}B$ when its free filtering parameter $\alpha=0$.
A general expression for the transfer function of a 1D binomial smoothing filter is given by
\begin{align}
    ^{(n,l)}\hat{B} = \left(1-\Big(1-\frac{1}{2}\big(1+\cos(\pi k)\big)\Big)^n\right)^l
\end{align}
The transfer function behavior of these smoothing filters has been illustrated in Figure (\ref{fig:tf}d). A more involved discussion of binomial and binomial smoothing filters is available in \cite{jahne1997digital}.

\subsection{Relaxation filtering and AD regularization}
Although deconvolution process can be used to compute SGS term, a stable LES would require a treatment for the energy transfer to the truncated scales \cite{mathew2006new}. An \emph{ad hoc} low order penalty term would be added with a free coefficient to provide additional dissipation \cite{vreman2003filtering,stolz2004approximatei}.
Another school of thought to address the problem of pile-up near the grid cut-off scales proposes the concept of relaxation filtering (RF), in which a secondary filtering process is used to provide a stabilization near the grid cut-off scale (e.g., see \cite{adams2002subgrid}). This secondary filtering may be explicitly used at the end of each time step in conjunction with AD-LES. In this study, the Pad\'{e} filter has been chosen as a relaxation filtering operator of choice and several numerical experiments have been carried out to assess the effect of this RF operator when used explicitly and along with AD-LES. In RF framework using a secondary filter, the free filtering parameter $\alpha$ should be adjusted to provide additional regularization at higher wavenumbers near grid cut-off scale. Figure (\ref{fig:tf}b) illustrates the transfer functions of this type of filters for various $\alpha$ parameters. An investigation on the choice of $\alpha$ will be provided in this study.

Alternatively, a coupled approximate deconvolution and dynamic mixed scale model has been proposed for large eddy simulation \cite{habisreutinger2007coupled}. In the present study, a performance study is also carried out for the behavior of AD-LES when combined with the concept of the Smagorinsky model wherein an additional dissipation is added to the system through an eddy viscosity. The Smagorinsky eddy viscosity concept has been described in Section \ref{sec:method5}. Combining Equation (\ref{eq:adm}) with Equation (\ref{eq:sma1}) the hybrid structural-functional model reads
\begin{align}
    \label{eq:adm}
    \partfrac{\bar{u}}{t} + G \ast R(\vartheta) = L(\bar{u}) + \partfrac{}{x}\big(\nu_e\partfrac{\bar{u}}{x}\big)
\end{align}
where $\nu_e$ is given by Equation (\ref{eq:sma2}). The potential pile-up at grid cut-off is thus addressed by this additional dissipation mechanism.

\subsection{Explicit filtering without AD process}

The relaxation filters can also be used without any approximate deconvolution process \citep{visbal2002use,lund2003use,bogey2006computation,bull2016explicit}. In this case, we assume that explicit filtering removes the frequencies higher than a selected cut-off threshold through the use of a custom specified filter illustrated in Figure (\ref{fig:tf}b). Indeed \cite{mathew2003explicit} interpreted AD as a nearly equivalent procedure of low-pass explicit filtering (EF) of the solution at each time step. Our analysis also includes this approach by referring it as EF model in the present paper.

\section{Implicit filtering models}
\label{sec:ILES1}

In implicit filtering models, the grid, or the numerical discretization scheme, is assumed to be LES low-pass filter, therefore usually called as implicit LES (ILES) \cite{grinstein2007implicit}. In this framework, we assume that the numerical scheme will dissipate energy in the same manner as the SGS model.  Although this takes full advantage of the grid resolution, and eliminates the computational cost of SGS model calculation, the control of numerical errors caused by truncation errors is not a trivial task in ILES \citep{thornber2007implicit,margolin2006modeling}.

Implicit filtering implies the use of the dissipative behavior of numerical discretizations for the prevention of an energy pile-up in the energy spectrum of the flow near its cut-off. It is well known that any upwind biased scheme has its own numerical dissipation. In effect, we assume the numerical discretization scheme to be a low-pass filter. The advantages of this approach are apparent in that an additional iterative step for deconvolution (or evaluation of SGS terms) is not necessary. In this section we detail the numerical formulation of the WENO and compact WENO schemes which have traditionally been used for shock capturing through added dissipation in a nonlinear fashion to damp oscillations near discontinuities or sharp gradients. We aim to utilize this damping characteristic of these schemes to act as an implicit filter preventing energy pile-up near the grid cut-off.

\subsection{Upwind schemes and flux splitting}
\label{sec:ILES2}
The conservative form we have employed for the 1D Burgers equation, Equation (\ref{BurgEq}), allows us to utilize the flux splitting method in the development of WENO and CRWENO schemes. Therefore, we rewrite Burgers equation in flux form.
\begin{align}
    \partfrac{u}{t} + \partfrac{f}{x} = L(u)
\end{align}
where $L(u)$ is the linear term given by Equation (\ref{lint}).
Our flux term is given by $f = \frac{1}{2} u^2$ and the nonlinear advective, $R(u) =\partfrac{f}{x}$ , term may be represented as:
\begin{align}
    \begin{split}
    \partfrac{f}{x}|_{x = x_j}  &= \frac{F_{j+1/2}-F_{j-1/2}}{\Delta x}\\
     &=  \frac{f^L_{j+1/2}-f^L_{j-1/2}}{\Delta x} + \frac{f^R_{j+1/2}-f^R_{j-1/2}}{\Delta x} + O(\Delta x^r)
    \end{split}
\end{align}
at a particular node $j$ and where $r$ is the desired order of the scheme. Here, $\Delta x = x_{j+1/2} - x_{j-1/2}$ is the mesh size (i.e., $\Delta x =h$ in this study), $F(x)$ refers the exact flux function, and superscripts $L$ and $R$ refer the positive and negative flux components in our discrete system, respectively.  This process is called flux splitting and the values of left and right biased values of the flux after splitting are calculated by
\begin{align}
    \begin{split}
        f^{L} &= \frac{1}{2}(f + \alpha u),\\
        f^{R} &= \frac{1}{2}(f - \alpha u);
    \end{split}
\end{align}
where $\alpha$ is called the absolute value of the flux Jacobian (i.e., $\alpha = |\partfrac{f}{u}| = |u|$). In the case of local pointwise flux splitting at location $x_j$, the value of $\alpha$ is given by
\begin{align}
\label{eq:ls}
\alpha  = |u_j|;
\end{align}
whereas $\alpha$ may also be chosen to be maximum over the local stencil, i.e.,
\begin{align}
\label{eq:ss}
    \alpha = \textnormal{max}(|u_{j-2}|,|u_{j-1}|,|u_{j}|,|u_{j+1}|,|u_{j+2}|),
\end{align}
in which case it is considered as local stencil flux splitting. A fifth order approximation of the flux $F(x)$ is given by:
\begin{align}
    \label{WENO5-1}
    f_{j+1/2}^L = \frac{1}{30}f_{j-2}^L - \frac{13}{60} f_{j-1}^L + \frac{47}{60} f_{j}^L + \frac{27}{60} f_{j+1}^L - \frac{1}{20} f_{j+2}^L,
\end{align}
which gives a truncation error of
\begin{align}
    f_{j+1/2} = F_{j+1/2} - \frac{1}{60}\frac{\partial ^5 f}{\partial x^5}|_{x=x_j} \Delta x^5 + O(\Delta x^6).
\end{align}
One can see here that there are more interpolation points to the left of the location where we need our approximate flux reconstruction. It should be noted, though, that the values $f_j$ used for the interpolation represent the cell averaged values at the nodal points $j$. The reader is directed to the description in \cite{shu2009high} for further information about such interpolations. The right biased flux (for a 1D wave travelling in the opposite direction) is developed symmetrically about the point $j$ in a similar manner, e.g.,
\begin{align}
    f_{j-1/2}^R = \frac{1}{30}f_{j+2}^R - \frac{13}{60} f_{j+1}^R + \frac{47}{60} f_{j}^R + \frac{27}{60} f_{j-1}^R - \frac{1}{20} f_{j-2}^R.
\end{align}
Superscripts for left and right biasing are omitted beyond this point. The aforementioned scheme will henceforth be denoted as UPWIND5.

A fifth order compact interpolation is given by (henceforth the CU5 scheme)
\begin{align}
    \label{cu5}
    \frac{3}{10}f_{j-1/2} + \frac{6}{10}f_{j+1/2} + \frac{1}{10}f_{j+3/2} = \frac{1}{30}f_{j-1}+\frac{19}{30}f_j+\frac{10}{30}f_{j+1};
\end{align}
which gives an error of
\begin{align}
    f_{j+1/2} = F_{j+1/2} - \frac{1}{600}\frac{\partial ^5 f}{\partial x^5}|_{x=x_j} \Delta x^5 + O(\Delta x^6),
\end{align}
One can note here that the leading error term in the compact interpolation scheme (CU5) is an order of magnitude less than its noncompact counterpart (UPWIND5).

\subsection{WENO reconstruction}
\label{sec:ILES3}
 The general form of the approximate flux reconstruction by the WENO scheme is \citep{jiang1996weno}
\begin{align}
    f_{j+1/2} = \sum_{k=1}^{r} w_k f_{j+1/2}^k,
\end{align}
where $r$ is the number of stencils of $r^{th}$ order, $f_{j+1/2}^k$ is the reconstructed flux using the $k^{th}$ stencil and $w_k$ is the weight of the $k^{th}$ stencil. These weights are optimally distributed in the event of a smooth field at $j+1/2$. If a stencil possesses a discontinuity, the WENO scheme forces the weight of that stencil to zero using weighting functions which are
\begin{align}
    \alpha_k = \frac{c_k}{(\beta_k + \epsilon)^m}.
\end{align}
Here, $c_k$ are the optimal weights (i.e., weights for the linear high-order scheme), $\beta_k$ is the smoothness indicator of the $k^{th}$ stencil, $m$ is chosen to ensure that the weights for discontinuous stencils approach zero quickly and $\epsilon$ is a small number to prevent a division by zero. It should be noted here that $m=2$ has been chosen as a standard value.

The aforementioned weights are then normalized for convexity to give
\begin{align}
    w_k = \frac{\alpha_k}{\sum_k \alpha_k}
\end{align}
It should be noted here that the order of a WENO scheme is given by (2$r$ - 1). As an example, a fifth order WENO scheme is developed by the combination of three third order stencils given by
\begin{align}
    \begin{split}
    f_{j+1/2}^1 &= \frac{1}{3} f_{j-2} -\frac{7}{6}f_{j-1} + \frac{11}{6} f_j\\
    f_{j+1/2}^2 &= -\frac{1}{6} f_{j-1} +\frac{5}{6}f_{j} + \frac{1}{3} f_{j+1}\\
    f_{j+1/2}^3 &= \frac{1}{3} f_{j} +\frac{5}{6}f_{j+1} + \frac{1}{6} f_{j+2}
    \end{split}
\end{align}
to give
\begin{align}
    f_{j+1/2} = \frac{w_1}{3} f_{j-2} - \frac{1}{6} (7 w_1 + w_2) f_{j-1} + \frac{1}{6}(11 w_1 + 5 w_2 + 2 w_3) f_j + \frac{1}{6} (2 w_2 + 5 w_3)f_{j+1} - \frac{w_3}{6} f_{j+2}.
\end{align}
For a smooth solution field, our weights are optimal at $c_1$ = 1/10, $c_2$ = 6/10, $c_3$ = 3/10 and we recover Equation (\ref{WENO5-1}). Smoothness indicators $\beta_k$ for the scheme (henceforth known as the WENO5 scheme) are given by
\begin{align}
    \label{SmoothFunc}
    \begin{split}
        \beta_1 & = \frac{13}{12}(f_{j-2}-2f_{j-1}+f_j)^2 +
        \frac{1}{4}(f_{j-2}-4f_{j-1}+3f_{j})^2,\\
        \beta_2 & = \frac{13}{12}(f_{j-1}-2f_{j}+f_{j+1})^2 + \frac{1}{4}(f_{j-1}-f_{j+1})^2,\\
        \beta_3 & = \frac{13}{12}(f_{j}-2f_{j+1}+f_{j+2})^2 + \frac{1}{4}(3 f_{j}-4f_{j+1}+f_{j+2})^2.
    \end{split}
\end{align}

\subsection{CRWENO Scheme}
\label{sec:ILES4}

Compact versions of the WENO scheme have also been developed to reduce stencil sizes without compromising accuracy \citep{ghosh2012compact}. They make use of implicit candidate interpolations resulting in smaller stencils providing a better resolution property. The fifth order compact reconstructed WENO scheme (henceforth CRWENO5) is given by:
\begin{align}
    \begin{split}
        \frac{2}{3}f_{j-1/2}^1 + \frac{1}{3} f_{j+1/2}^1 & = \frac{1}{6}(f_{j-1}+5 f_j),\\
        \frac{1}{3}f_{j-1/2}^2 + \frac{2}{3} f_{j+1/2}^2 & = \frac{1}{6}(5 f_{j}+ f_{j+1}),\\
        \frac{2}{3}f_{j+1/2}^3 + \frac{1}{3} f_{j+3/2}^3 & = \frac{1}{6}(f_{j}+5 f_{j+1}).
    \end{split}
\end{align}
A fifth order compact interpolation may thus be devised:
\begin{align}
    \begin{gathered}
    (\frac{2}{3}w_1 + \frac{1}{3}w_2)f_{j-1/2} +\left[\frac{1}{3}w_1 + \frac{2}{3}(w_2 + w_3)\right]f_{j+1/2} + \frac{1}{3}f_{j+3/2} =\\ \frac{w_1}{6}f_{j-1} + \frac{5(w_1 + w_2) + w_3}{6} f_j + \frac{w_2 + 5 w_3}{6}f_{j+1},
    \end{gathered}
\end{align}
which under the optimal weights of $c_1 = 1/5$, $c_2 = 1/2$, $c_3 = 3/10$ result in our previously defined fifth order compact interpolation (the CU5 scheme) in Equation (\ref{cu5}).

\begin{figure}[!t]
\centering
\mbox{
\subfigure[Dispersion]{\includegraphics[width=0.45\textwidth]{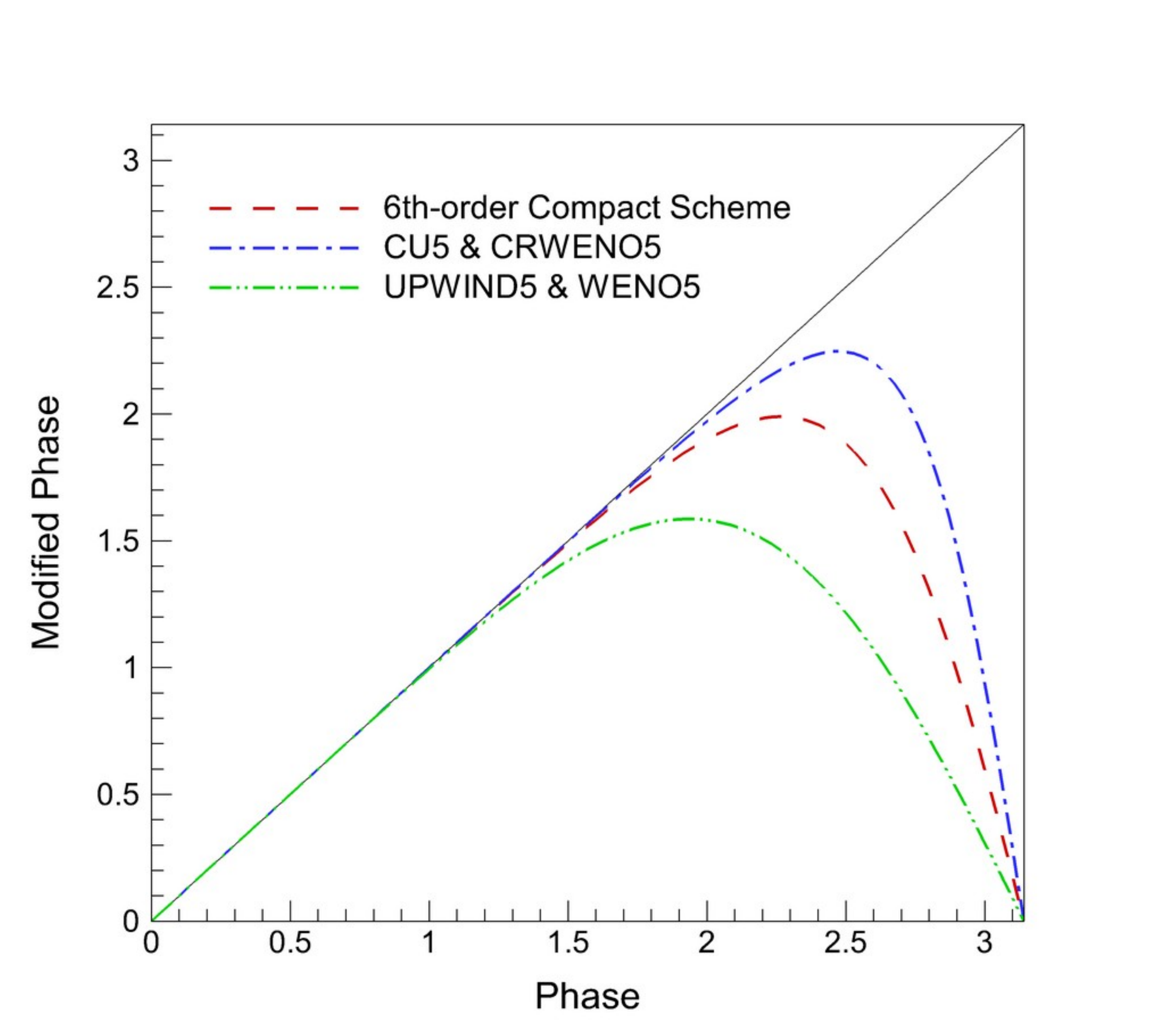}}
\subfigure[Dissipation]{\includegraphics[width=0.45\textwidth]{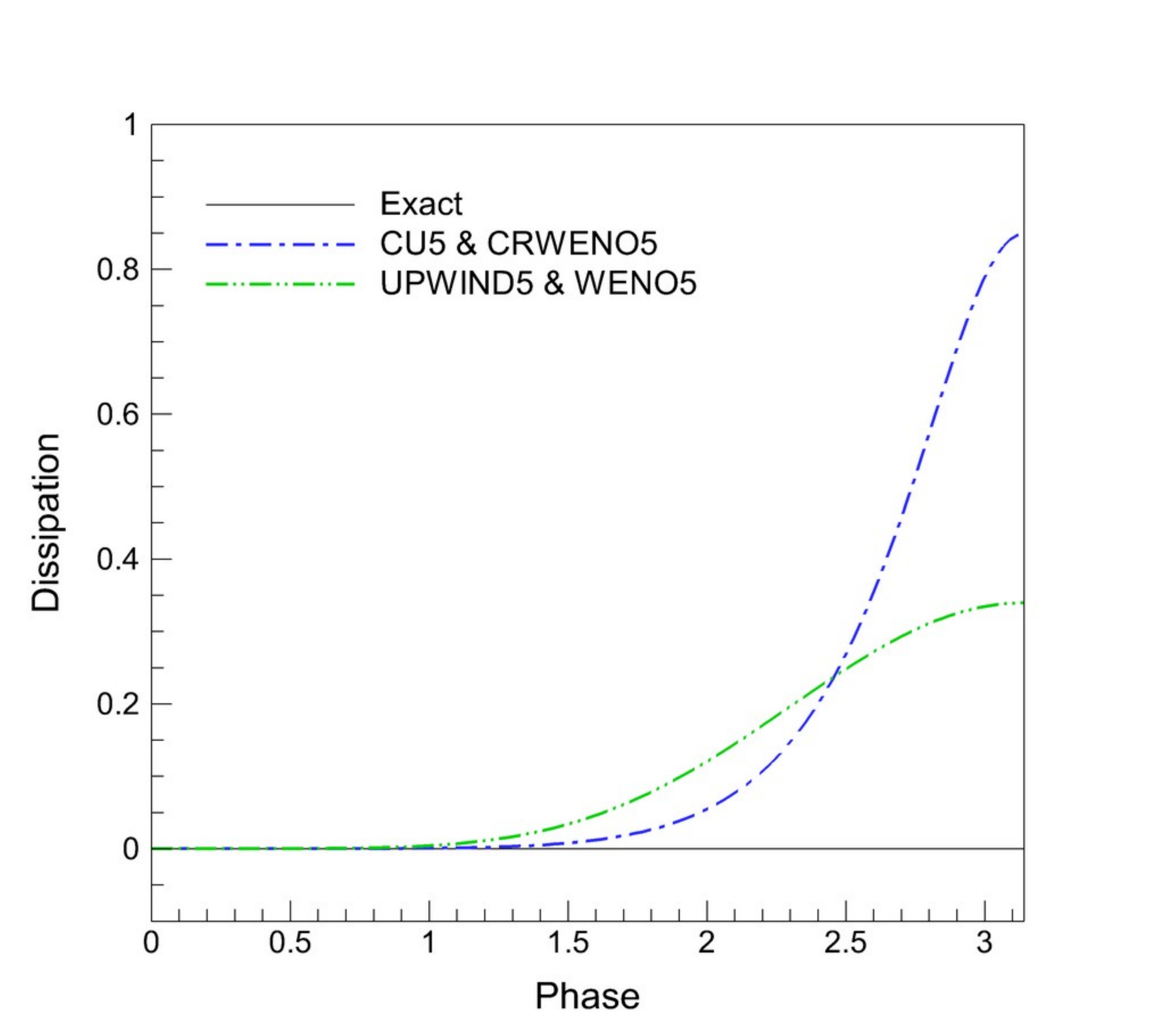}}
}
\caption{Spectral characteristics of the fifth-order upwind (WENO5 with linear weights) and compact upwind (CRWENO5 with linear weights) schemes.}
\label{fig:tf1}
\end{figure}

The smoothness indicators for the CRWENO5 and WENO5 schemes are obtained in the same manner (i.e., using Equation (\ref{SmoothFunc})). An illustration of the dispersion and dissipation of both the WENO5 and CRWENO5 schemes is shown in Figure (\ref{fig:tf1}) using the optimal linear weights. An in-depth discussion of compact reconstructed WENO schemes and their dispersion and dissipation characteristics may be found in \cite{peng2015improvement}.

\section{Results}
\label{sec:results}

This section covers the results of the AD-LES and ILES methods outlined in the previous sections for the 1D viscous Burgers equation. Using the standard LES methodology, coarse grained LES resolutions were compared to high fidelity DNS data obtained from much higher resolutions which capture all physical scales.
In this study, the efficacy of eddy viscosity models (EV-LES, DEV-LES, ADEV-LES), AD-LES (used with and without secondary relaxation filtering) equipped with various low-pass filters, AD regularization with the Smagorinsky eddy viscosity, EF model (i.e., applying explicit relaxation filters with no use of AD procedure), and various ILES models (UPWIND5, CU5, WENO5, and CRWENO5) were considered. The binomial and Pad\'{e} class of filters were investigated for a wide range of parameters. All LES models were compared to the underresolved DNS (UDNS) data at the same (coarse) resolution to assess their modeling characteristics in a systematic manner.

The decaying Burgers turbulence problem is considered in a  domain of $x \in [0,2\pi]$ with periodic boundary conditions. The initial conditions of this problem are given by an initial energy spectrum as follows:
\begin{align}
    E(k) = Ak^4 \textnormal{exp} (-(k/k_0)^2),
\end{align}
where the constant A is set to the following value
\begin{align}
    A = \frac{2k_0^{-5}}{3\sqrt{\pi}}
\end{align}
which ensures a total energy $\int E(k) dk = 1/2$ at the initial condition. The parameter $k_0$ is assumed to be 10 and is the wavenumber at which the peak value of the energy spectrum is obtained. The velocity magnitudes in Fourier space can be expressed as a function of the initial energy spectrum by
\begin{align}
    |\hat{u}(k)| = \sqrt{2 E(k)}.
\end{align}
Several realization of an initial velocity field are then obtained by incorporating a random phase:
\begin{align}
    \hat{u}(k) = \sqrt{2 E(k)} \textnormal{exp} (i 2\pi \Psi (k)),
\end{align}
where $\Psi(k)$ is a uniform random number distribution between 0 and 1 at each wavenumber. This distribution also has to satisfy the $\Psi(k) = -\Psi(-k)$ conjugate relationship in order to obtain a real velocity field in physical space \citep{san2012high}. Inversions from Fourier space are computed using a Fast Fourier transform algorithm (FFT) given by \cite{press1992numerical}. Randomly selected 64 sample fields were constructed with different phases and simulated till the energy content of the flow was considerably lower than its initial value. Ensemble averaged results for the sample simulations were computed and presented in the following. It must be noted that the initial conditions remained identical for all inter-comparative simulations through the use of identical random number seeds. An investigation is carried out on the energy spectrum which is a function of wavenumber and time and is defined as
\begin{align}
    E(k,t) = \frac{1}{2}  |\hat{u}(k,t)|^2,
\end{align}
and the total energy can be computed as
\begin{align}
    E(t) = \int_{-k_m}^{k_m} E(k,t) dk.
\end{align}
It is also possible to analyze the performance of the numerical methods using other measures such as the total dissipation rate, $-dE(t)/dt$, (i.e., also given by $D(t) = \nu \int_{-k_m}^{k_m} k^2 |\hat{u}(k,t)|dk$). A discussion on the use of various forms of dissipation rate formulas can be found in \cite{debonis2013solutions} and \cite{san2016analysis}.

\begin{figure}[!t]
\centering
\mbox{
\subfigure[Energy spectra]{\includegraphics[width=0.45\textwidth]{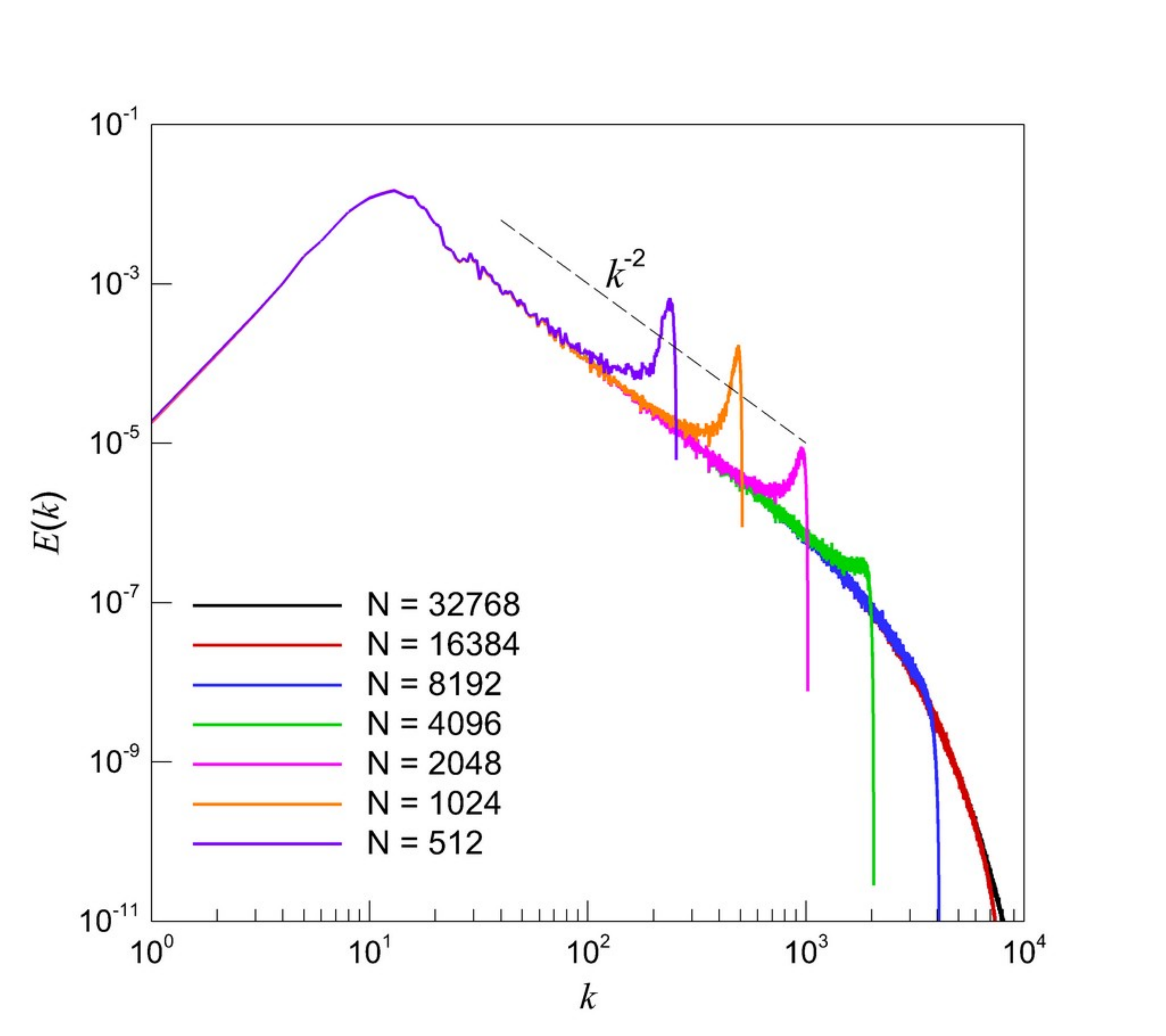}}
\subfigure[Dissipation rates]{\includegraphics[width=0.45\textwidth]{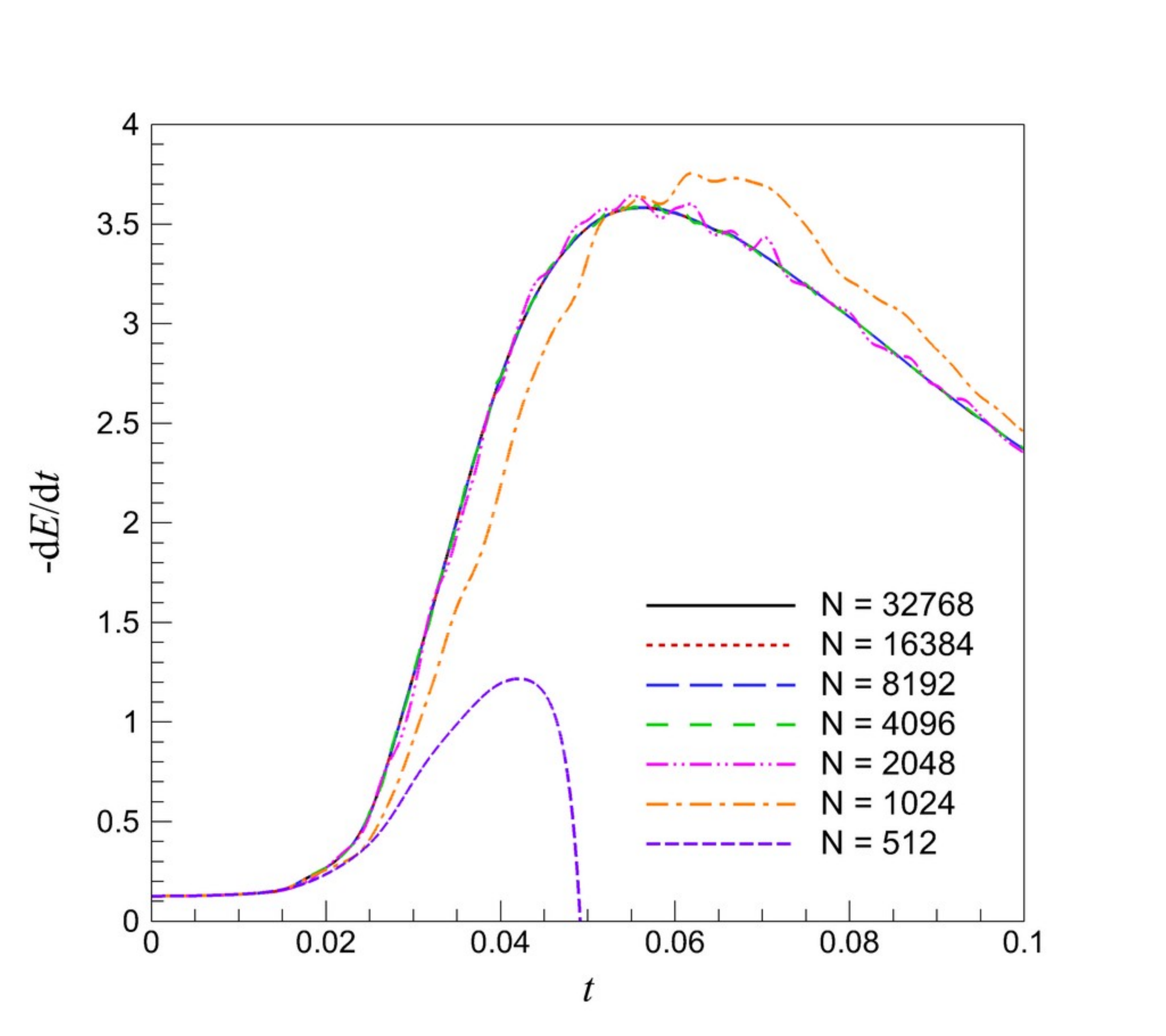}}
}
\caption{Comparison of energy spectra (at time $t=0.05$) and time evolution of dissipation rates for the decaying Burgers turbulence problem at $\nu=5 \times 10^{-4}$ with varying grid resolution from very coarse mesh ($N=512$) to fully resolving DNS mesh ($N=32768$). Ideal scaling $k^{-2}$ is also included in the energy spectra plot.}
\label{fig:udns}
\end{figure}

\subsection{DNS and UDNS results}

Prior to analyzing the different numerical methods presented in the previous sections, it is important to carry out a grid convergence investigation to determine DNS resolution requirements for our chosen viscosity scale of $\nu = 5 \times 10^{-4}$. The numerical experiment was carried out till a time of 0.1 beyond which the energy dissipation rates reaches its maximum value. Resolutions of $2^9$ to $2^{15}$ were tested and it is seen that the highest resolution can be considered to be a fully resolved DNS data. Figure (\ref{fig:udns}) shows the energy spectra for DNS and UDNS computations of different resolutions and it can be seen that the ideal $k^{-2}$ scaling is seen at inertial range. One must note that $t=0.05$ corresponds to approximately the time for the maximum dissipation rate of the problem. The problem of energy pile-up close to grid cut-off scale is also evident at lower resolution cases, where one can notice a considerable deviation from the fully resolved spectrum. Therefore, we perform our LES analysis by using a set of resolutions of 512, 1024 and 2048 grid points.

\subsection{EV-LES results}
\begin{figure}[!t]
\centering
\mbox{
\subfigure[$C_s = 0.2 $]{\includegraphics[width=0.45\textwidth]{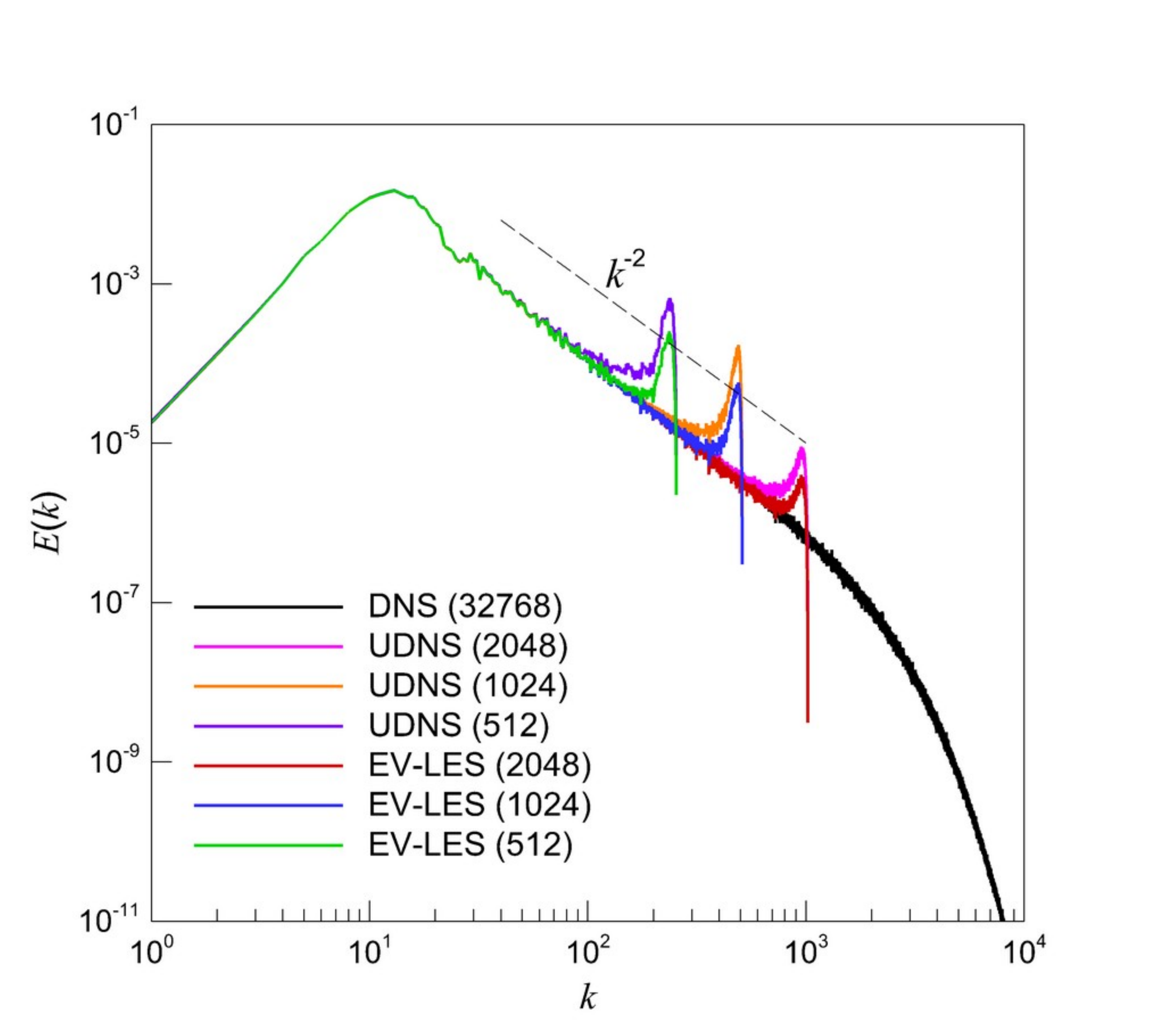}}
\subfigure[$C_s = 0.3 $]{\includegraphics[width=0.45\textwidth]{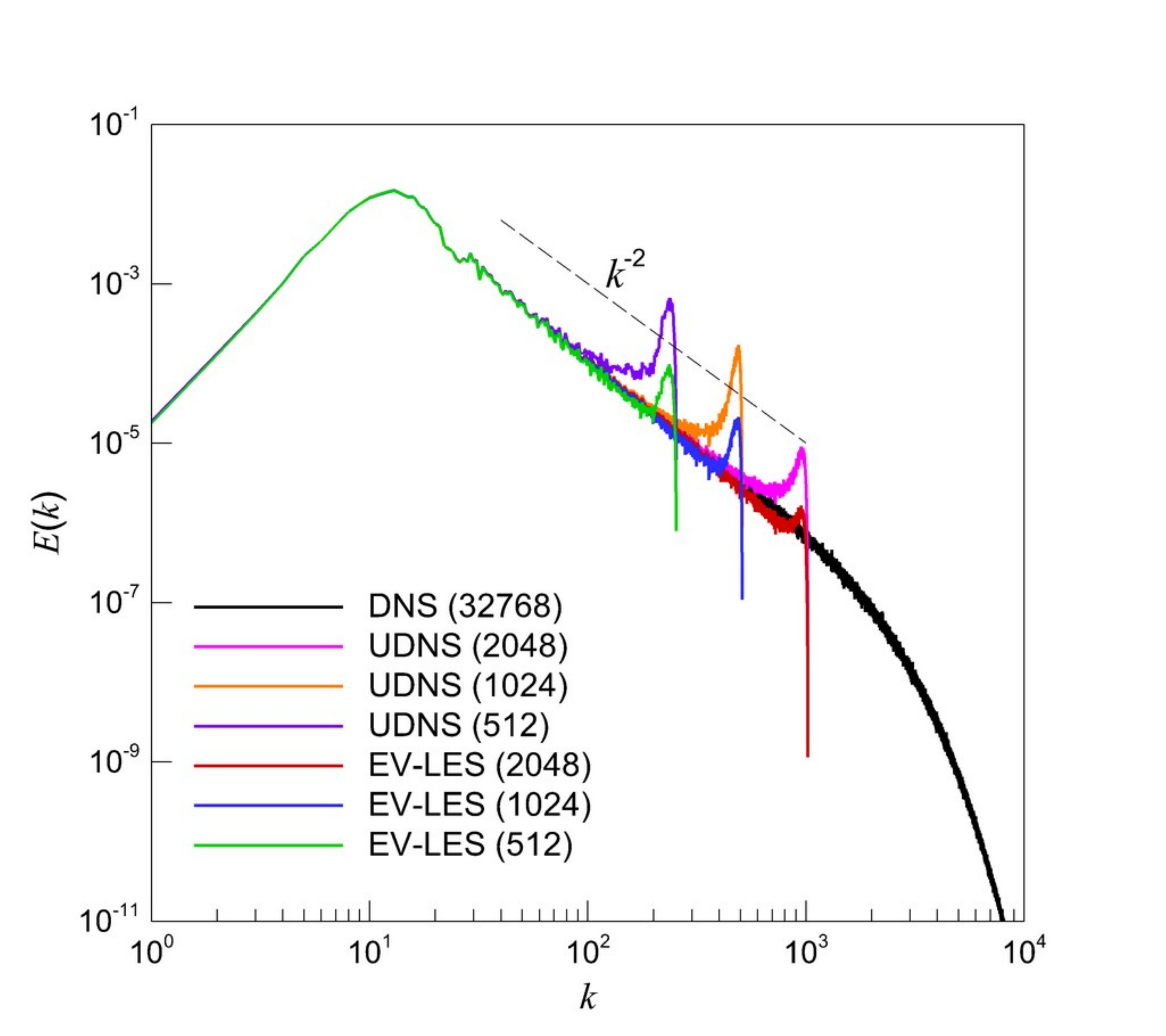}}
}\\
\mbox{
\subfigure[$C_s = 0.4 $]{\includegraphics[width=0.45\textwidth]{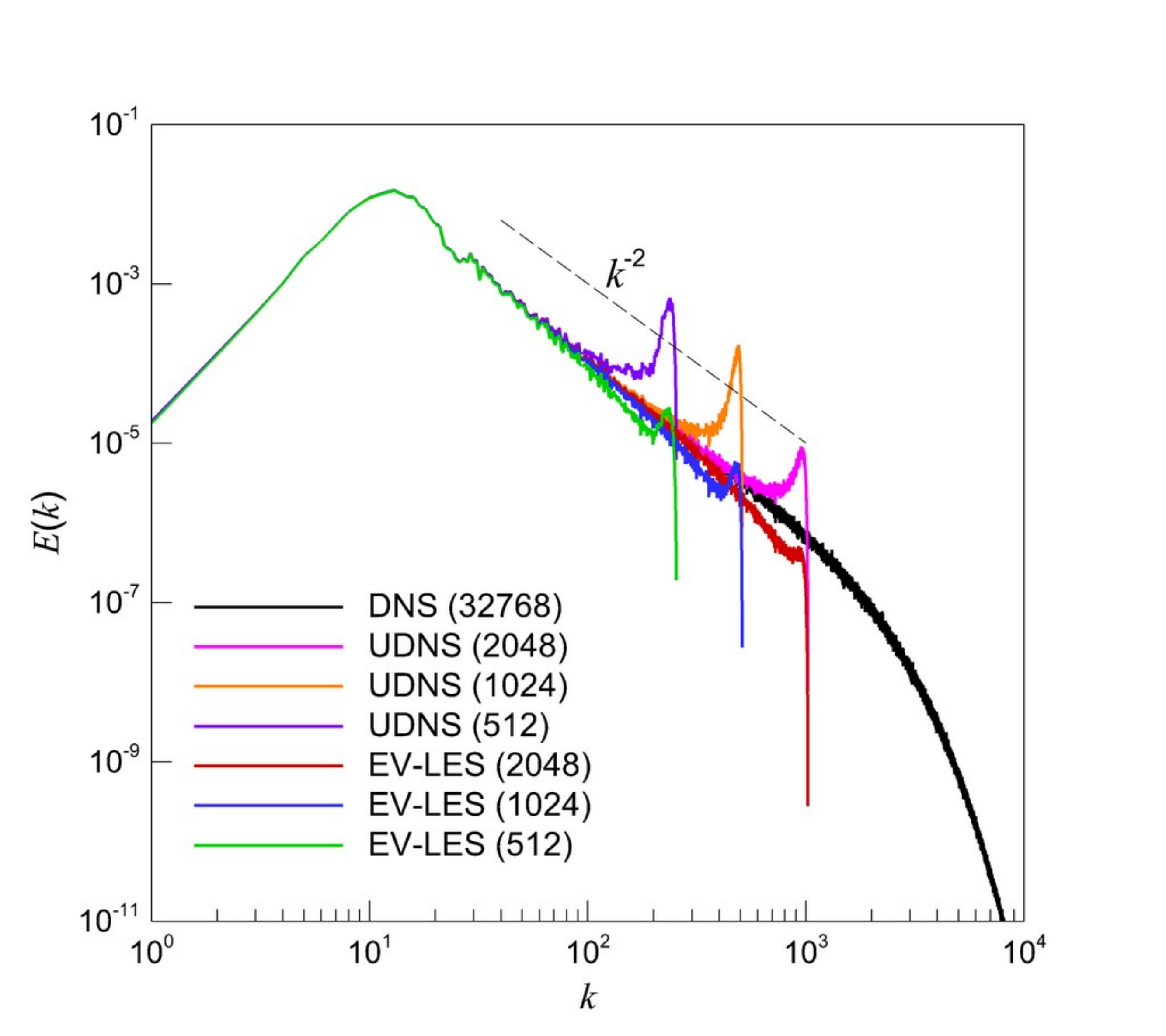}}
\subfigure[$C_s = 0.5 $]{\includegraphics[width=0.45\textwidth]{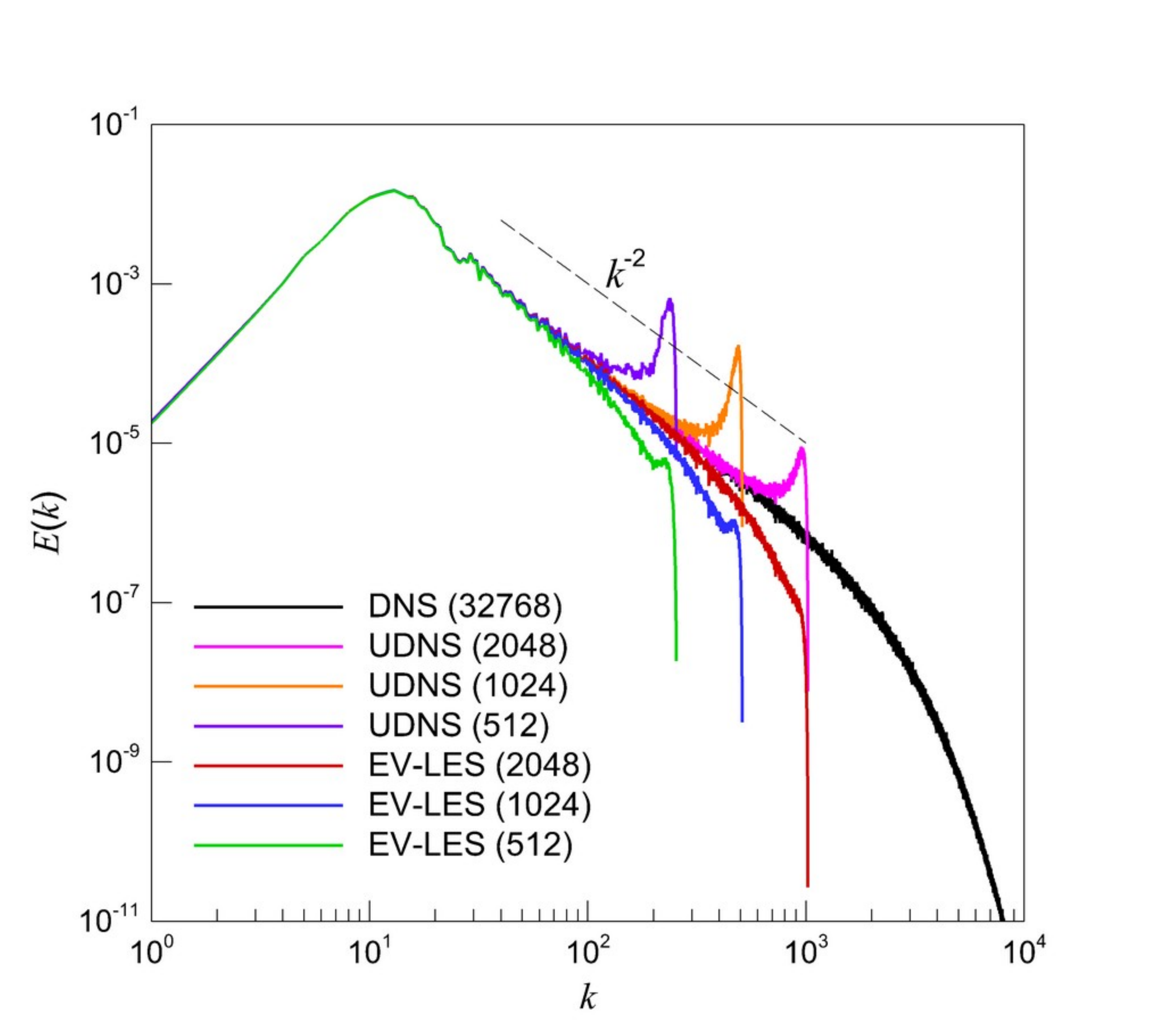}}
}
\caption{Energy spectra obtained by the eddy viscosity (EV) model for various Smagorinsky constant.}
\label{fig:a1}
\end{figure}

First, the well established Smagorinsky LES model was used to perform coarse simulations and compared to the the resolved DNS computations as shown in Figure (\ref{fig:a1}). For the purpose of comparison, underresolved DNS (UDNS) results are also shown. As mentioned earlier, the process of adding dissipation in this model is through the choice of the Smagorinsky coefficient which controls the point-wise eddy viscosity. As shown in the figure, the established value of the Smagorinsky coefficient ($C_s$ = 0.2) was unable to prevent the phenomenon of pile-up. However, we note that this value has been suggested for homogeneous three-dimensional (3D) Kolmogorov turbulence. One can also argue that the absolute value of the strain rate tensor in 3D setting has a scaling factor of $\sqrt{2}$ in its standard form. Therefore, we perform additional numerical experiments with increasing $C_s$ values for Burgers turbulence where shocks dominate the flow field. Thus, in order to add more dissipation, the coefficient was increased at steps of 0.1 but it is seen that the most accurate prediction for the energy spectrum is given at $C_s = 0.3$ with a resolution of $N = 2048$. A small amount of energy accumulation is still seen at this optimal configuration for the model. A higher value of $C_s$ adds more than the requisite dissipation required and produces steeper scaling in energy spectrum at almost all inertial length scales for $C_s = 0.5$.

\begin{figure}[!t]
\centering
\mbox{
\subfigure[Test filter: $B^2$]{\includegraphics[width=0.45\textwidth]{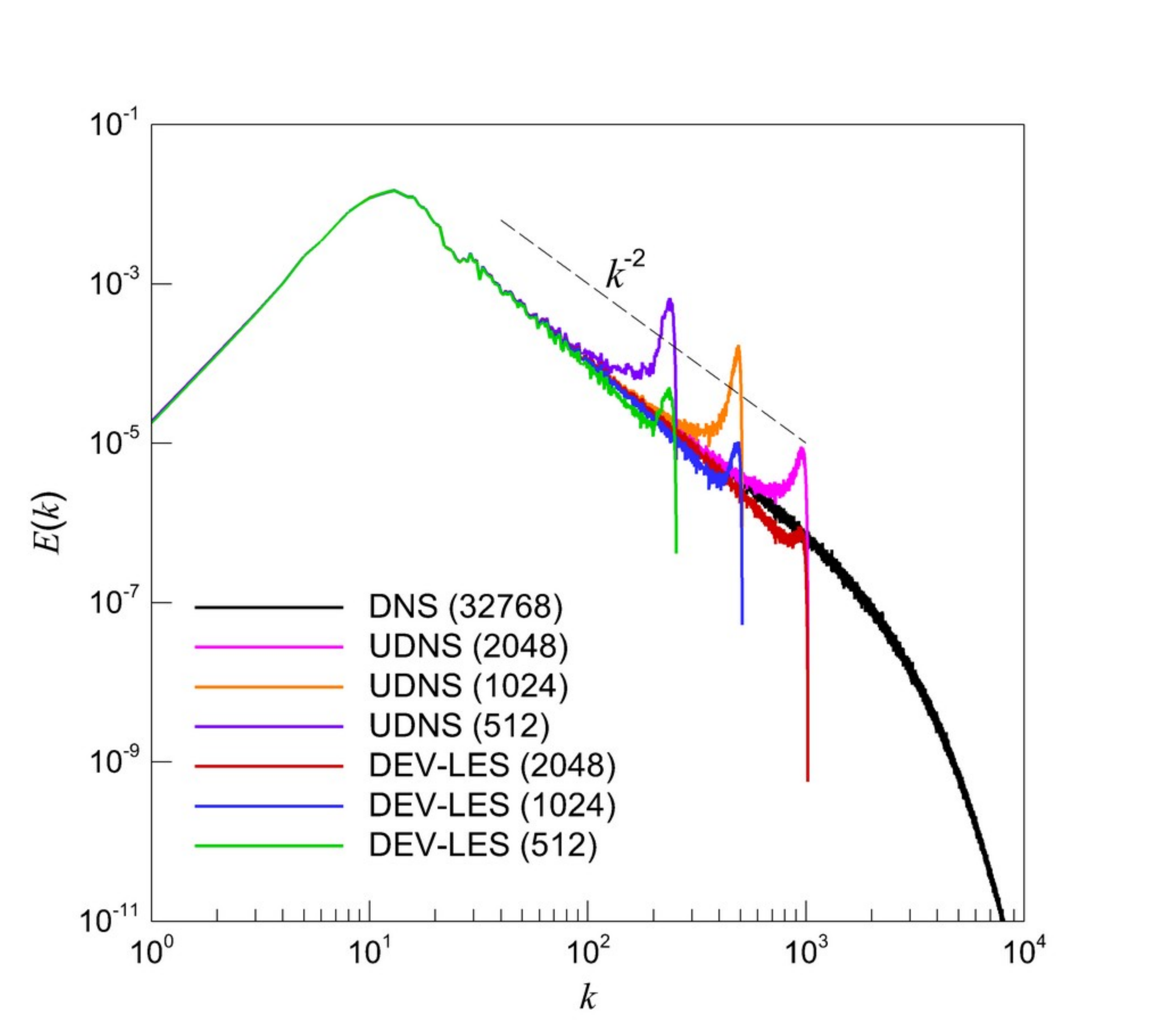}}
\subfigure[Test filter: $B^2$ ]{\includegraphics[width=0.45\textwidth]{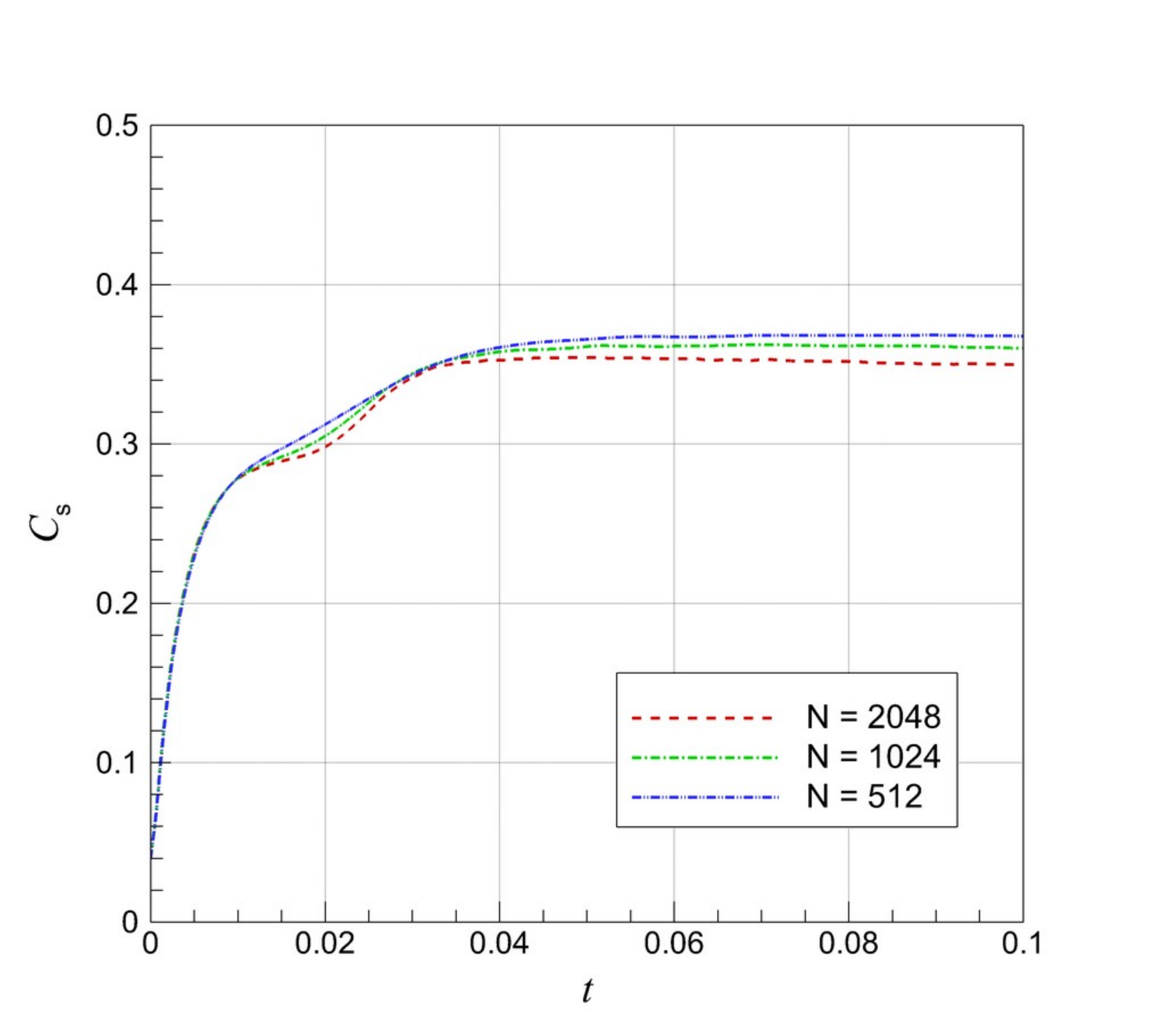}}
}\\
\mbox{
\subfigure[Test filter: $^{(3,1)}B$]{\includegraphics[width=0.45\textwidth]{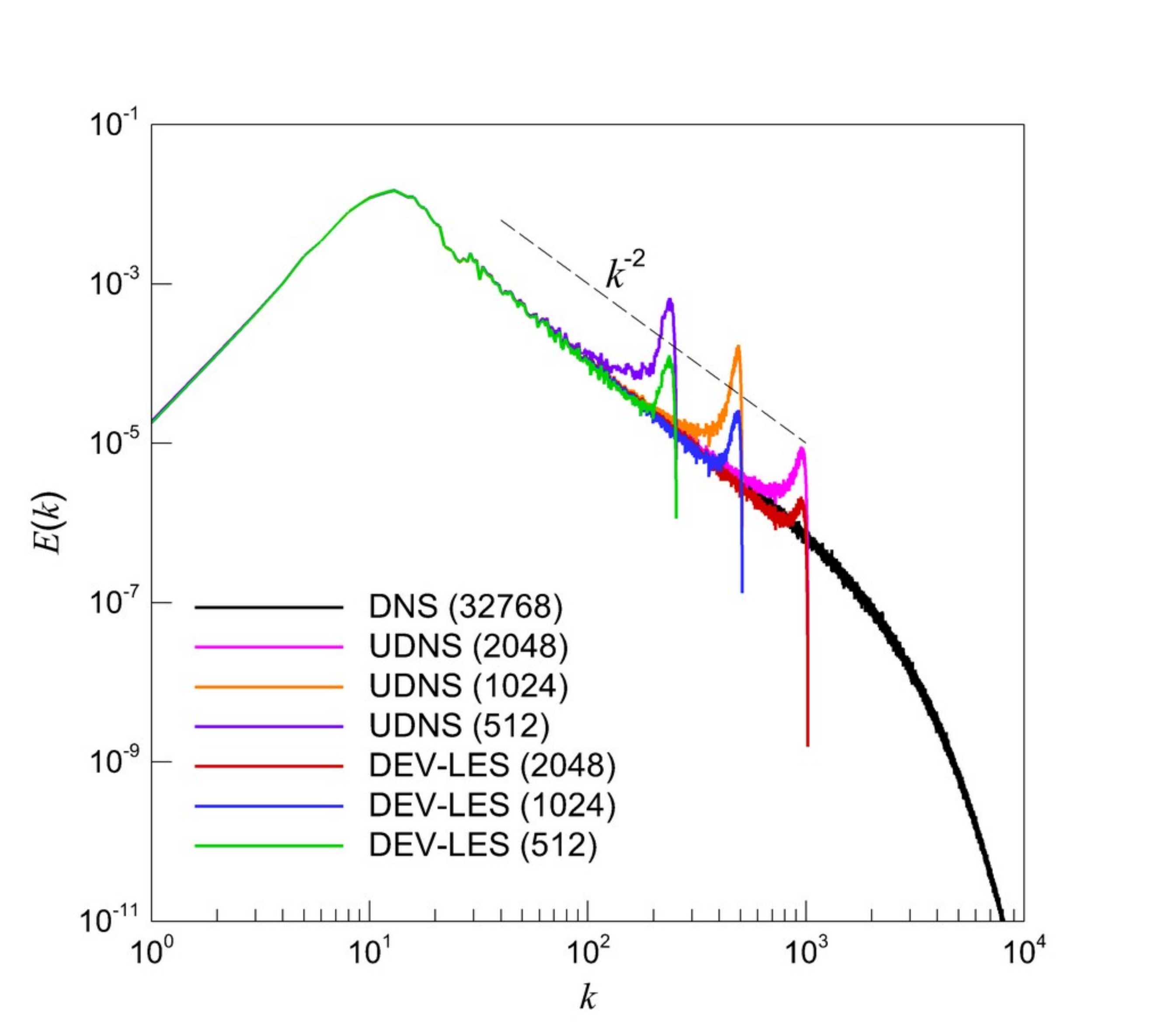}}
\subfigure[Test filter: $^{(3,1)}B$]{\includegraphics[width=0.45\textwidth]{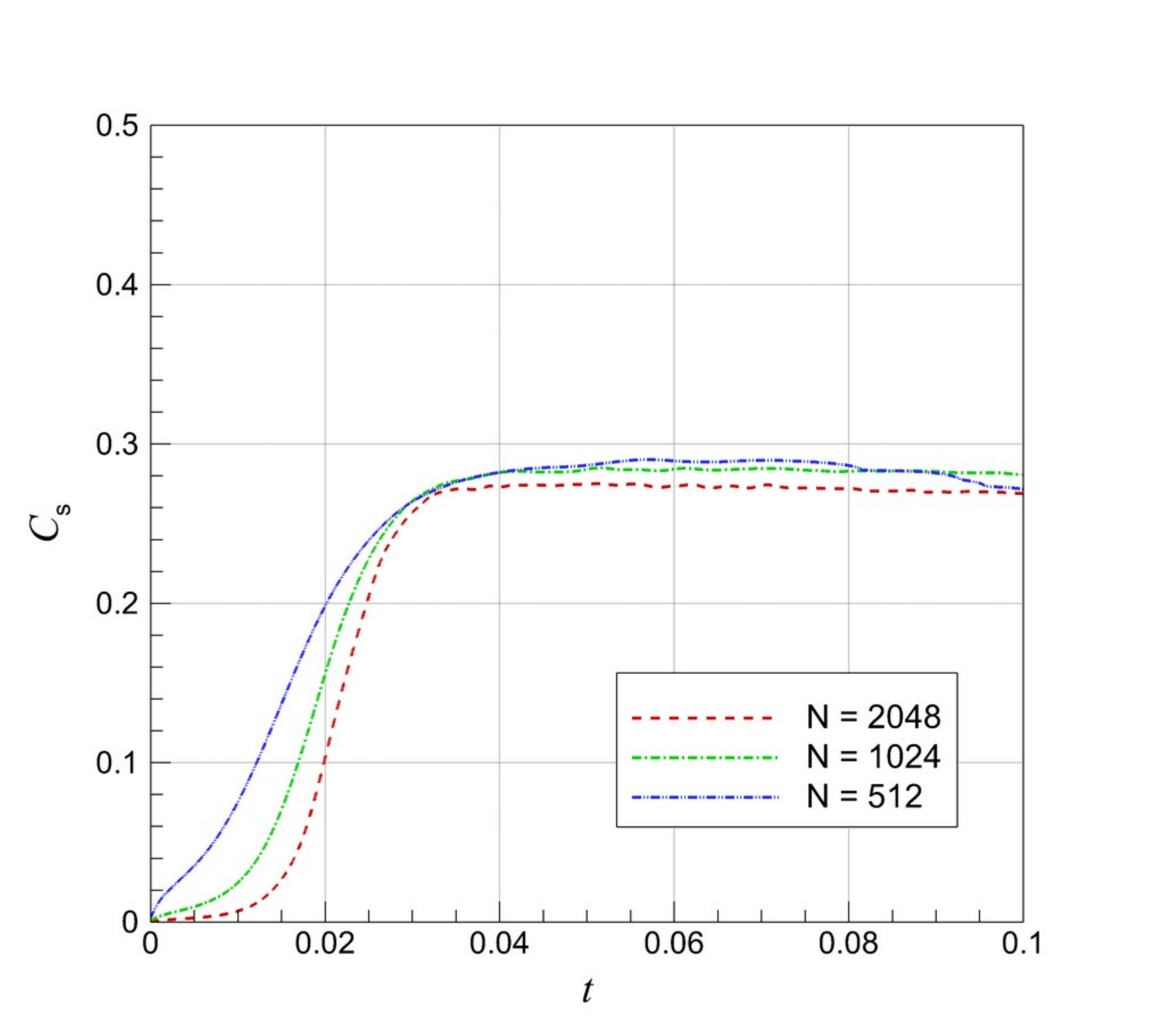}}
}
\caption{Energy spectra (left) and evolution of the Smagorinsky constant (right) obtained by the dynamic eddy viscosity (DEV) model using various test filters.}
\label{fig:a1b}
\end{figure}

\begin{figure}[!t]
\centering
\mbox{
\subfigure[Test filter: $B^2$]{\includegraphics[width=0.45\textwidth]{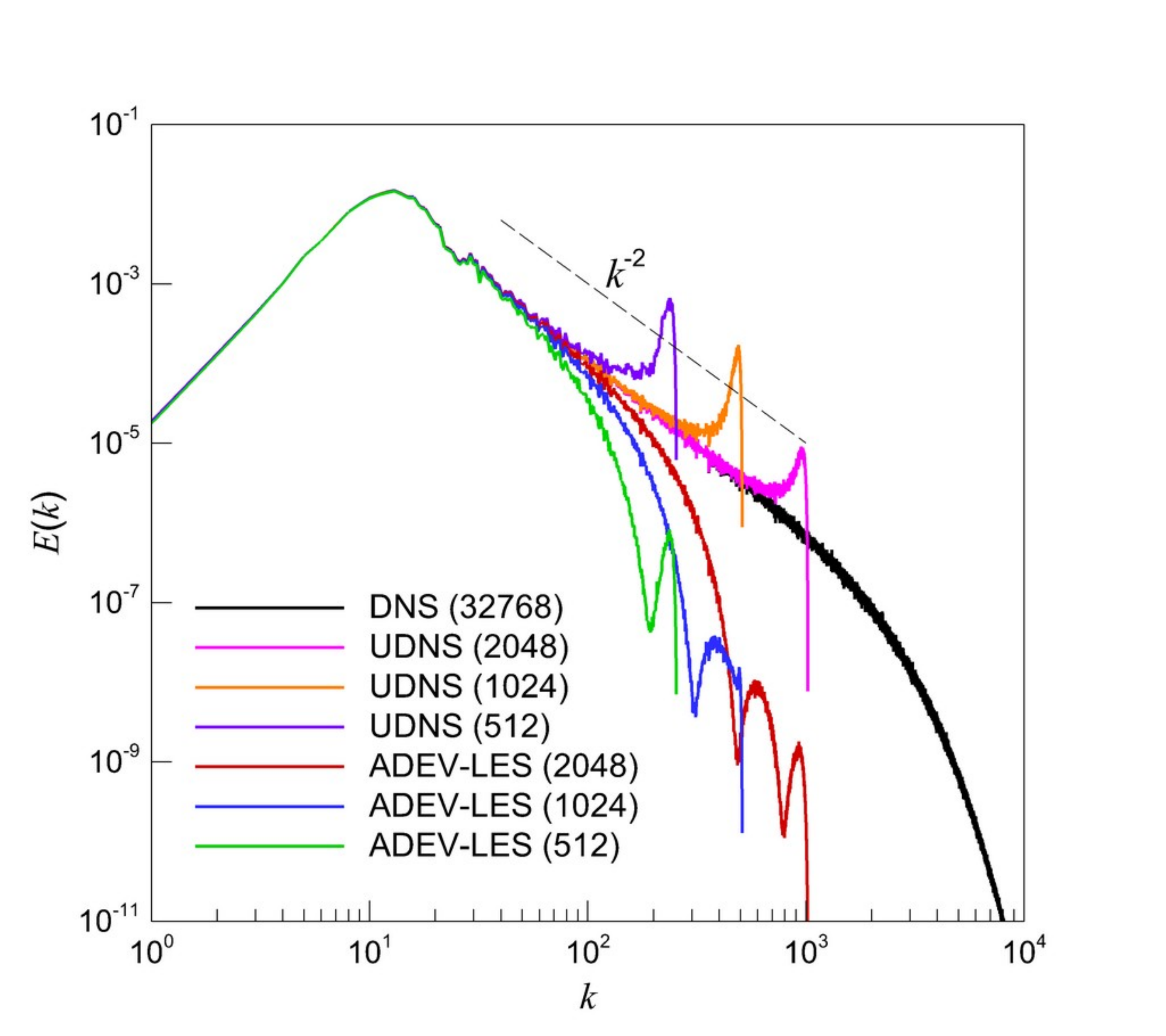}}
\subfigure[Test filter: $B^2$ ]{\includegraphics[width=0.45\textwidth]{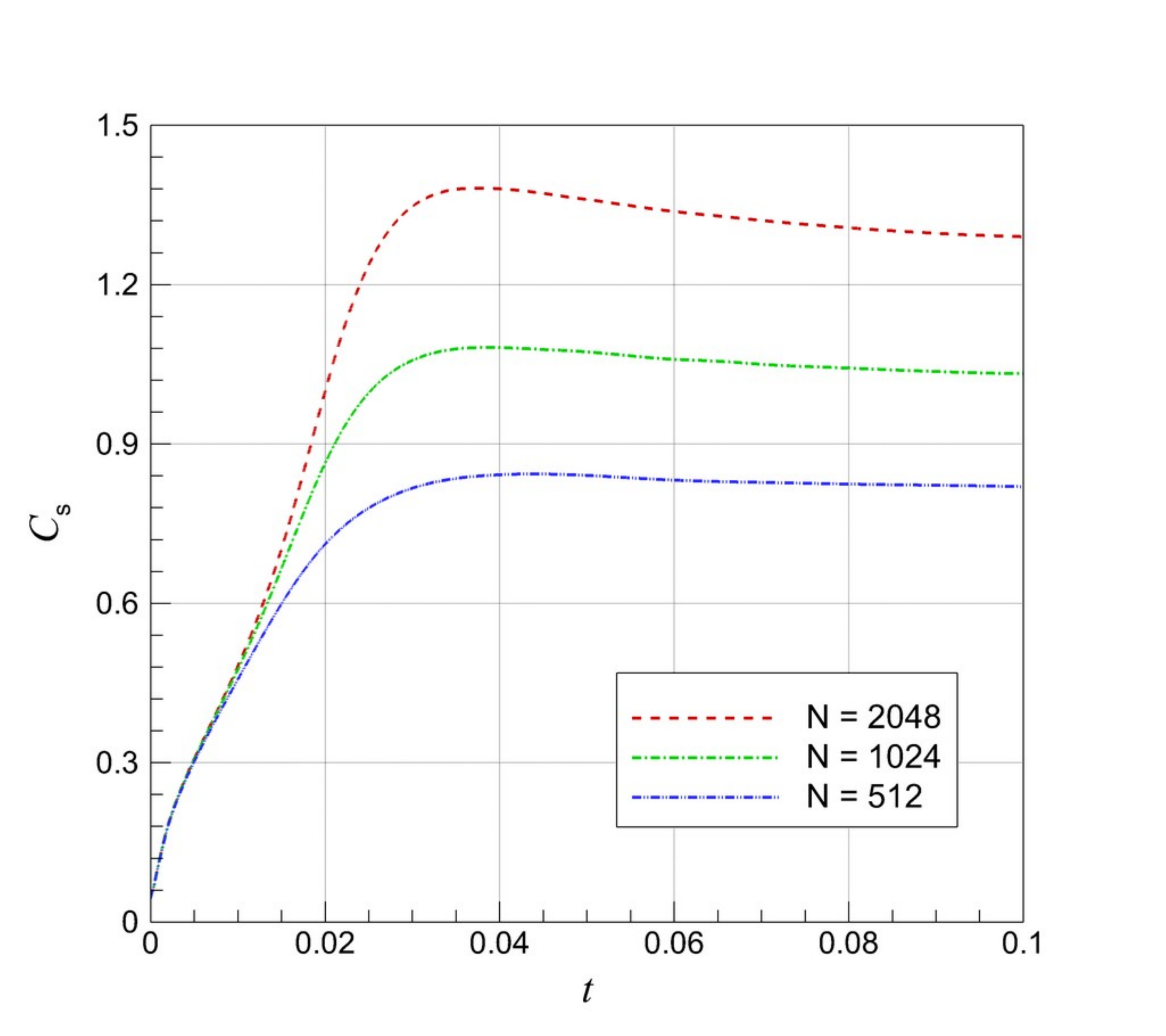}}
}\\
\mbox{
\subfigure[Test filter: $^{(3,1)}B$]{\includegraphics[width=0.45\textwidth]{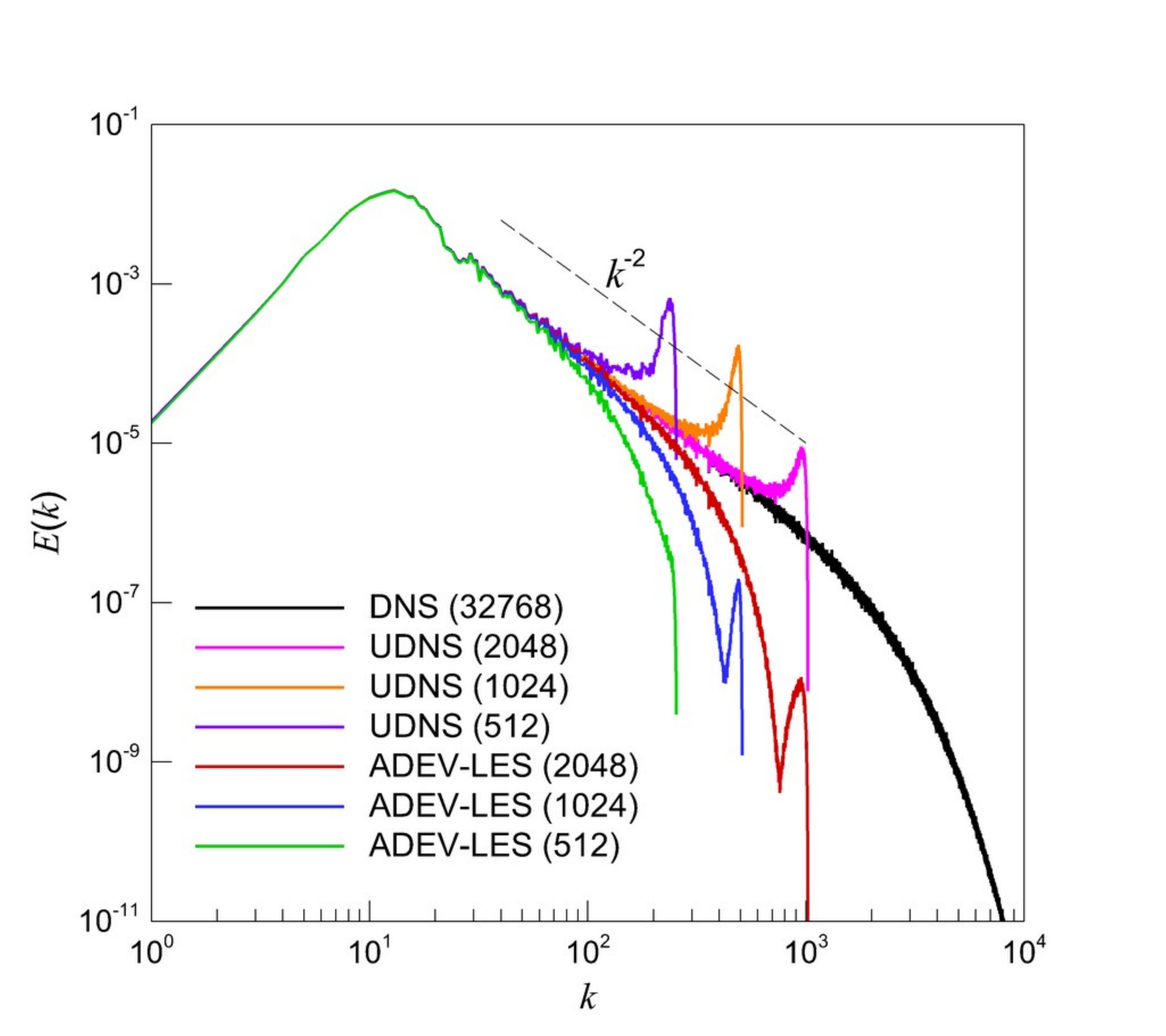}}
\subfigure[Test filter: $^{(3,1)}B$]{\includegraphics[width=0.45\textwidth]{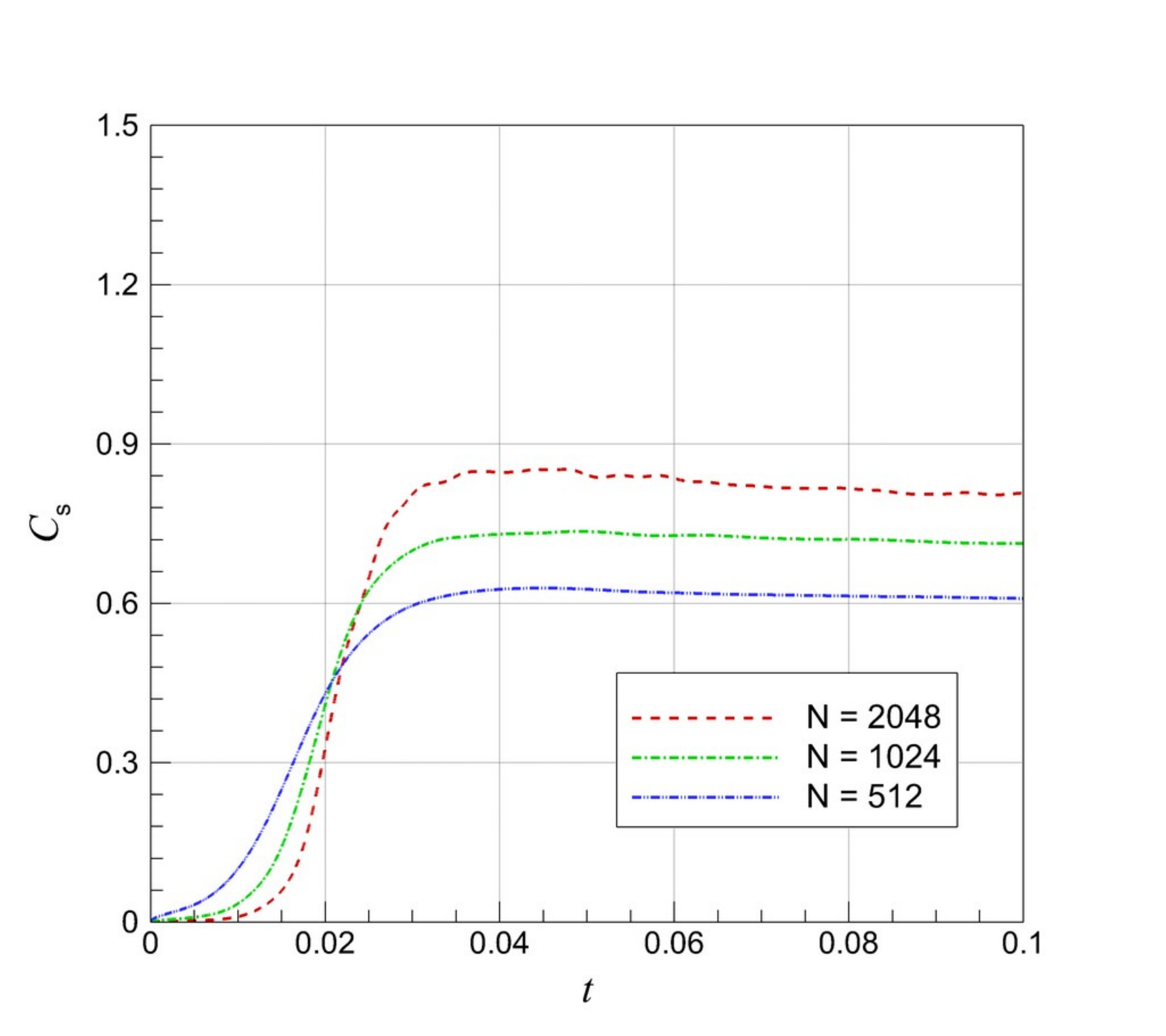}}
}
\caption{Energy spectra (left) and evolution of the Smagorinsky constant (right) obtained by the averaged dynamic eddy viscosity (ADEV) model using various test filters.}
\label{fig:a1c}
\end{figure}

Next, we evaluate the dynamic version of the EV model. The numerical experiments are conducted by utilizing two different test filters: $B^2$ binomial filter (i.e., also known as trapezoidal box filter), and $^{(3,1)}B$ binomial shooting filter (i.e., equivalent to the Pad\'{e} filter when $\alpha=0$). Energy spectra and time histories of the Smagorinsky constant obtained by the DEV-LES model are illustrated in Figure (\ref{fig:a1b}) showing a significant improvement compared to the EV-LES. It is clear that DEV-LES equipped with $^{(3,1)}B$ test filter yields less dissipative results than the model equipped with $B^2$ filter. This can be interpreted that the filter $B^2$ removes a larger amount of energy from the large
scales (e.g., see Figure (\ref{fig:tf}) for filters transfer functions). This manifest itself in evolution of the the Smagorinsky constant (i.e., dynamic constant) showing some dependance on the choice of the filter. As also demonstrated in the figure that the variation of the dynamic constant slightly depends on the resolution making the dynamic model robust to the grid resolution. An interested reader is referred to the study of \cite{najjar1996study} where various discrete test filters and finite difference approximations for the dynamic subgrid-scale stress model has been investigated for turbulent channel flow simulations.

We conduct additional computations with a variant of the dynamic model (ADEV-LES) using an averaged value for the absolute strain rate term as described in Section \ref{sec:dyn}. This type of averaging procedure has been utilized by \cite{fauconnier2009family}. The results for ADEV-LES are shown in Figure (\ref{fig:a1c}) in terms of energy spectra and time histories of the dynamic constant using the same test filters. It is evident on examination of Figure (\ref{fig:a1c}) that the averaging procedure in dynamic model effectively eliminates energy pile-up at all resolution but provides an excessive dissipation to the large part of the inertial range.

\subsection{AD-LES and RF results}

\begin{figure}[!t]
\centering
\mbox{
\subfigure[$B^2 $]{\includegraphics[width=0.45\textwidth]{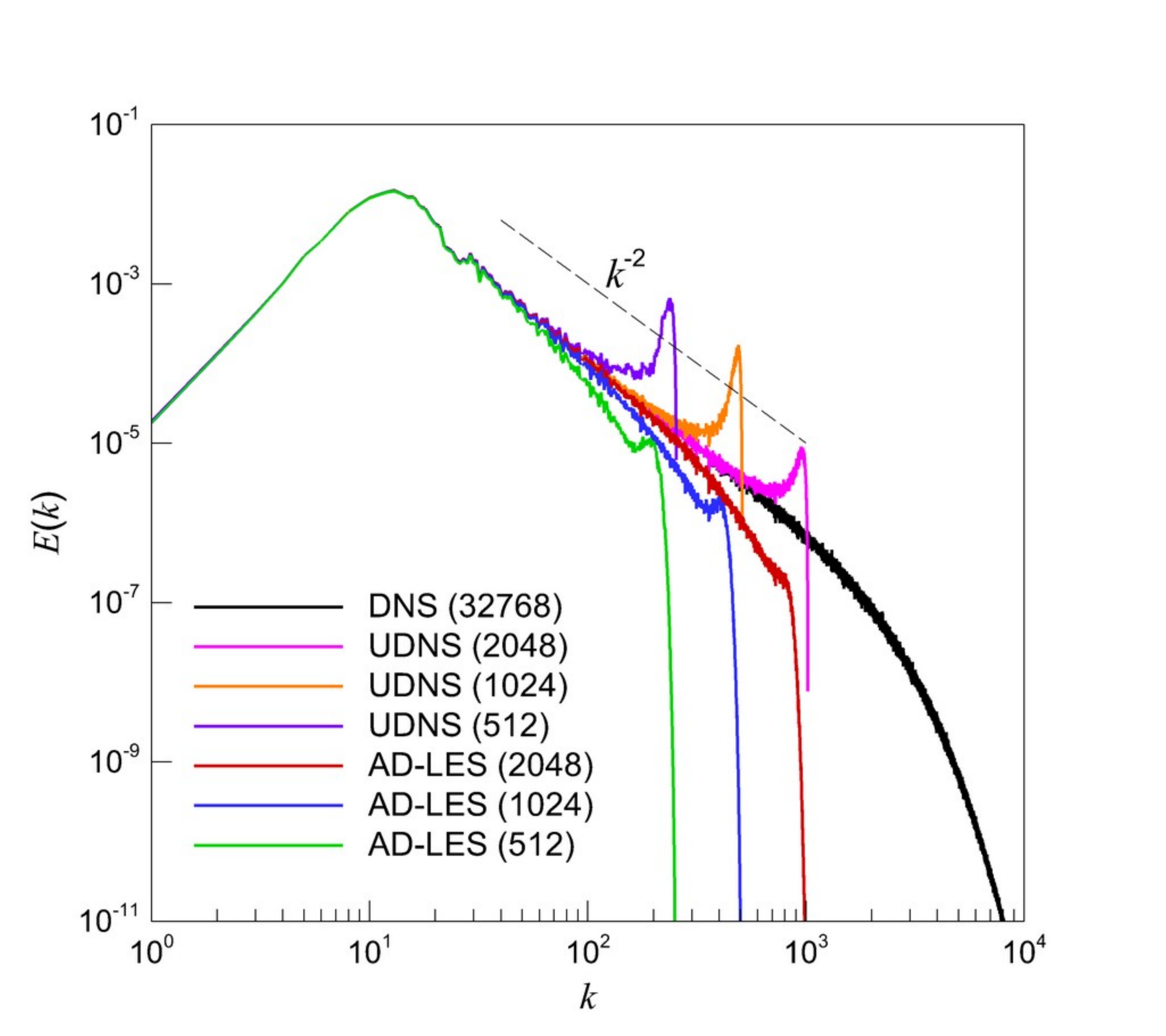}}
\subfigure[$B^4 $]{\includegraphics[width=0.45\textwidth]{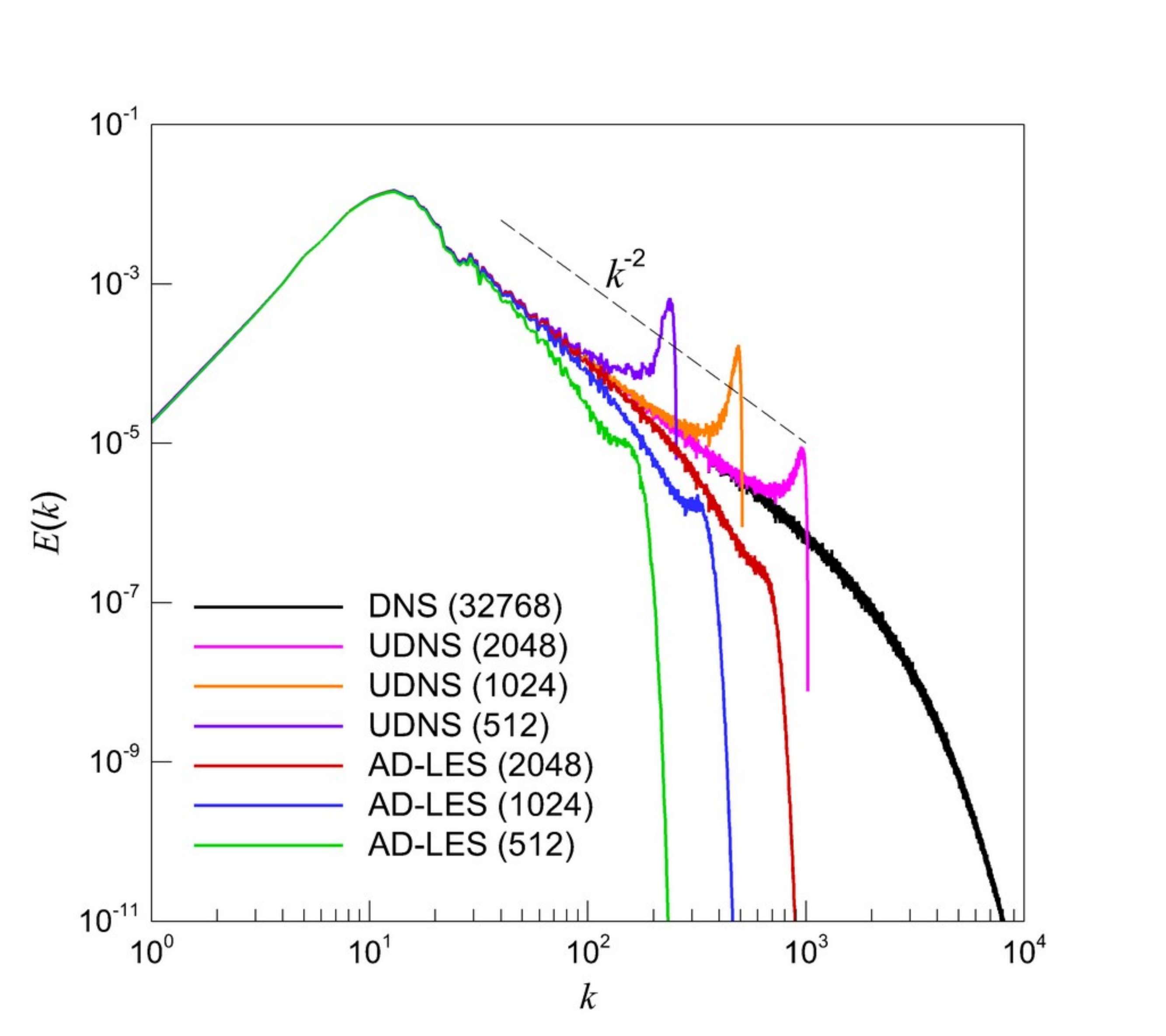}}
}\\
\mbox{
\subfigure[$B^6 $]{\includegraphics[width=0.45\textwidth]{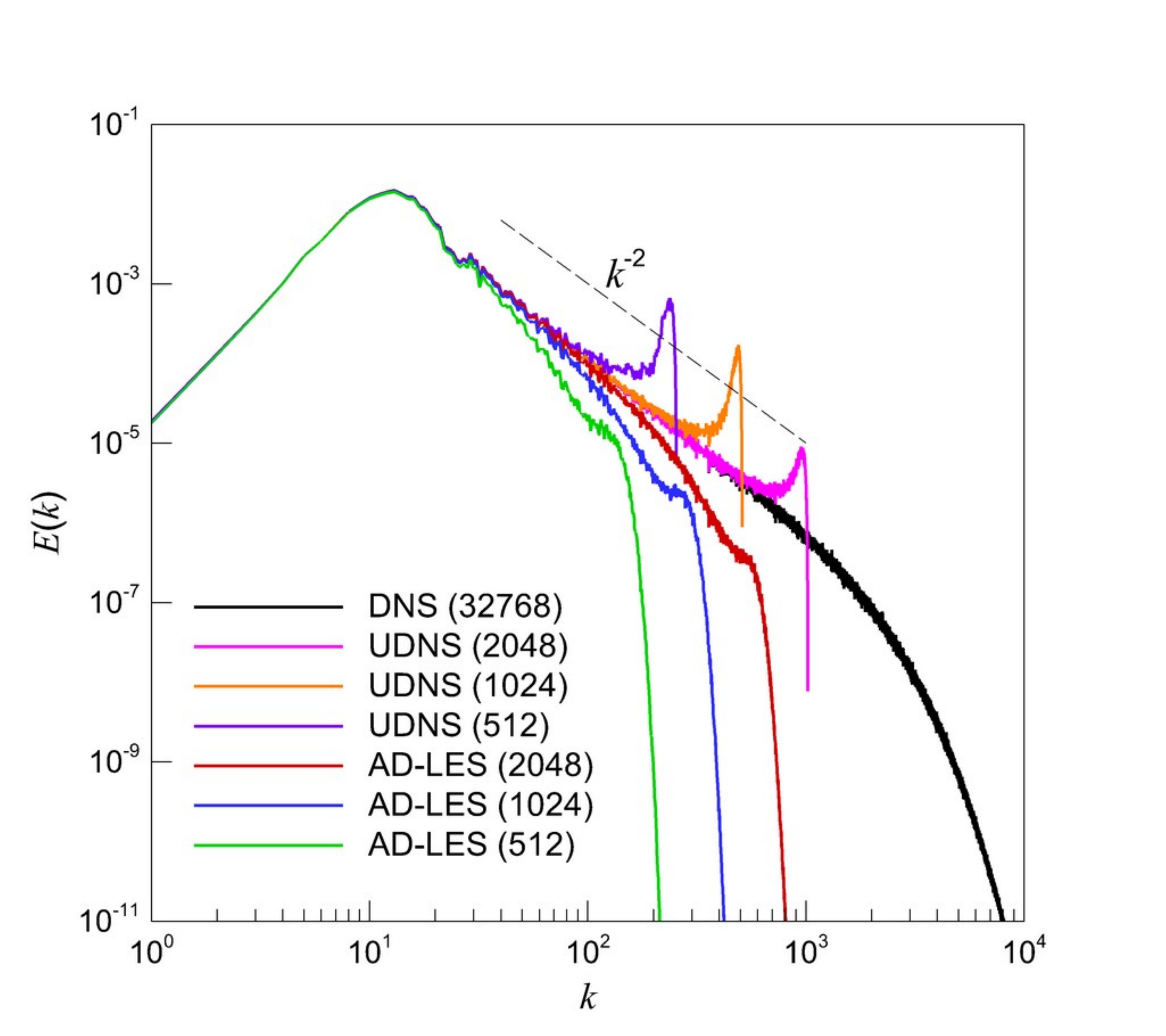}}
\subfigure[$B^8 $]{\includegraphics[width=0.45\textwidth]{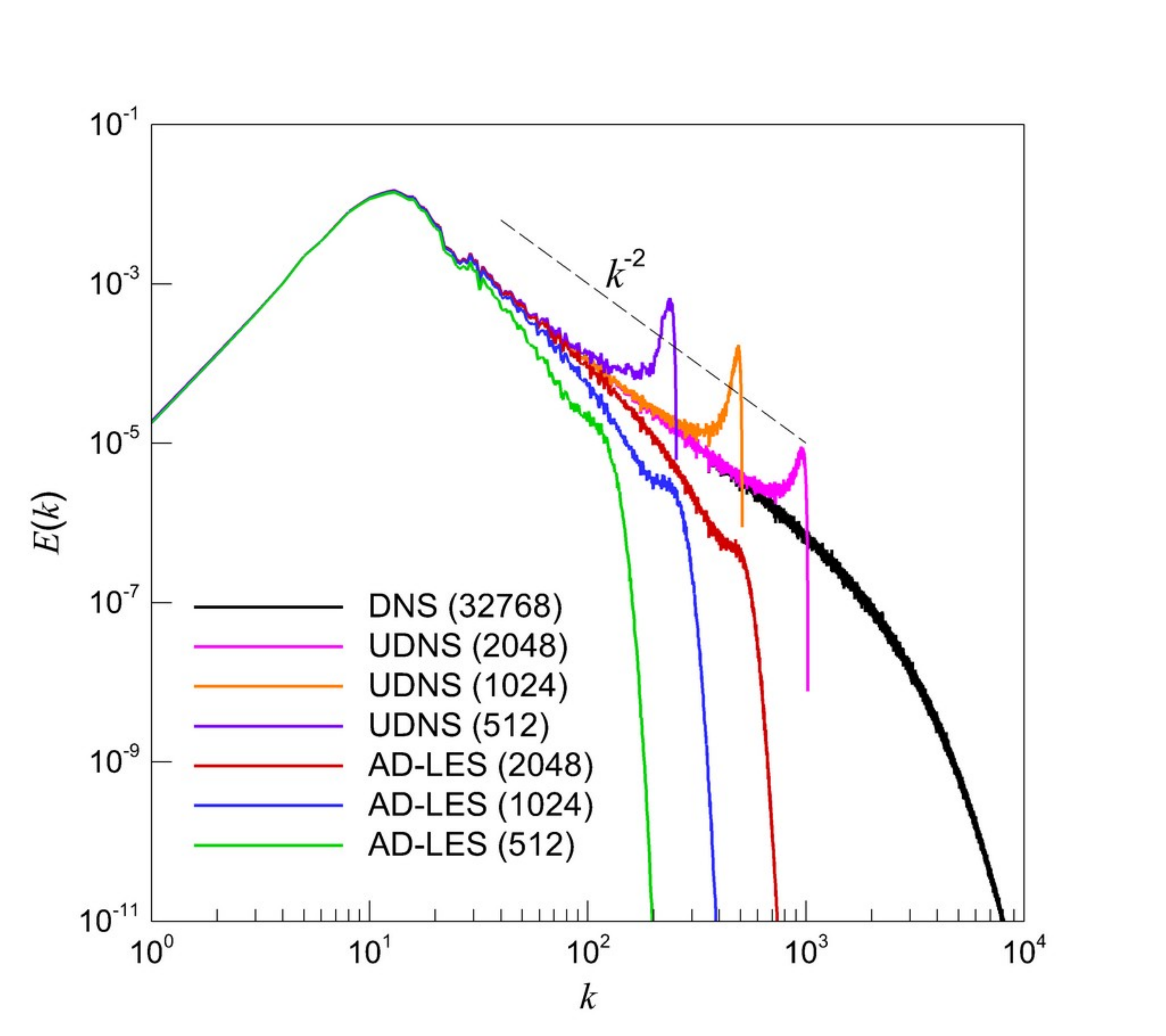}}
}
\caption{Energy spectra obtained by AD-LES equipped with binomial filters.}
\label{fig:a2}
\end{figure}

Coarse grid resolution results for the newly introduced binomial filters are depicted in Figure (\ref{fig:a2}) and were compared to the fully resolved DNS and UDNS simulations. It is evident that the binomial spatial averaging operators are too dissipative in nature and cause a significant error in the inertial scales as the power law scaling behavior is lost. This can be seen from their transfer functions given by Equation (\ref{BinoNonSmoothTF}). Increased stencil sizes (i.e., could be interpreted as an increase in variance of the Gaussian) were seen to add large amount of dissipation. This is consistent with the behavior of their transfer functions in Fourier space given in Figure (\ref{fig:tf}c) showing large amounts of attenuation even in the inertial region. The binomial filters have been designed to mimic the Gaussian filter and the observed dissipative behavior was consistent with our predictions for the energy spectrum.

\begin{figure}[!t]
\centering
\mbox{
\subfigure[$^{(1,1)}B $]{\includegraphics[width=0.45\textwidth]{B2.pdf}}
\subfigure[$^{(2,1)}B $]{\includegraphics[width=0.45\textwidth]{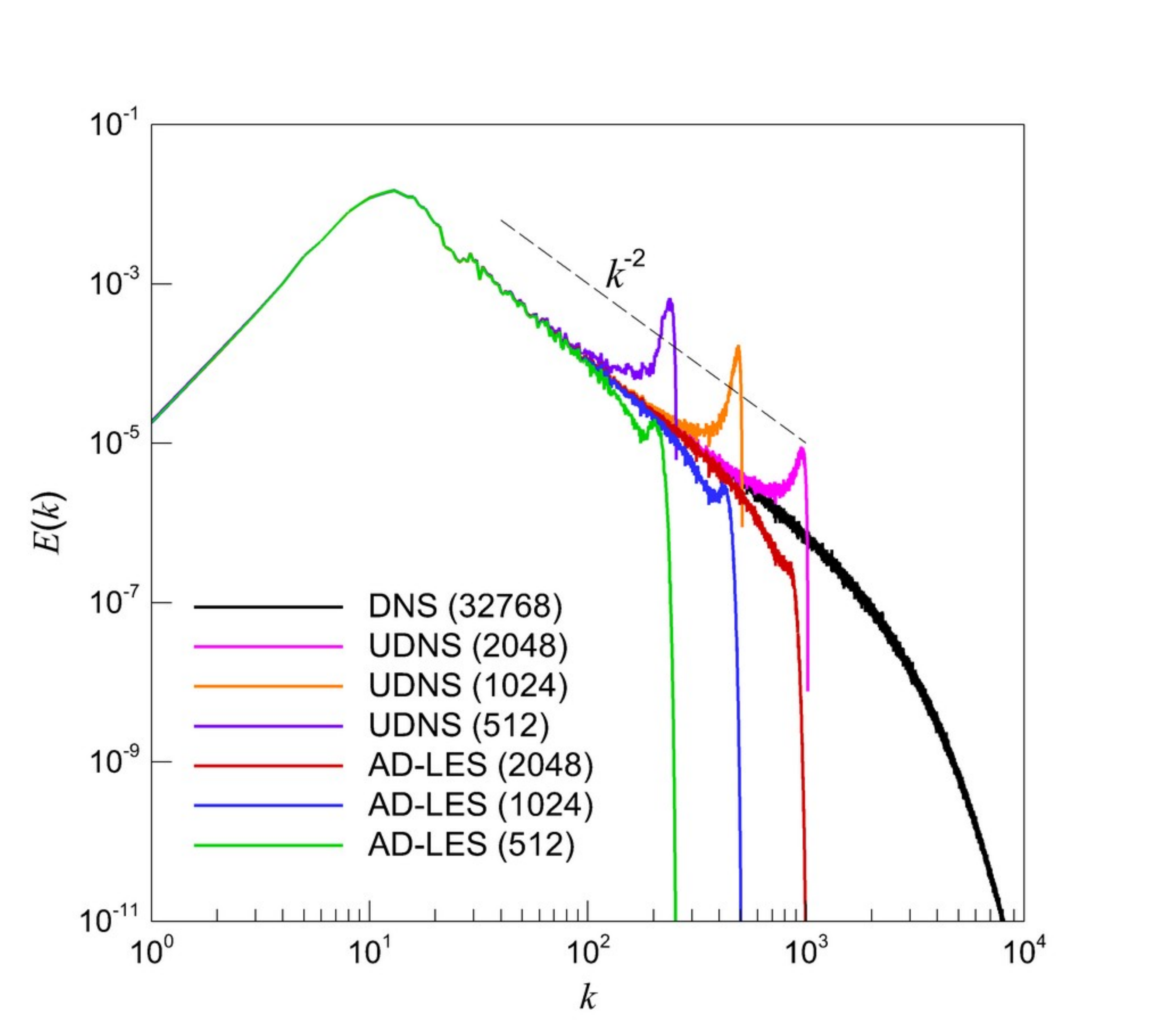}}
}\\
\mbox{
\subfigure[$^{(3,1)}B $]{\includegraphics[width=0.45\textwidth]{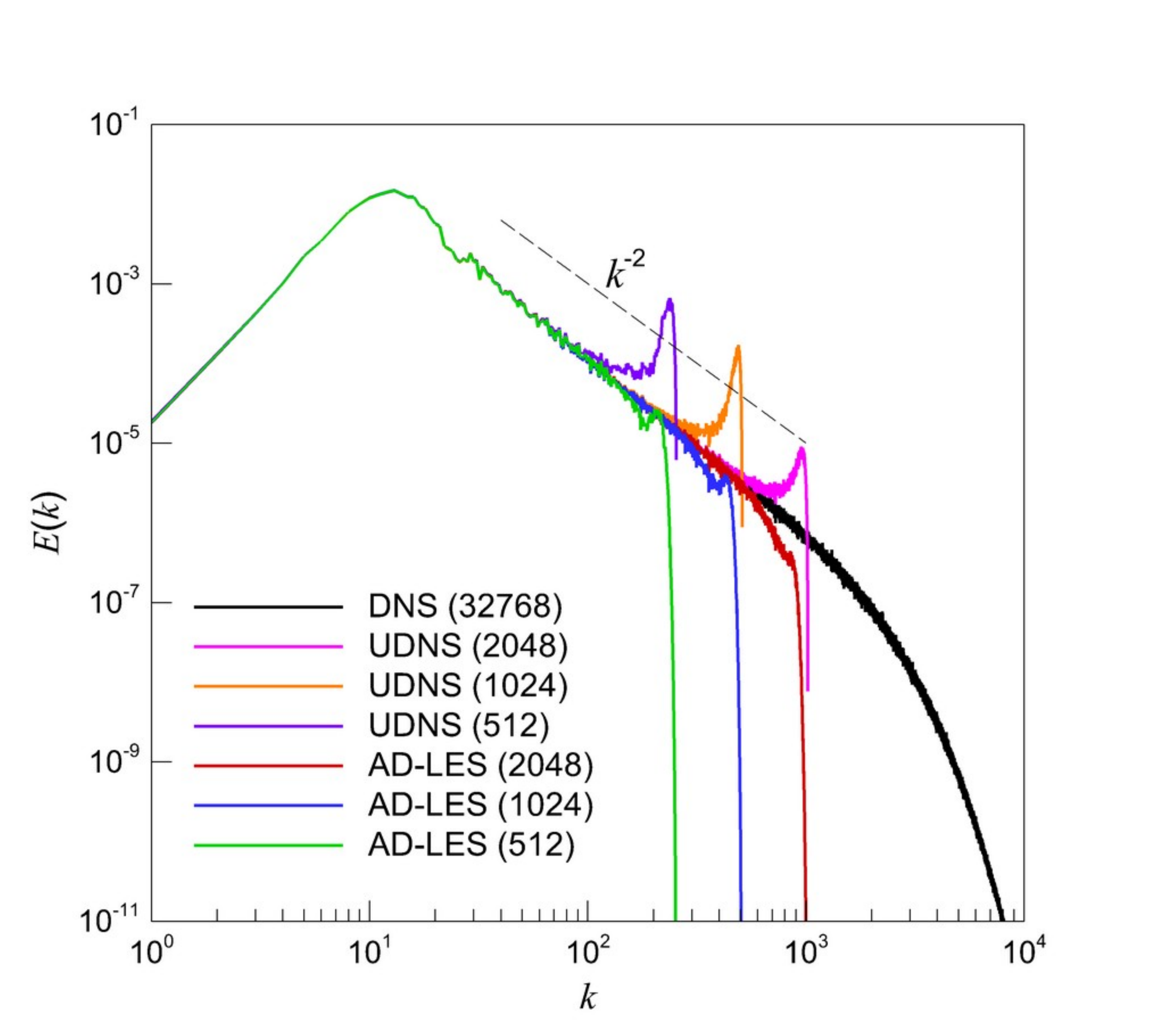}}
\subfigure[$^{(4,1)}B $]{\includegraphics[width=0.45\textwidth]{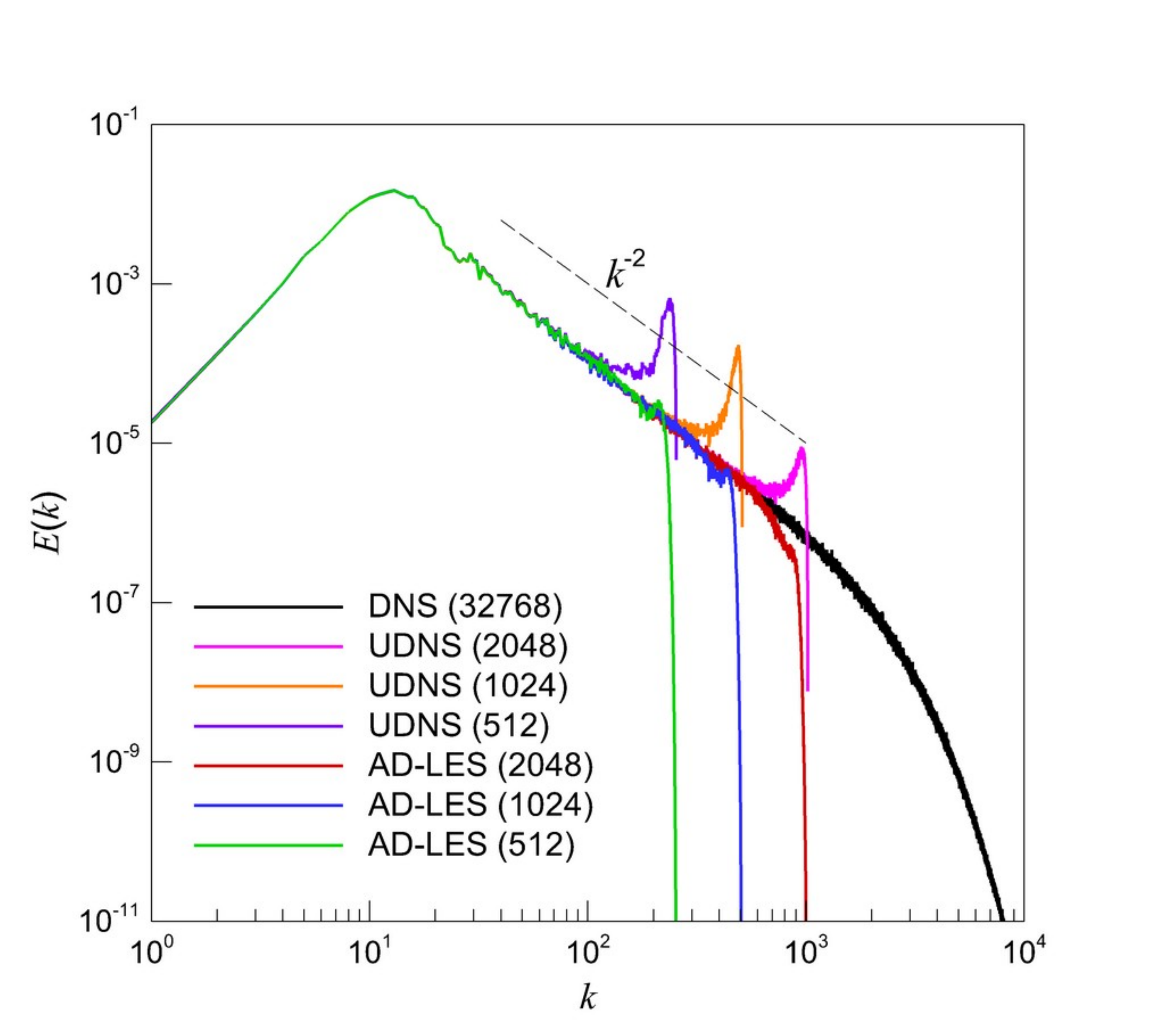}}
}
\caption{Energy spectra obtained by AD-LES equipped with smoothing filters.}
\label{fig:a3}
\end{figure}

The modified binomial filters (referred to as the binomial smoothing filters) were also used for the filtering operation in the AD-LES method used and comparisons for coarse resolutions are shown in Figure (\ref{fig:a3}). A noticeable improvement was seen when compared with the results given by binomial filters. Increasing stencil sizes led to better predictions of the energy spectrum in the inertial ranges and it was noticed that the improvement in results was marginal beyond filters of order $n = 3$. The transfer function for smoothing filters do not show large increases in steepness (i.e., their capacity to pass lower spatial frequencies) as $n$ is increased beyond three. The binomial smoothing filters, however, were able to yield considerably better results than the UDNS results for the same coarse meshes. Performance was exceptional for the $^{(3,1)}B$ and $^{(4,1)}B$ at coarse resolutions while the energy spectrum was slightly under-predicted at higher resolutions.

\begin{figure}[!t]
\centering
\mbox{
\subfigure[$\alpha = 0.0 $]{\includegraphics[width=0.45\textwidth]{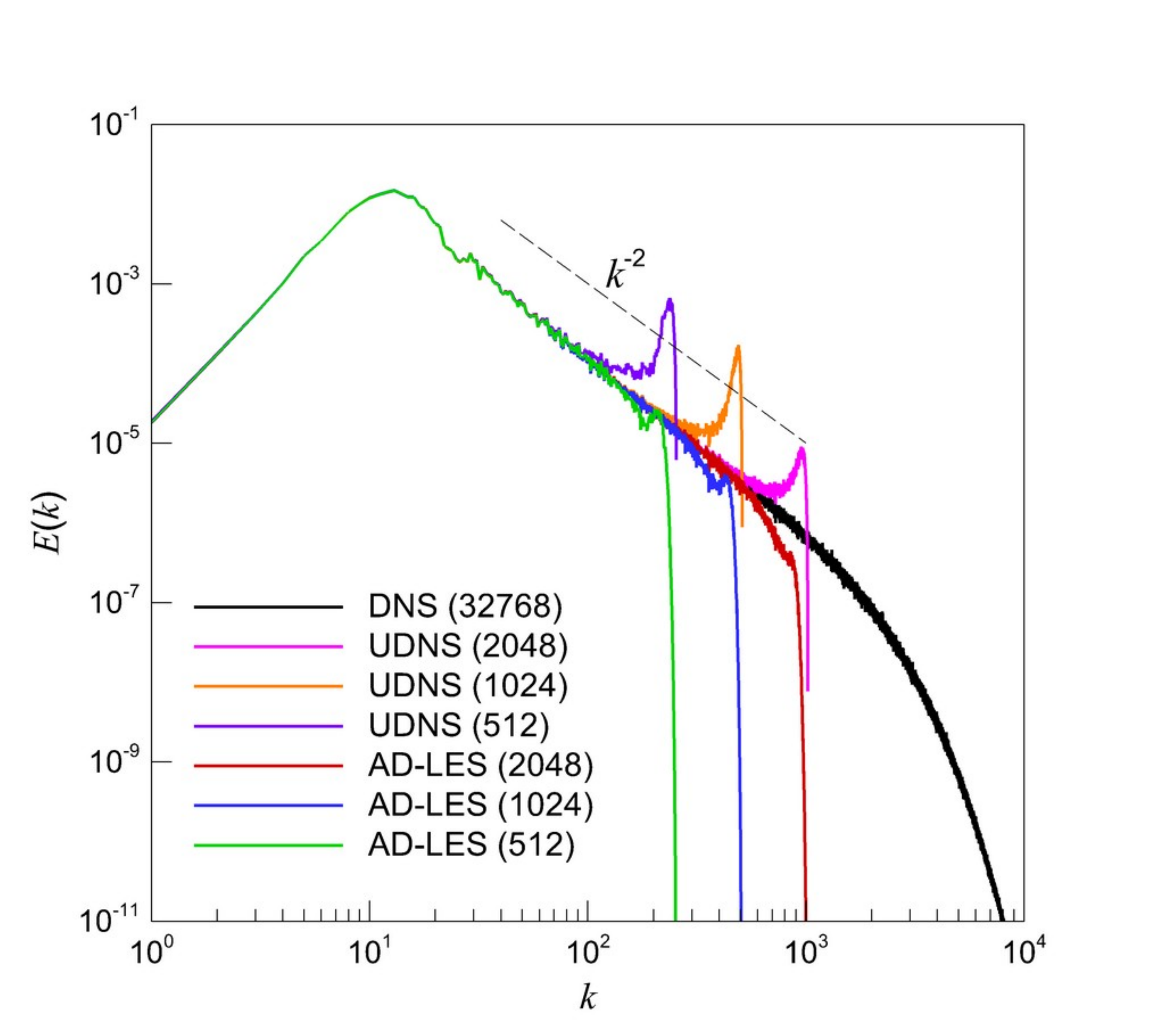}}
\subfigure[$\alpha = 0.2 $]{\includegraphics[width=0.45\textwidth]{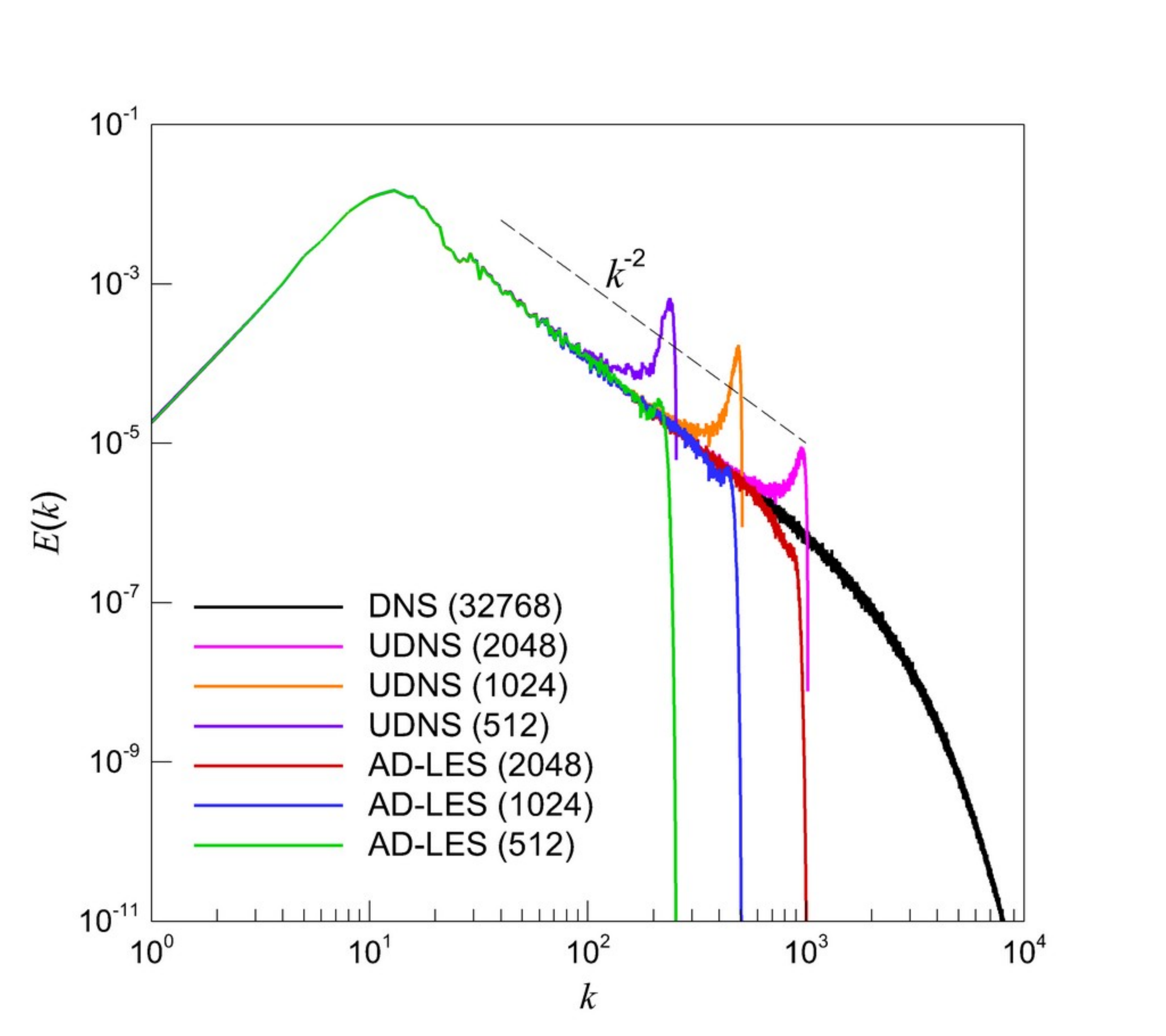}}
}\\
\mbox{
\subfigure[$\alpha = 0.3 $]{\includegraphics[width=0.45\textwidth]{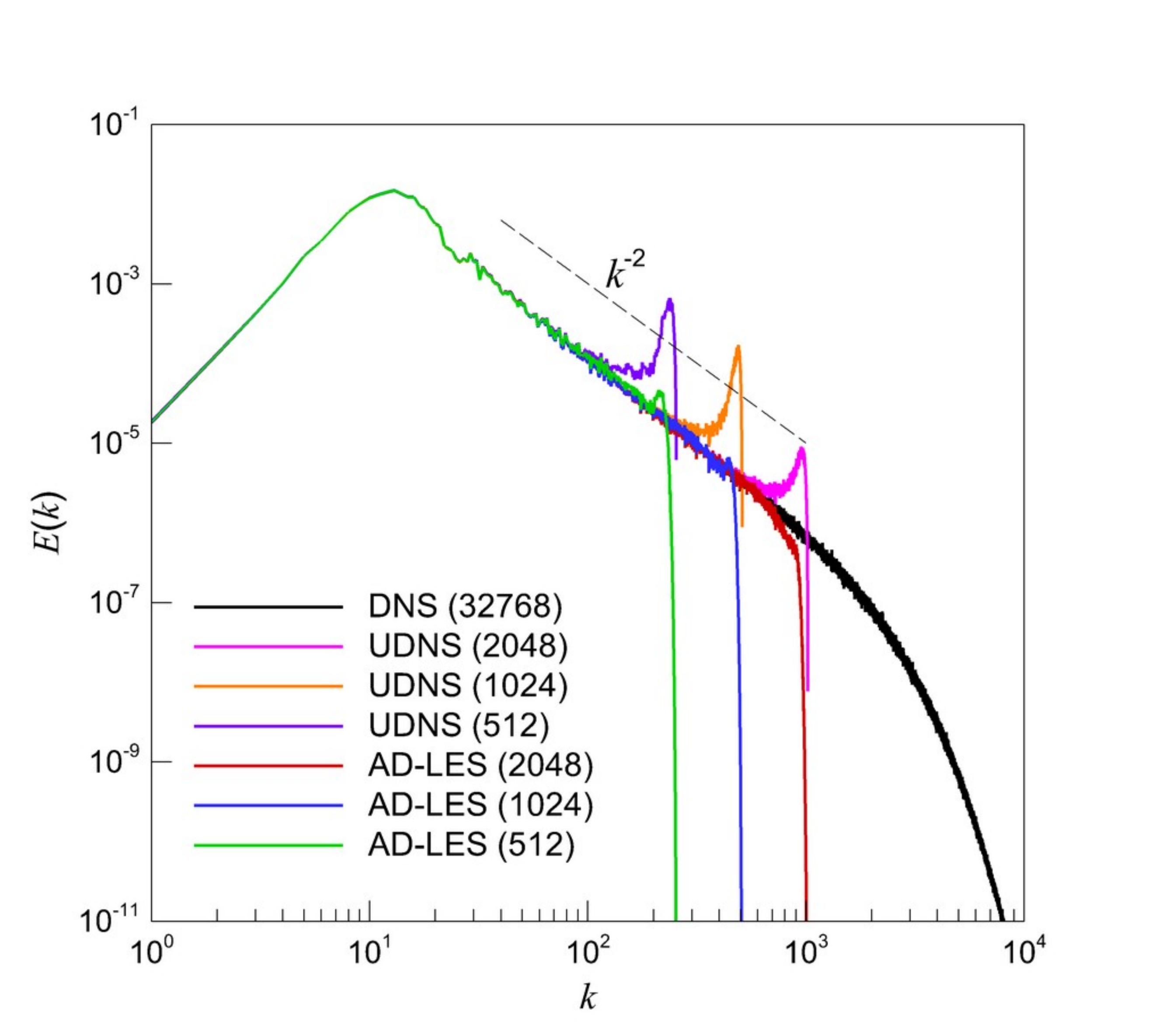}}
\subfigure[$\alpha = 0.4 $]{\includegraphics[width=0.45\textwidth]{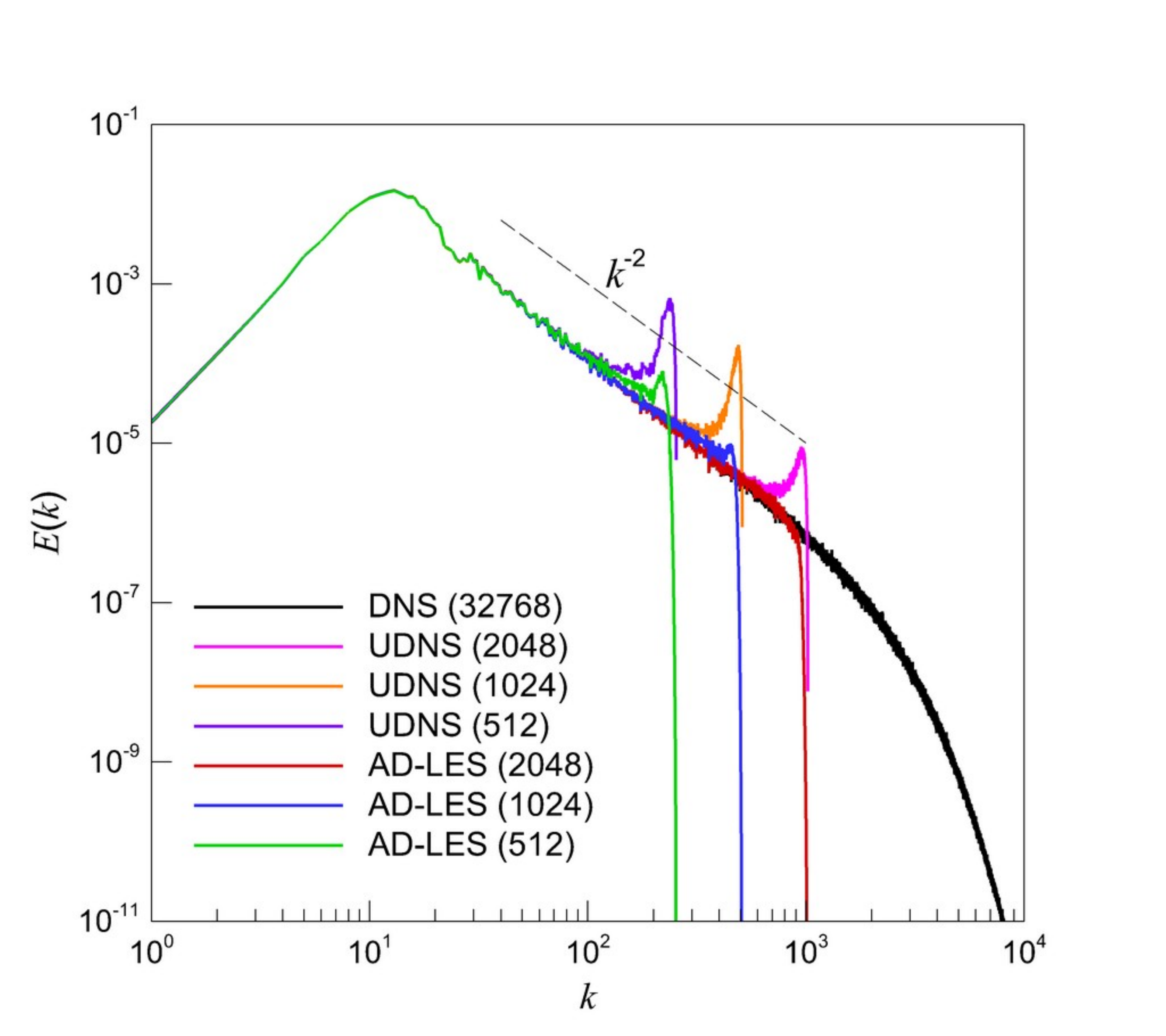}}
}
\caption{Energy spectra obtained by AD-LES equipped with Pad\'{e} filters.}
\label{fig:a4}
\end{figure}

In a previous work by \cite{san2016analysis}, the Pad\'{e} filter was used with the approximate deconvolution method for solving Burgers equation with skew-symmetric form. However, in the current work, we conduct our analysis by using the conservative form. Therefore our analysis here can be used to assess the performance of various formulations of the Burgers equation. In Figure (\ref{fig:a4}), a comparison of the results obtained using this filter and UDNS is shown. It can be seen that the Pad\'{e} filter yields significantly better results in removing the energy content at smaller scales with increasing values of $\alpha$. Decrease in the $\alpha$ value leads to increased dissipation at the grid cut-off. Values of $\alpha$ between 0.2 and 0.4 yield balanced results. An important point to note here is that the process of dissipation is through the effect of the repeated filtering procedure in AD process (through the shape of the transfer function of the low-pass operator) and the physical viscosity (taken constant at $\nu = 5 \times 10^{-4}$). We also note that the Pad\'{e} filter with $\alpha = 0$ corresponds to the case of the $^{(3,1)} B$ binomial filter.

\begin{figure}[!t]
\centering
\mbox{
\subfigure[$\alpha = 0.48 $]{\includegraphics[width=0.45\textwidth]{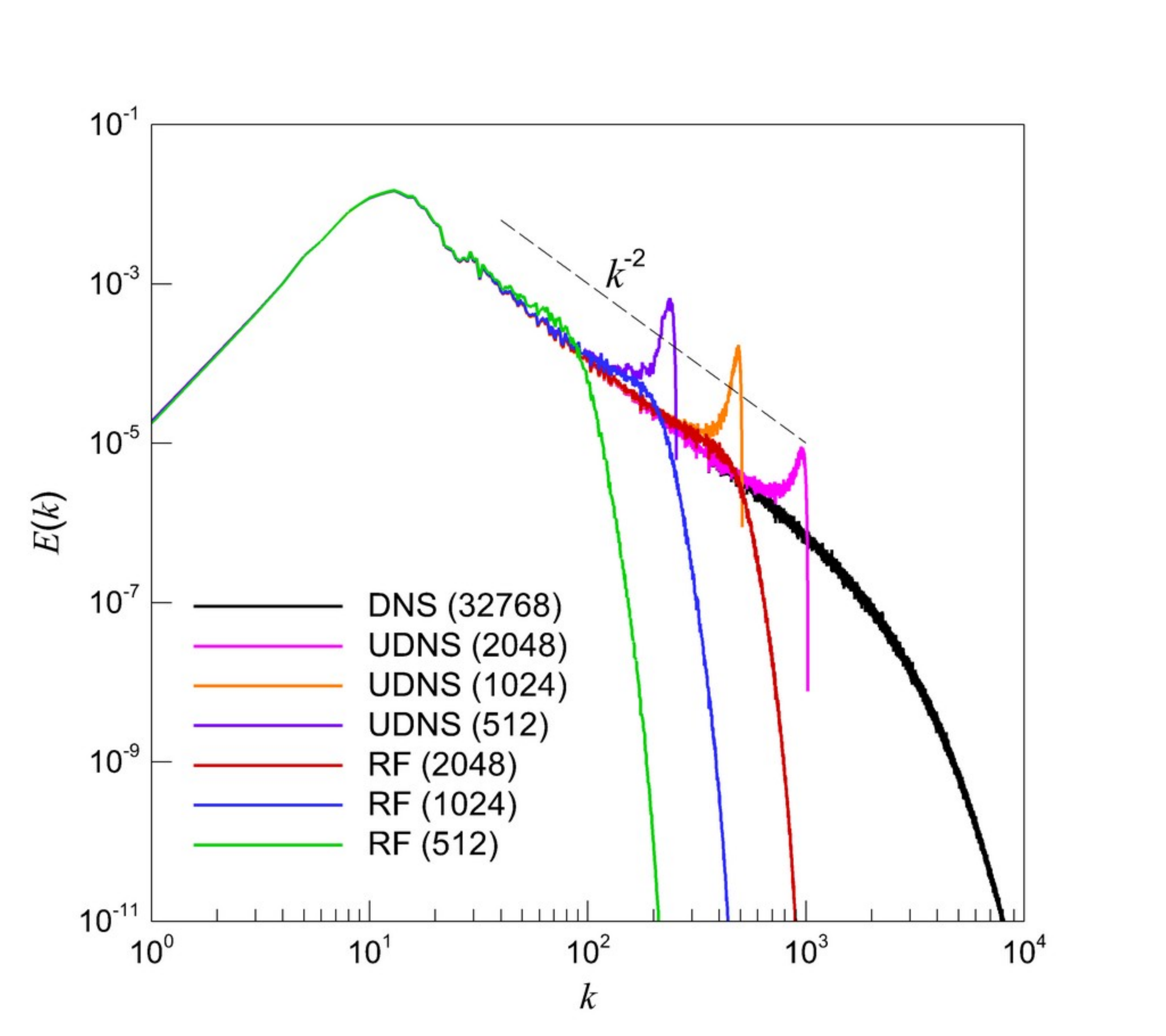}}
\subfigure[$\alpha = 0.49 $]{\includegraphics[width=0.45\textwidth]{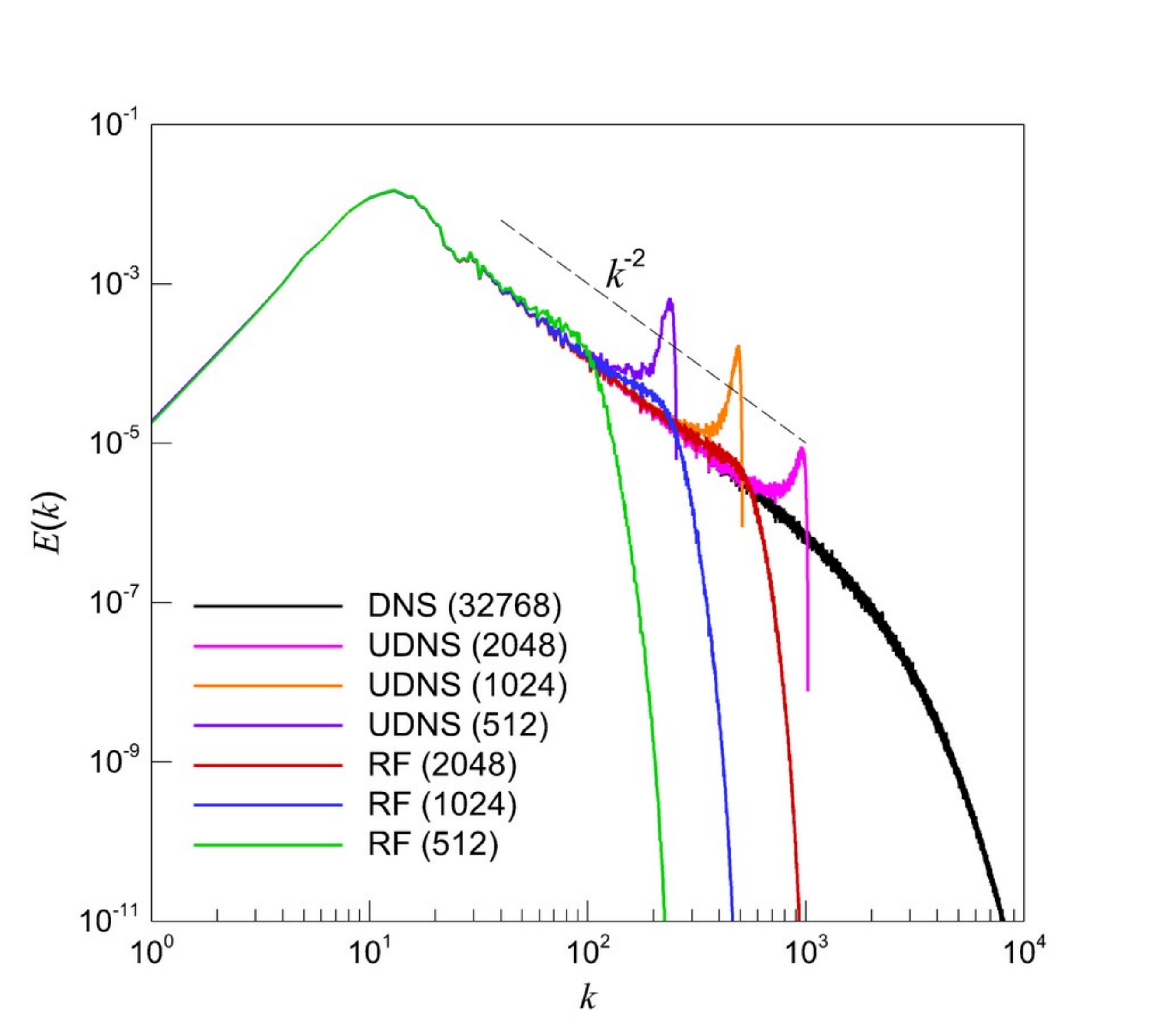}}
}\\
\mbox{
\subfigure[$\alpha = 0.495 $]{\includegraphics[width=0.45\textwidth]{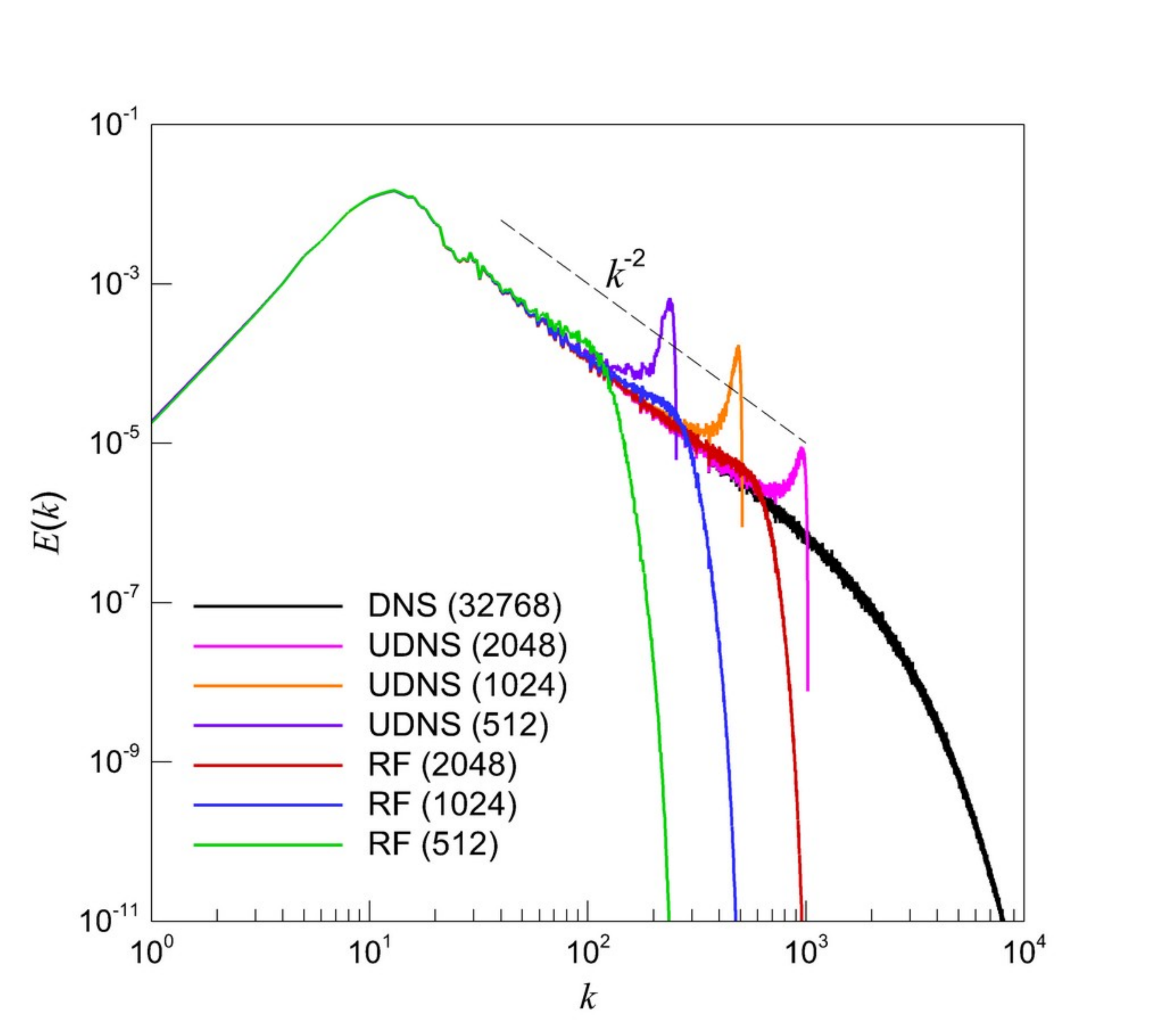}}
\subfigure[$\alpha = 0.499 $]{\includegraphics[width=0.45\textwidth]{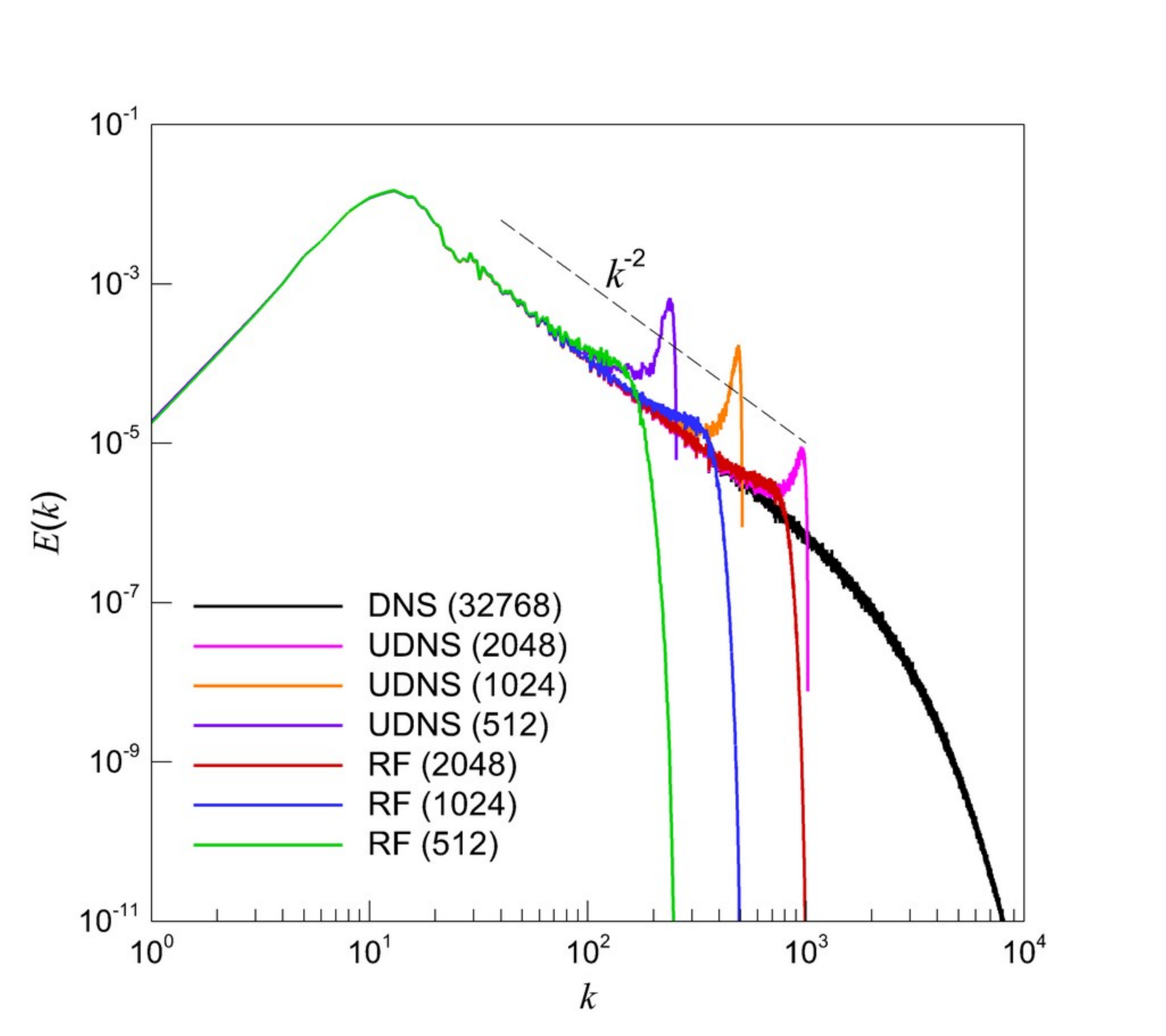}}
}
\caption{Relaxation filtering (RF) applied at the end of each time step by using the Pad\'{e} filter for various $\alpha$ values.}
\label{fig:a5}
\end{figure}

\begin{figure}[!t]
\centering
\mbox{
\subfigure[$\alpha = 0.48 $]{\includegraphics[width=0.45\textwidth]{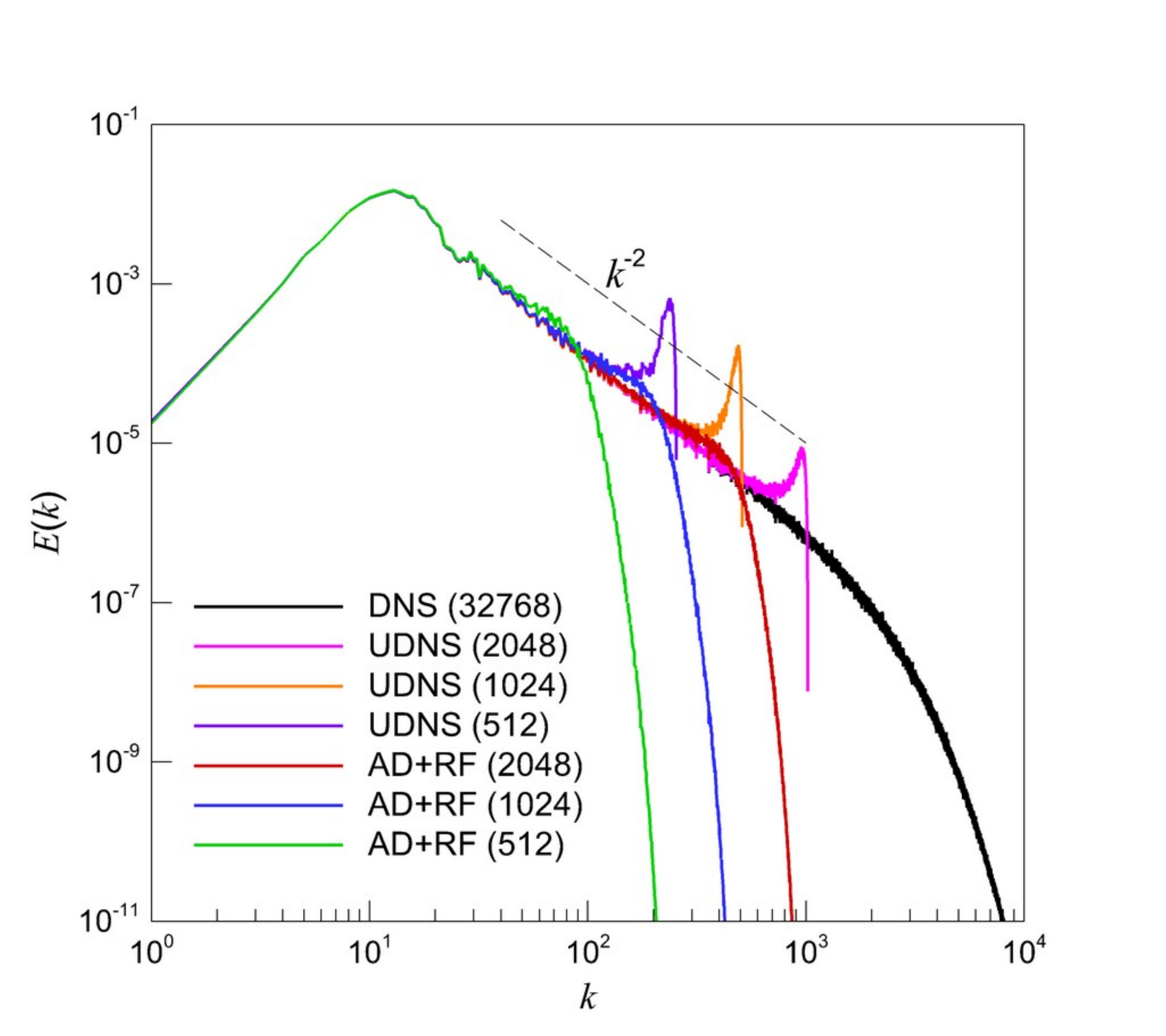}}
\subfigure[$\alpha = 0.49 $]{\includegraphics[width=0.45\textwidth]{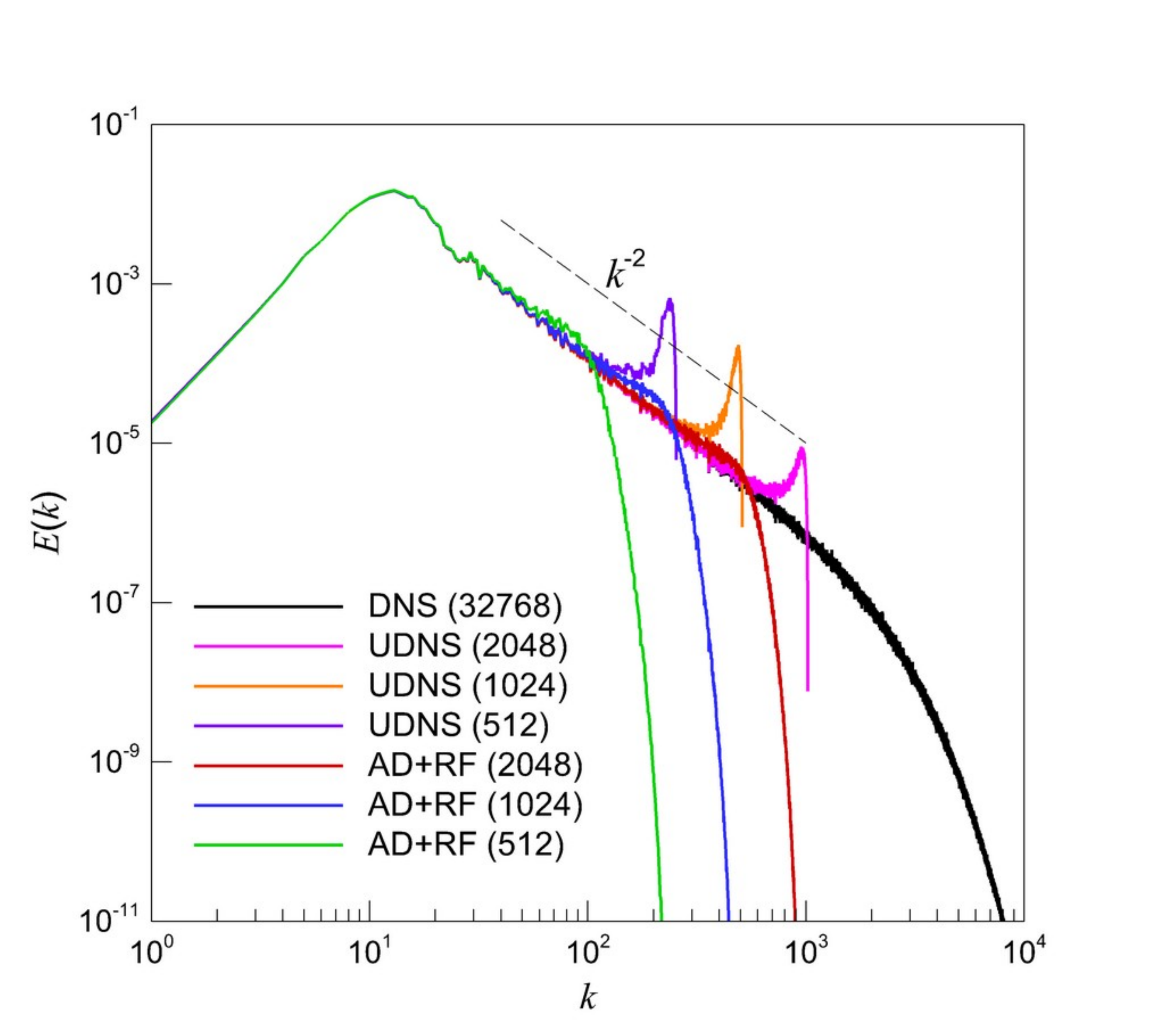}}
}\\
\mbox{
\subfigure[$\alpha = 0.495 $]{\includegraphics[width=0.45\textwidth]{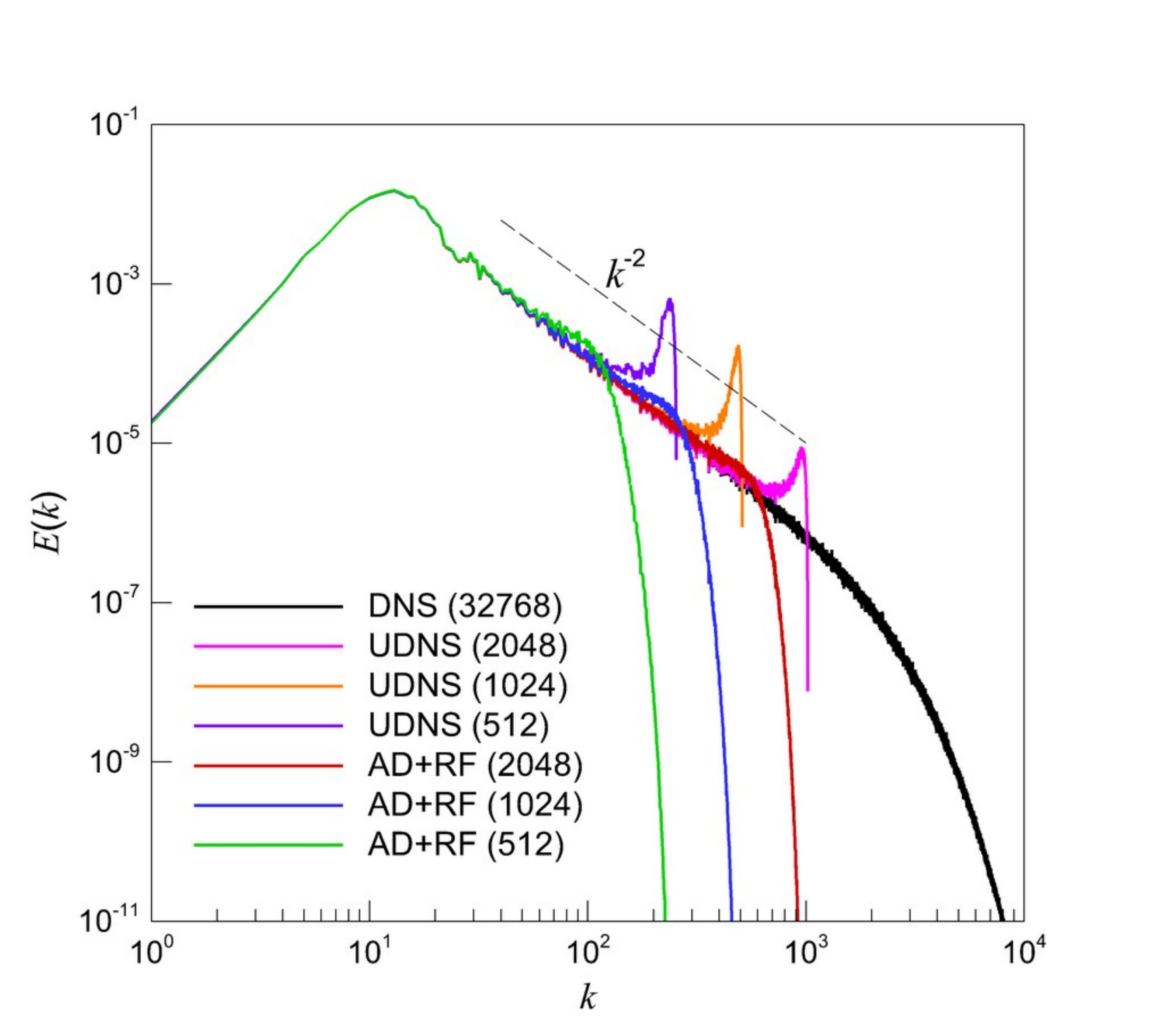}}
\subfigure[$\alpha = 0.499 $]{\includegraphics[width=0.45\textwidth]{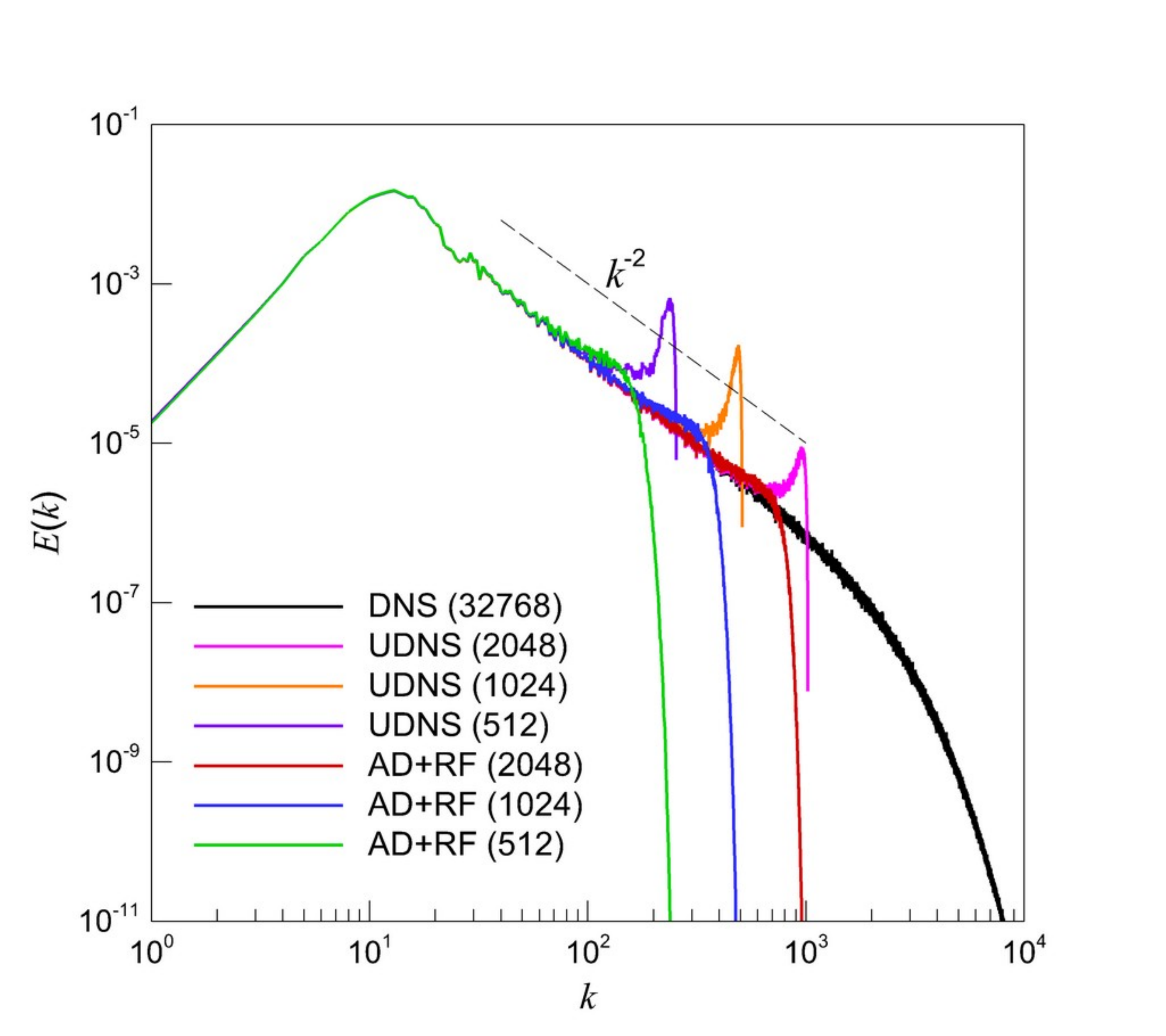}}
}
\caption{Relaxation filtering (RF) combined with AD-LES equipped by the Pad\'{e} filter with $\alpha = 0.4$ (i.e., primary filter in AD). As a secondary filter, the sixth-order Pad\'{e} filter is applied at the end of each time step varying $\alpha$ values between 0.48 and 0.499.}
\label{fig:a6}
\end{figure}

The arrest of numerical oscillations can also be achieved by the use of an explicit low-pass filtering process at the end of each time step. This process is known as relaxation filtering (RF) and may be accompanied with an AD-LES method as well (in which case it is known as secondary filtering) to regularize it near the grid cut-off scales. Without using the AD process, the application of a low-pass explicit filter at the end of each time step prevented energy pile-up at the grid cut-off as shown in Figure (\ref{fig:a5}). However, it was also seen that a considerable amount of inertial range information is lost due to its application. The energy spectrum predicted by this process becomes steeper than the ideal $k^{-2}$ scaling and shows a considerable loss of some scales in the inertial range supported by grid. The Pad\'{e} filter has been used explicitly for relaxation filtering and increasing values of the $\alpha$ parameter are seen to provide less amount of dissipation at the grid cut-off which is consistent with our understanding of the transfer function of the Pad\'{e} filters. When used in the context of secondary filtering (i.e., combined with AD-LES), it was seen that the secondary filter dominated the iterative AD process as shown by the similarity in results obtained between Figure (\ref{fig:a5}) and Figure (\ref{fig:a6}). The addition of the repeated filtering associated with the approximate deconvolution process was seen to be insignificant when we use RF at the end of each time step. We also emphasize that the level of attenuation due to the RF depends on the free secondary filtering parameter.  However, it is also seen here, that energy pile-up no longer remains an issue near the grid cut-off scale as compared to the UDNS spectrum.

\begin{figure}[!t]
\centering
\mbox{
\subfigure[$\alpha = 0.0 $]{\includegraphics[width=0.45\textwidth]{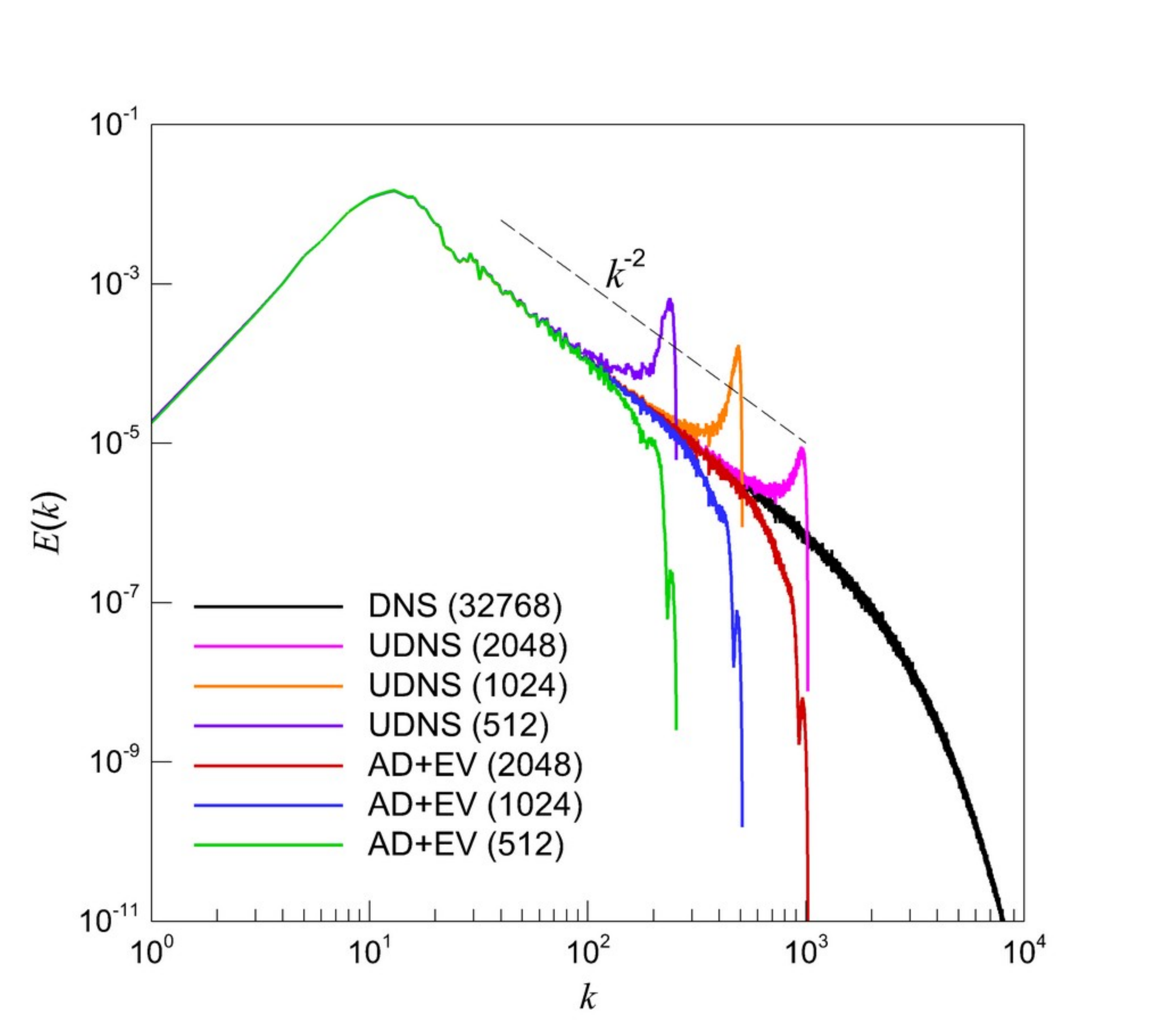}}
\subfigure[$\alpha = 0.2 $]{\includegraphics[width=0.45\textwidth]{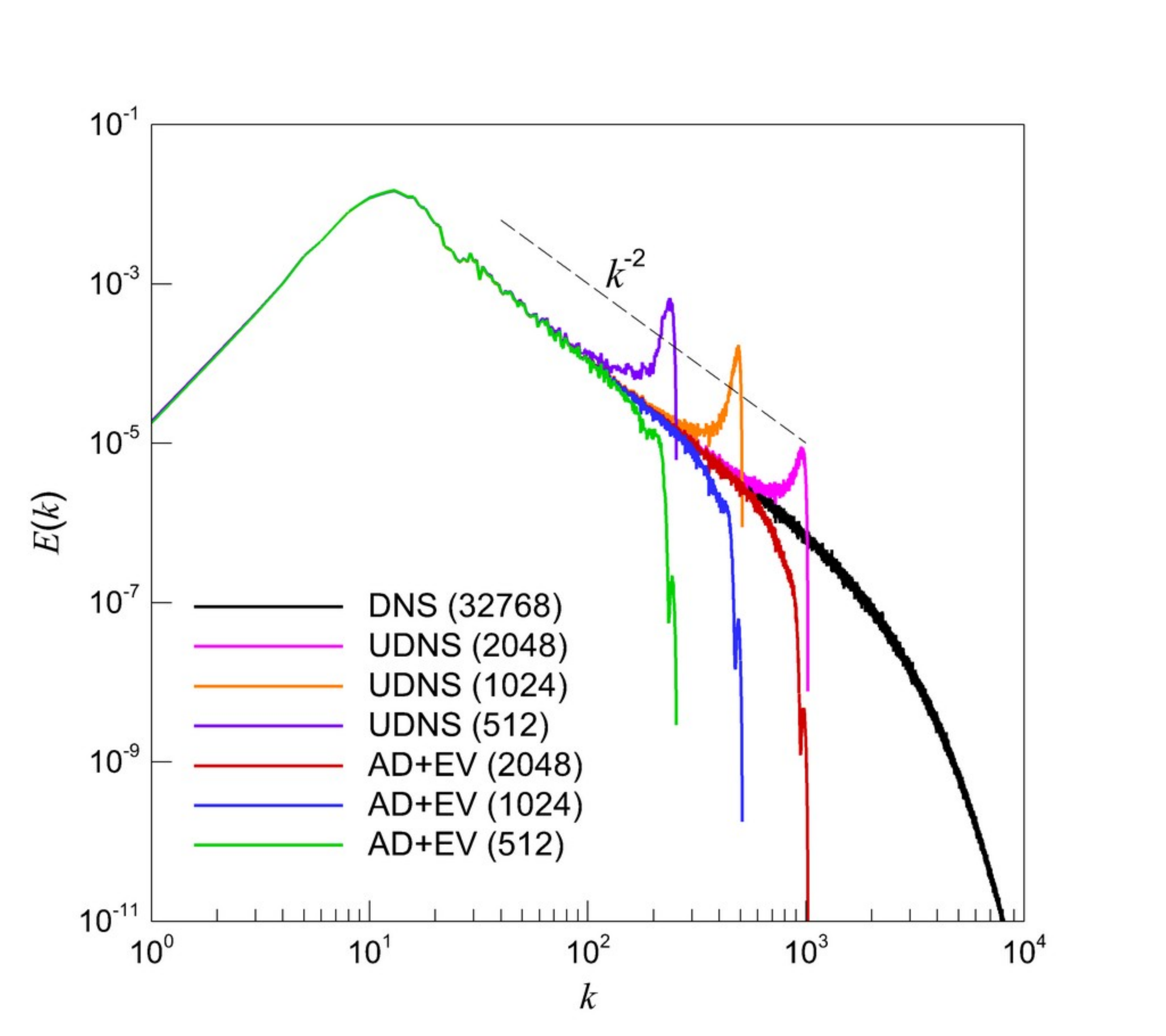}}
}\\
\mbox{
\subfigure[$\alpha = 0.3 $]{\includegraphics[width=0.45\textwidth]{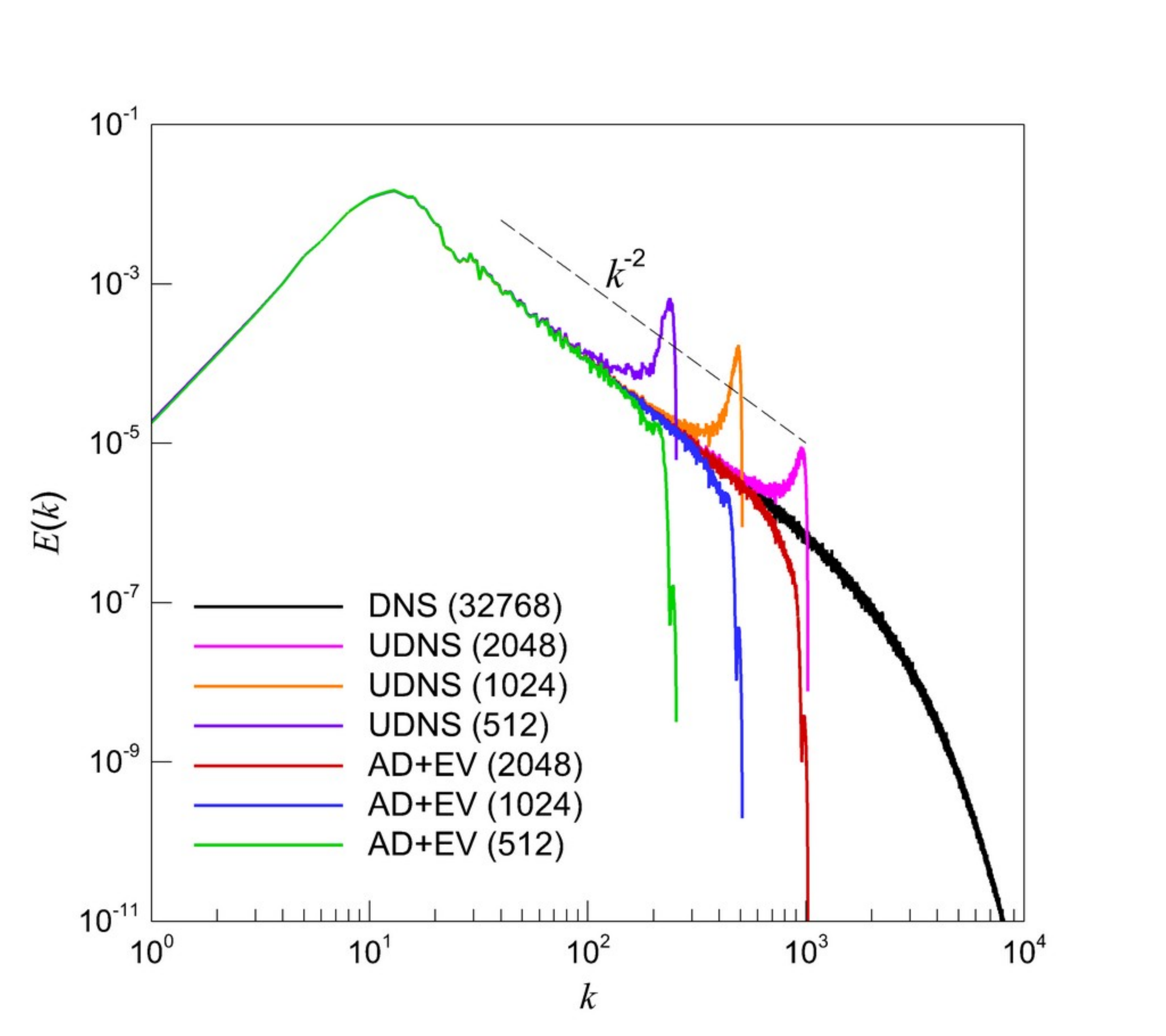}}
\subfigure[$\alpha = 0.4 $]{\includegraphics[width=0.45\textwidth]{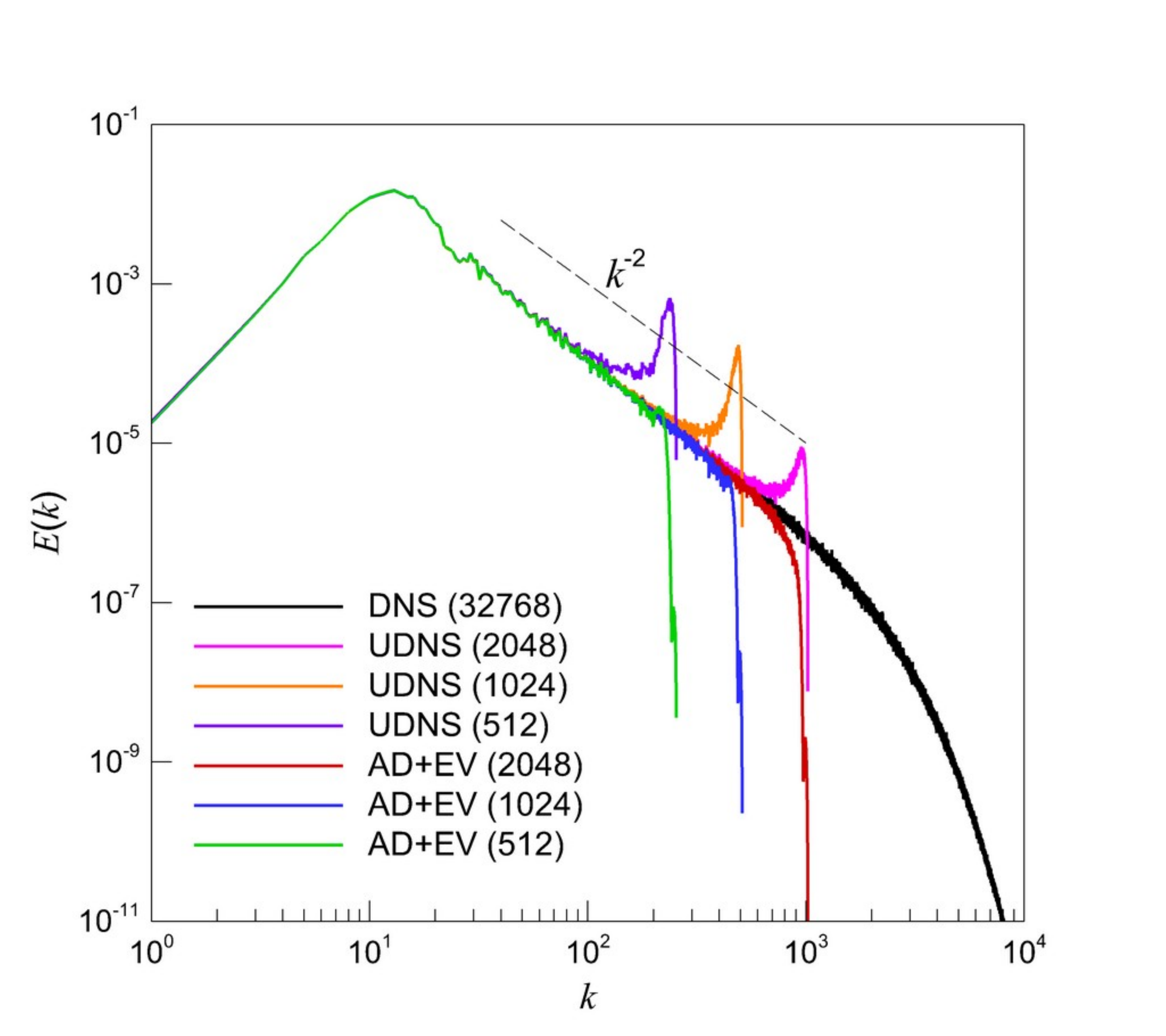}}
}
\caption{Regularization of AD-LES (using the Pad\'{e} filter with various $\alpha$ values) with the Smagorinsky type eddy viscosity model using $C_s = 0.2$.}
\label{fig:a7}
\end{figure}

A regularization of the AD-LES method was carried out using a Smagorinsky type eddy viscosity for additional dissipation as shown in Figure (\ref{fig:a7}). UDNS and fully resolved DNS results are also shown for the purpose of comparison. As expected, the regularization process is seen to add additional dissipation to the energy spectrum. The Smagorinksy coefficient of $C_s$ = 0.2 was chosen to introduce the artificial dissipation. On comparison with Figure (\ref{fig:a1}), it is evident that the combination of both functional and structural concepts here has led to an effective prevention of pile-up at the finer resolutions. The effects of $\alpha$ from the primary filter are still seen with decreased dissipation with an increase in its values.

\begin{figure}[!t]
\centering
\mbox{
\subfigure[$\beta = 1.0 $]{\includegraphics[width=0.45\textwidth]{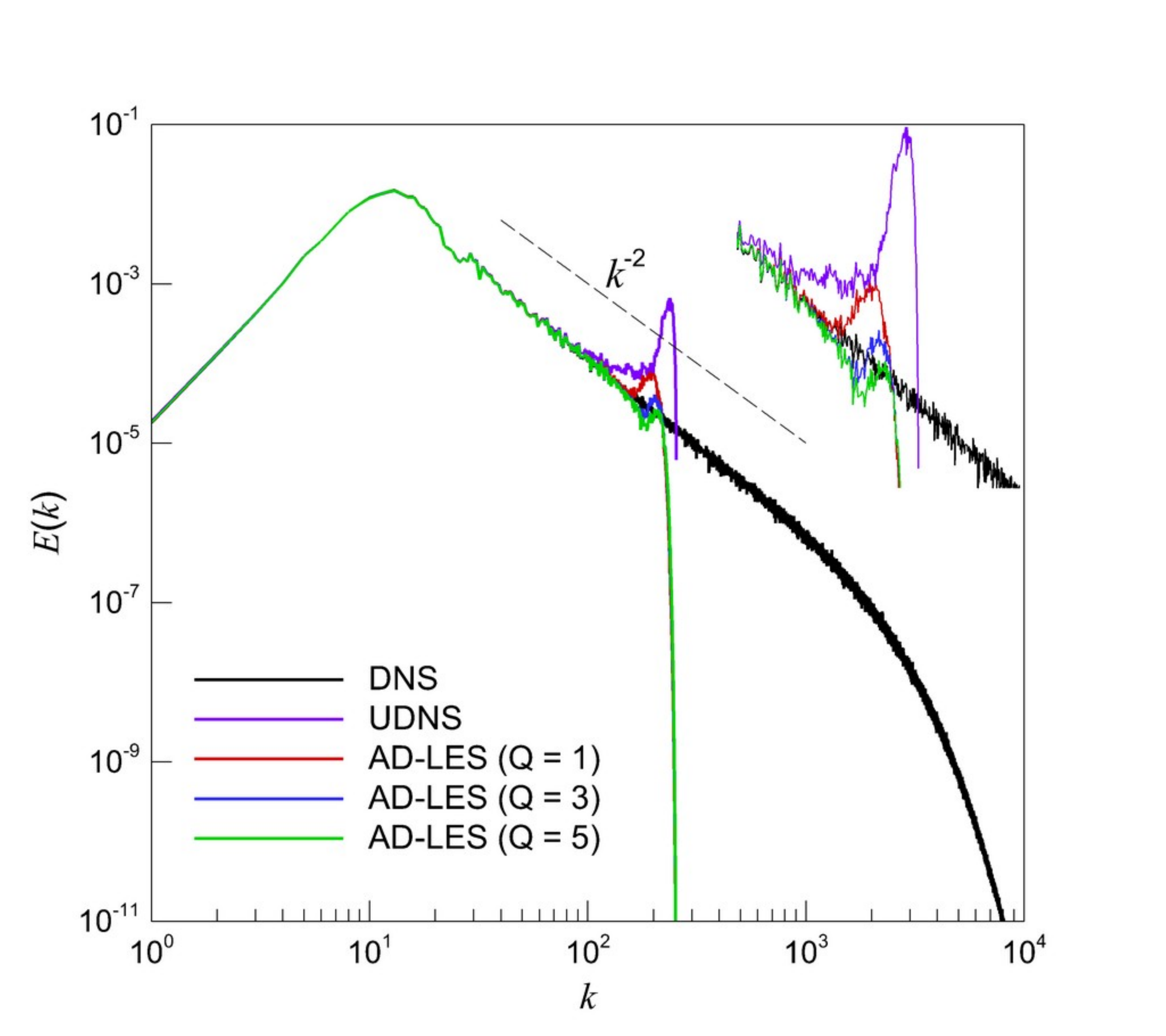}}
\subfigure[$Q = 1 $]{\includegraphics[width=0.45\textwidth]{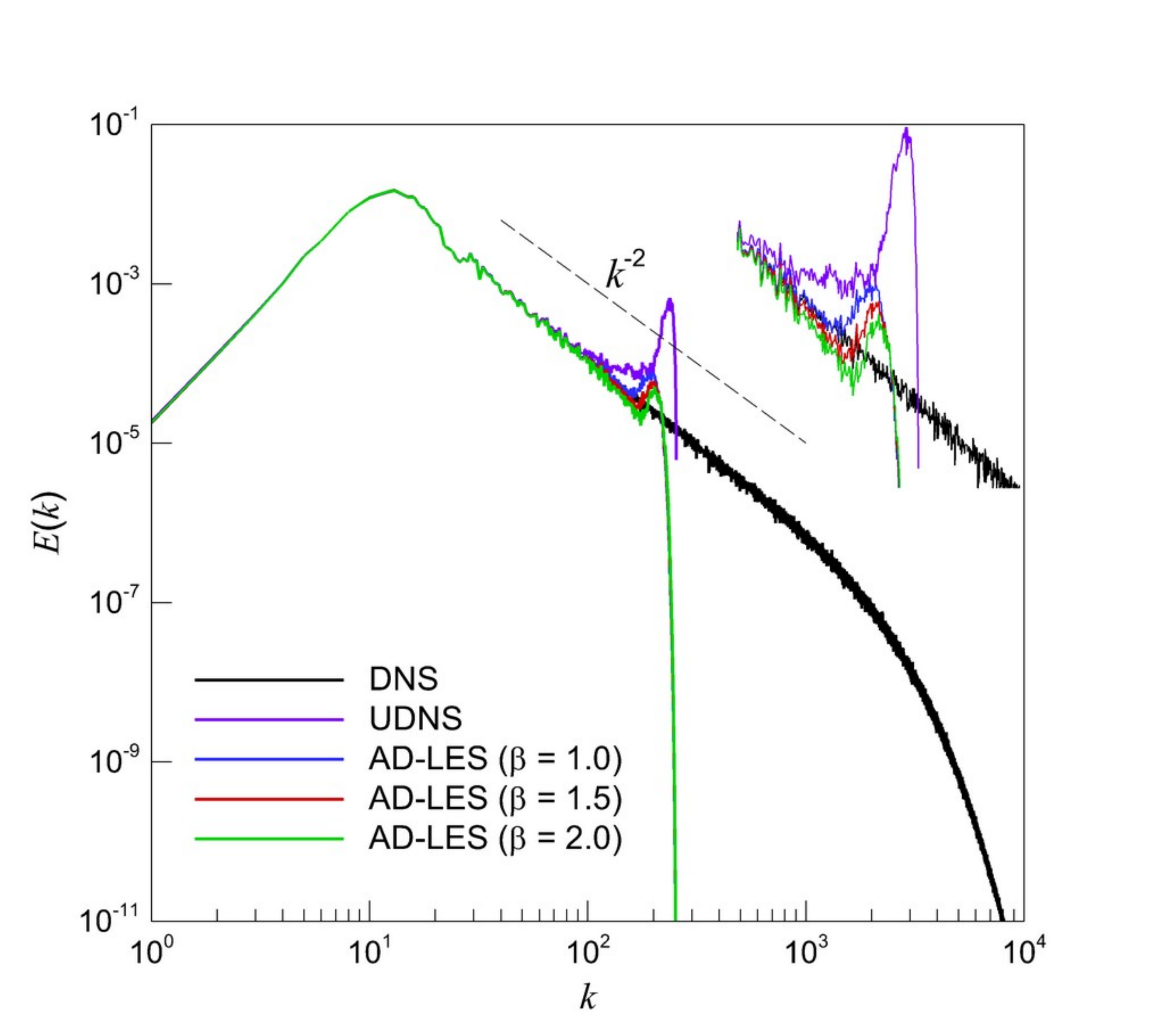}}
}\\
\mbox{
\subfigure[$Q = 3 $]{\includegraphics[width=0.45\textwidth]{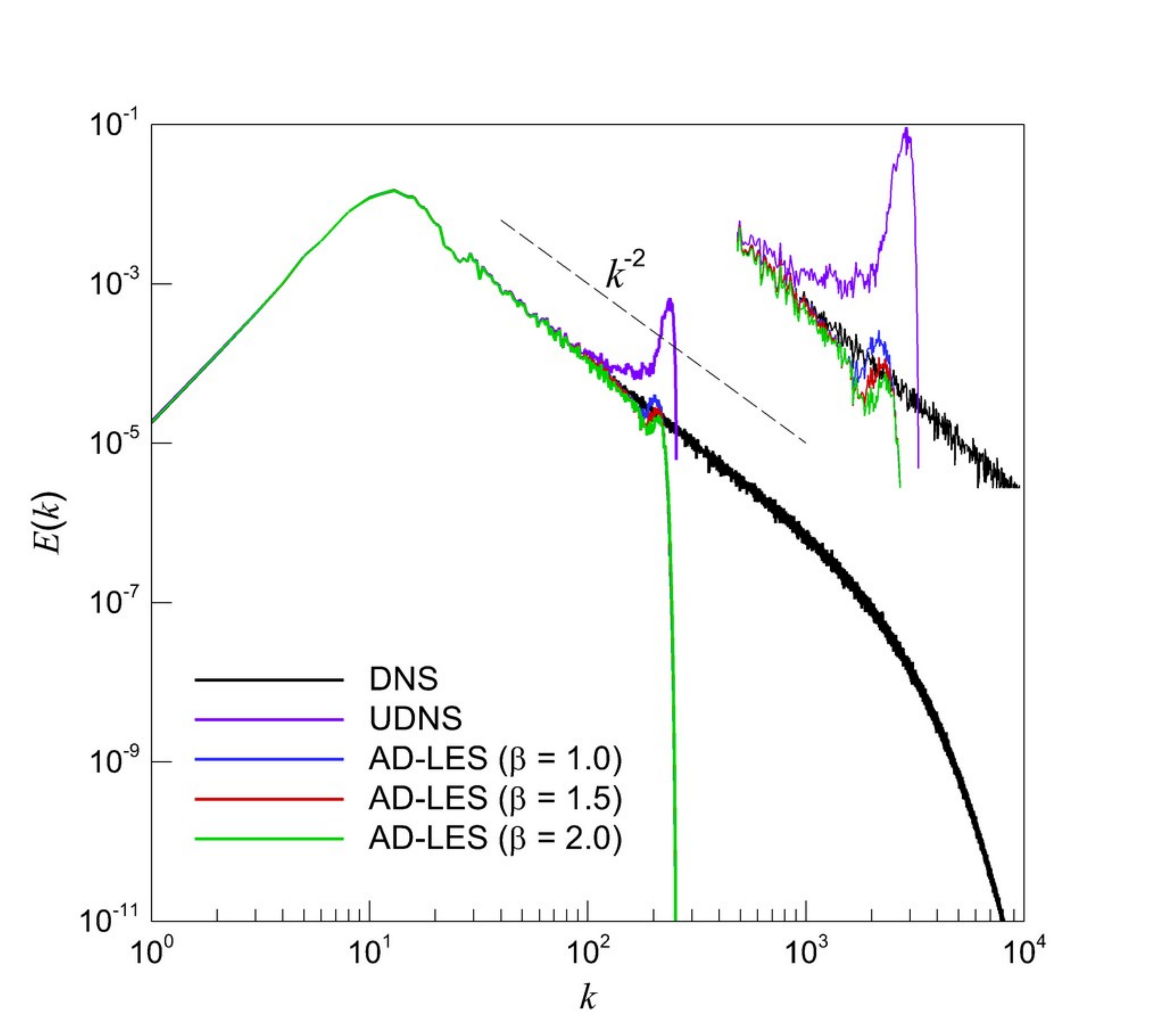}}
\subfigure[$Q = 5 $]{\includegraphics[width=0.45\textwidth]{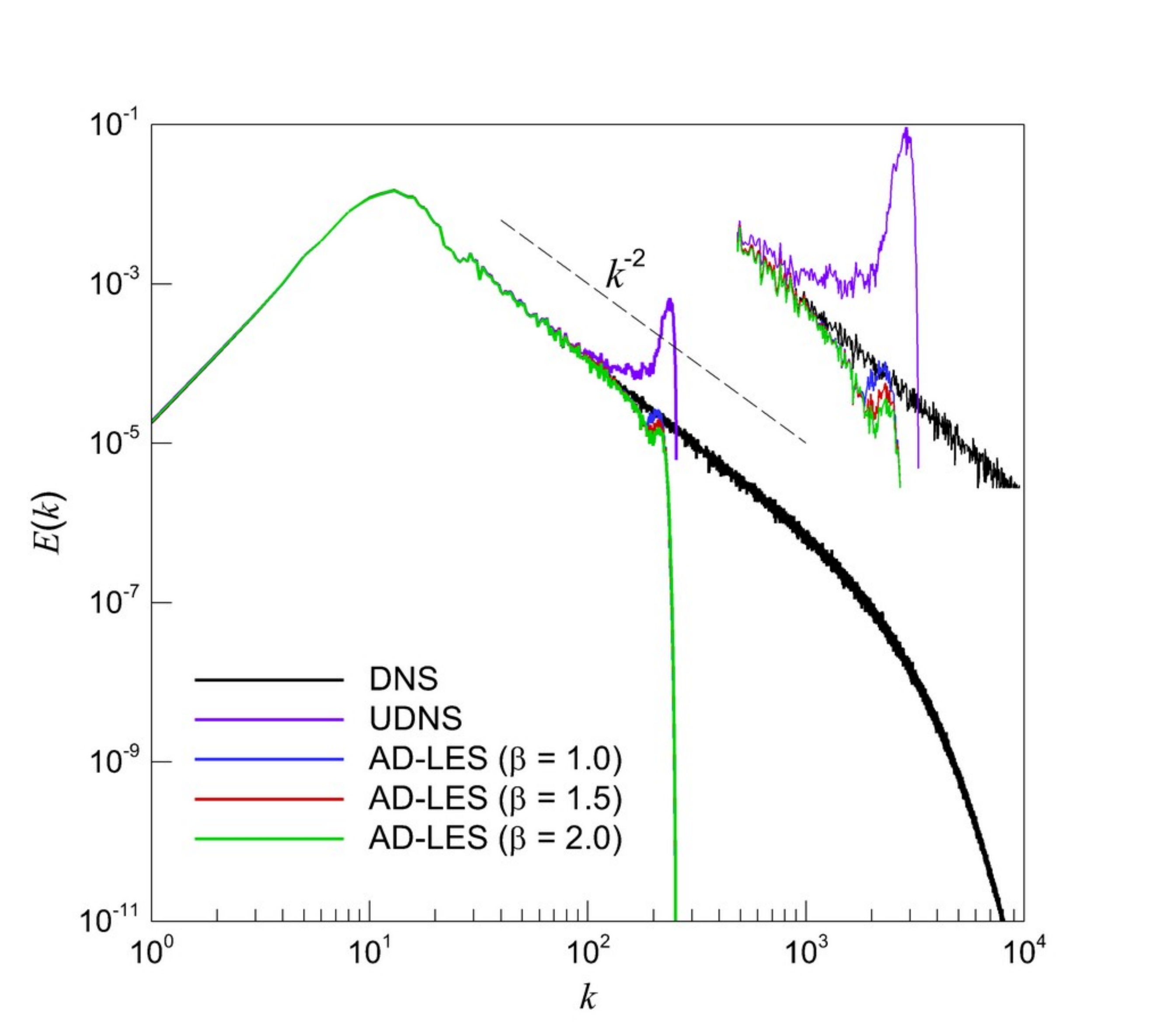}}
}
\caption{Sensitivity analysis with respect to $\beta$ and $Q$ in AD-LES modelling with $N=512$ grid points.}
\label{fig:a8}
\end{figure}

\begin{figure}[!t]
\centering
\mbox{
\subfigure[$\beta = 1.0 $]{\includegraphics[width=0.45\textwidth]{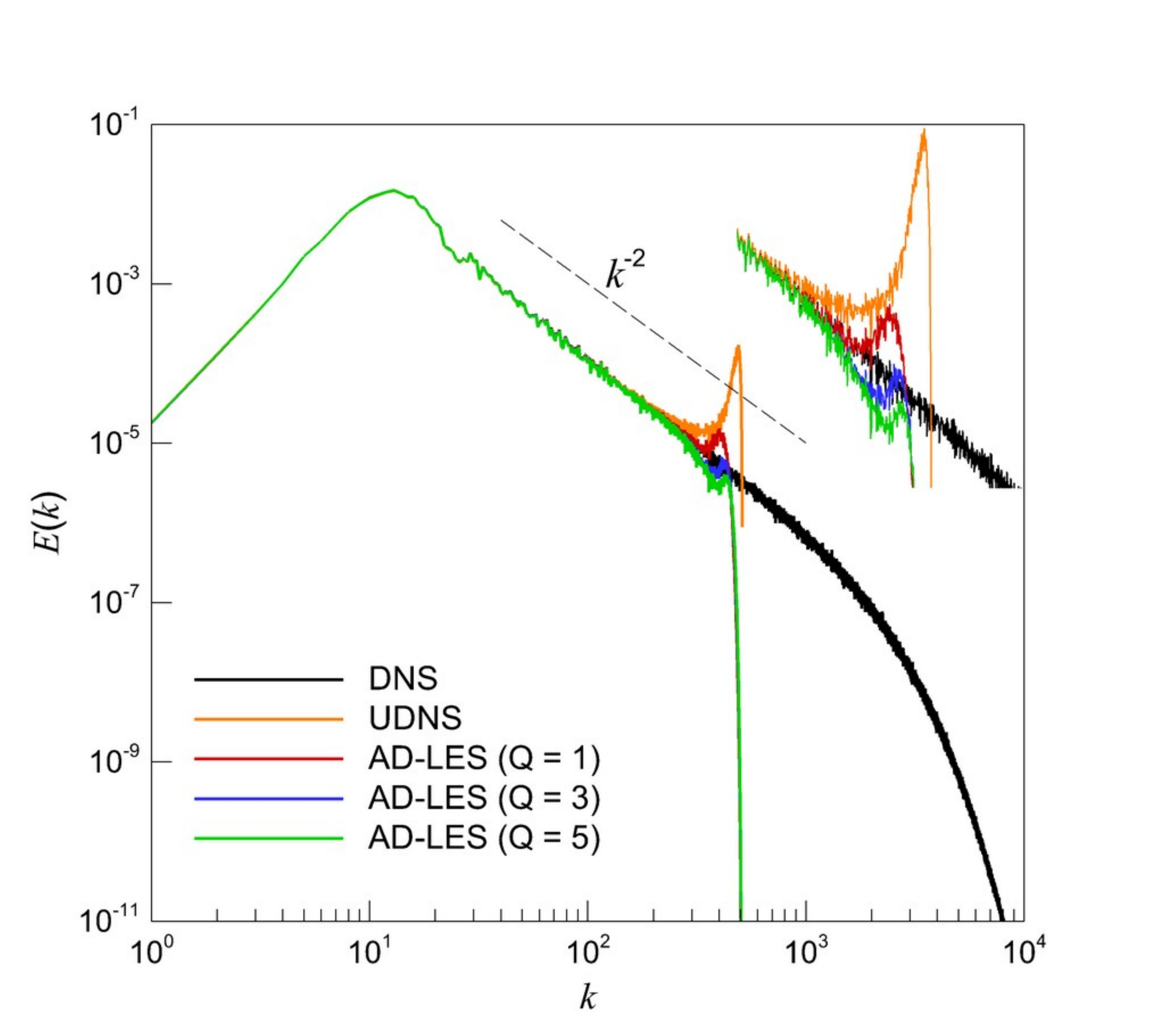}}
\subfigure[$Q = 1 $]{\includegraphics[width=0.45\textwidth]{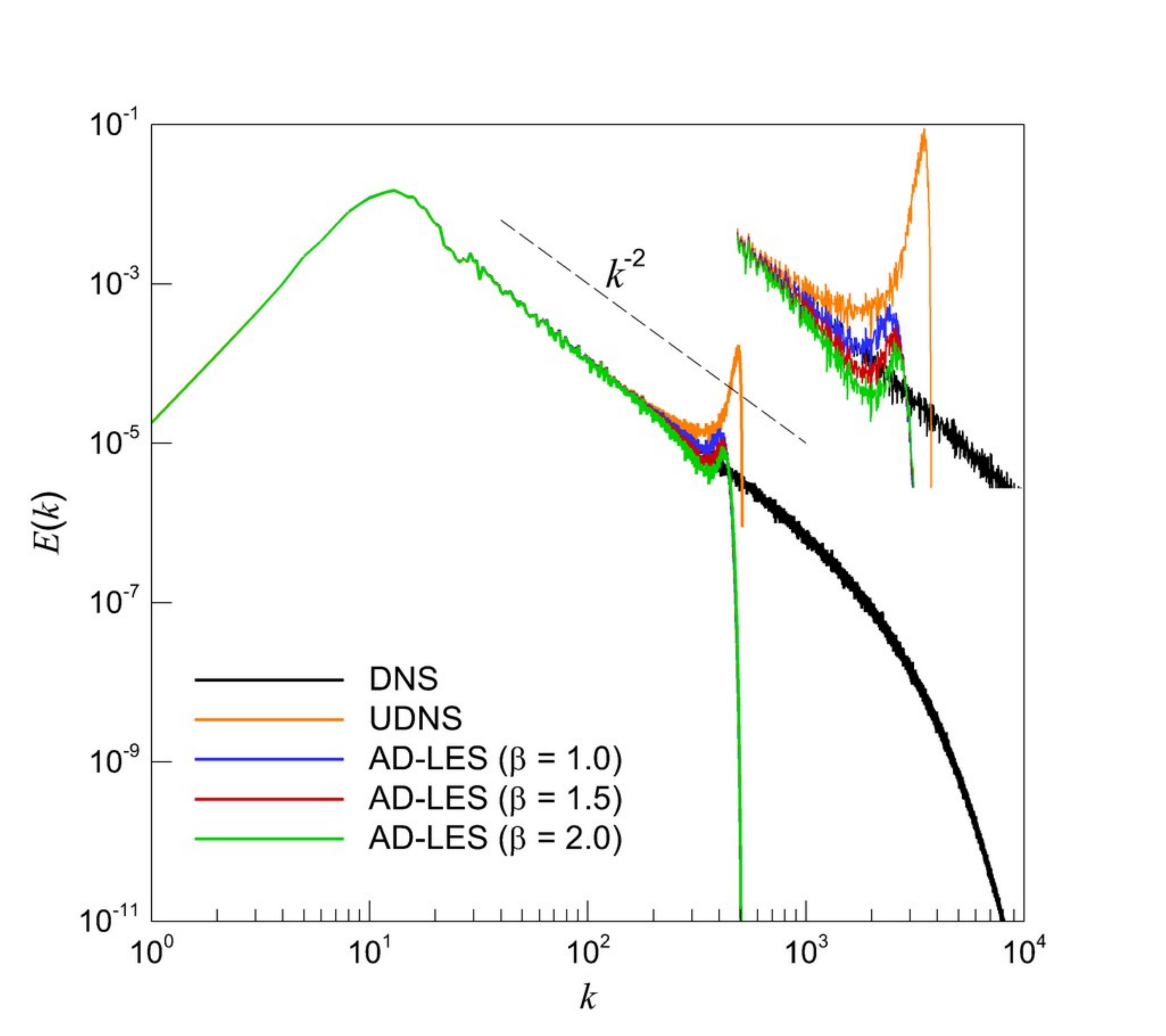}}
}\\
\mbox{
\subfigure[$Q = 3 $]{\includegraphics[width=0.45\textwidth]{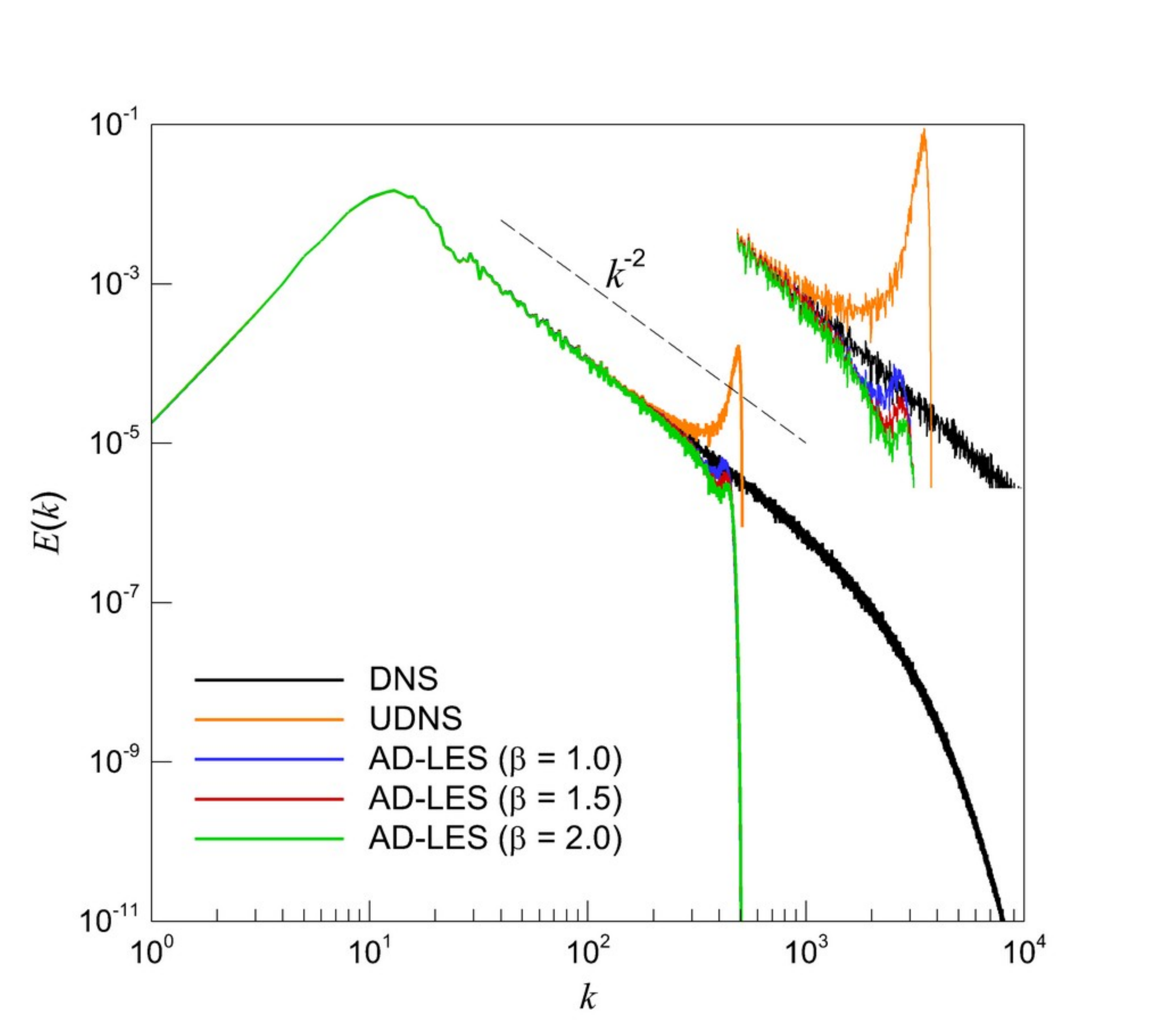}}
\subfigure[$Q = 5 $]{\includegraphics[width=0.45\textwidth]{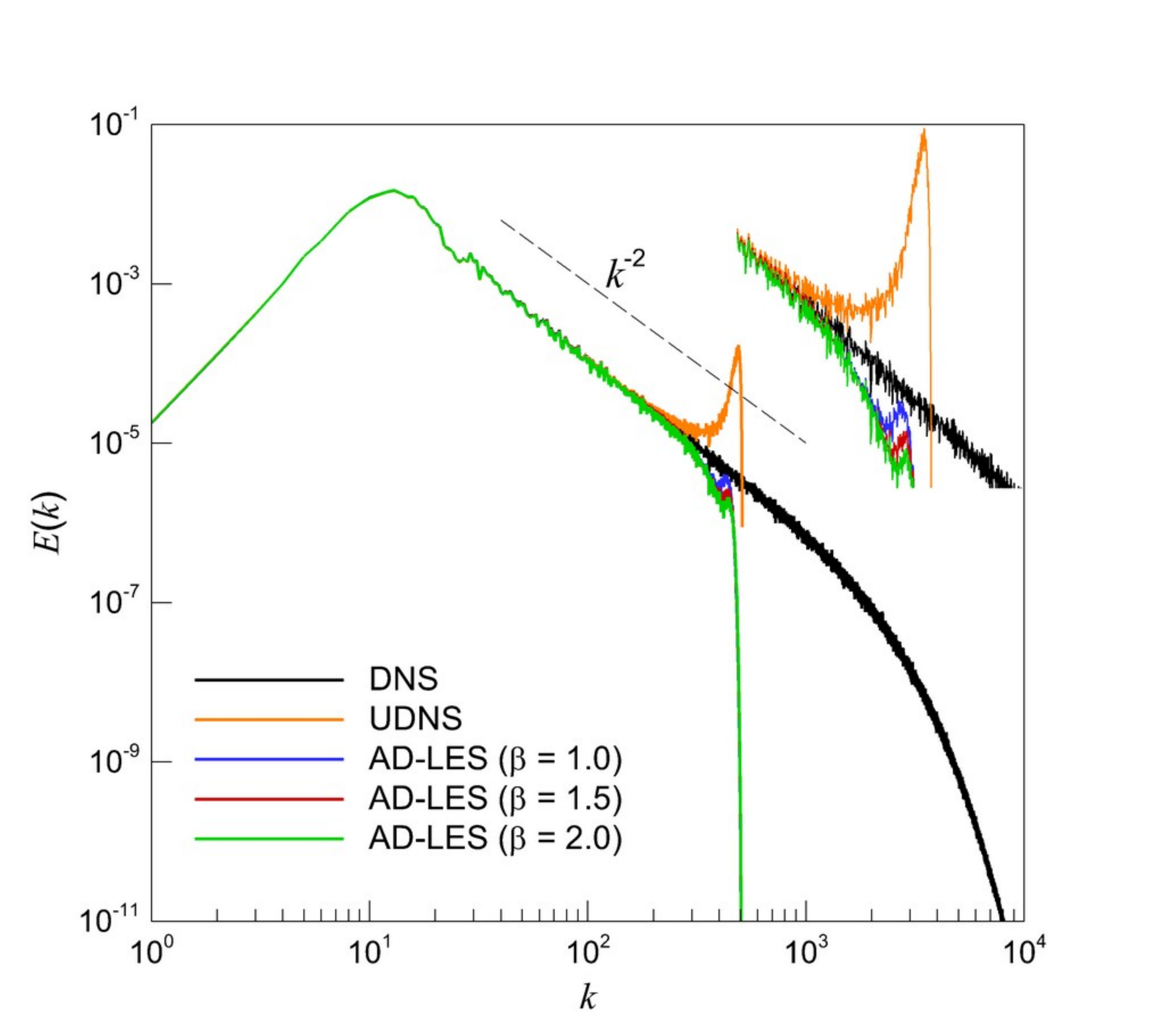}}
}
\caption{Sensitivity analysis with respect to $\beta$ and $Q$ in AD-LES modelling with $N=1024$ grid points.}
\label{fig:a9}
\end{figure}

In order to understand the effect of the parameters $\beta$ and $Q$ which are the over relaxation parameter and the number of iterative substitutions in the deconvolution process, a sensitivity analysis was carried out on the aforementioned parameters to understand how they affected convergence by using the $^{(3,1)} B$ filter. As shown in Figures (\ref{fig:a8}) and (\ref{fig:a9}) on different resolutions, increasing the $Q$ value predictably increased the accuracy in the approximation of the recovered field $\vartheta$ from the filtered field $\bar{u}$. Increasing $\beta$ within its acceptable range of [0,2] (to ensure convergence) from the standard value of 1 caused improved accuracy for a fixed value of $Q$. A similar trend was observed in both cases for the spatial resolution ($N = 512$ and $N = 1024$). This study chosen to terminate further iterative substitution in the deconvolution processes at $Q = 5$, although our analysis shows that more accurate results can be obtained with by using a different low-pass filter. The results presented in Figures (\ref{fig:a8}) and (\ref{fig:a9}) clearly indicate that $\beta=2$ can be used to accelerate the AD iterative process.

\subsection{ILES results}

\begin{figure}[!t]
\centering
\mbox{
\subfigure[WENO5]{\includegraphics[width=0.45\textwidth]{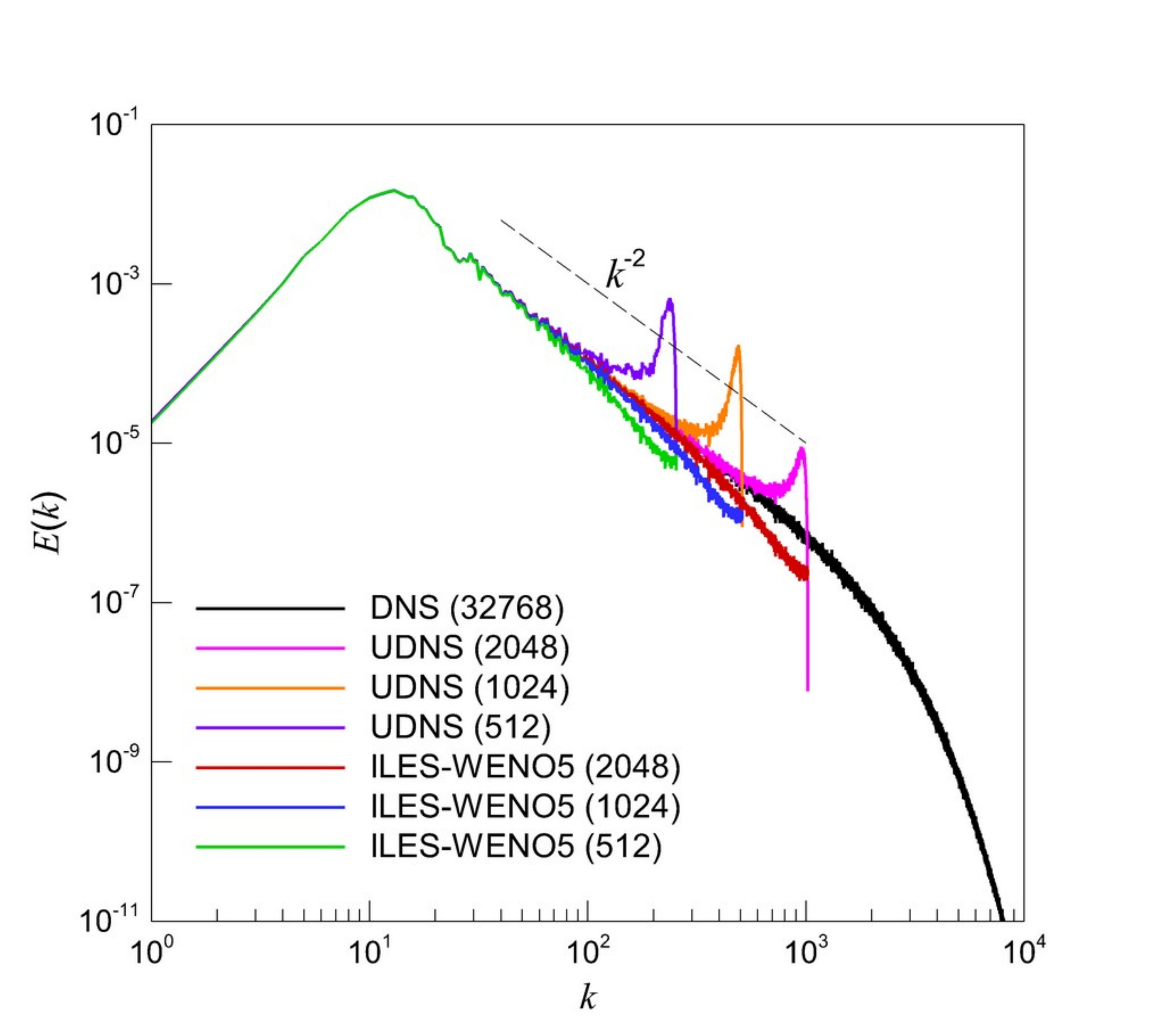}}
\subfigure[CRWENO5]{\includegraphics[width=0.45\textwidth]{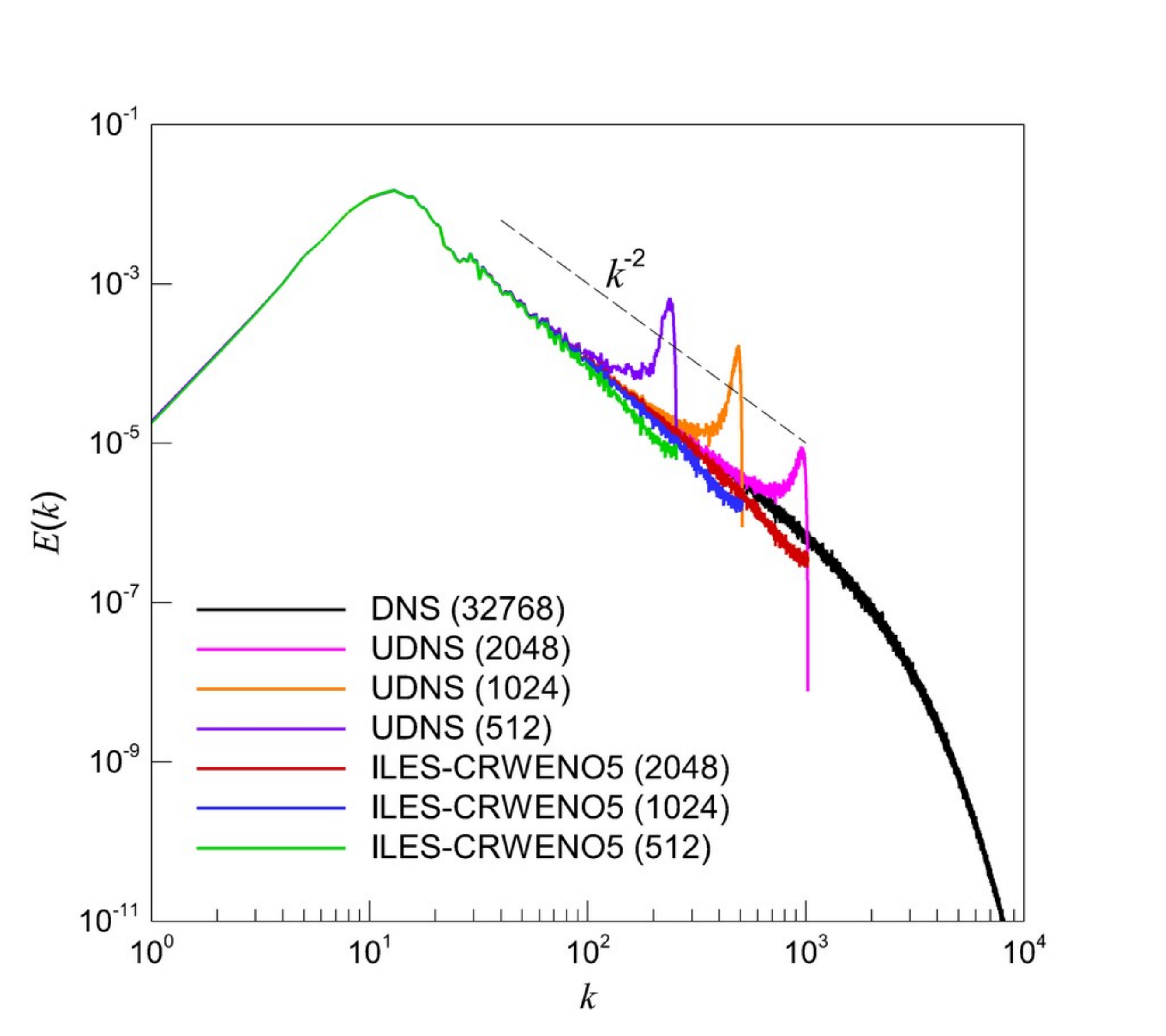}}
}\\
\mbox{
\subfigure[UPWIND5]{\includegraphics[width=0.45\textwidth]{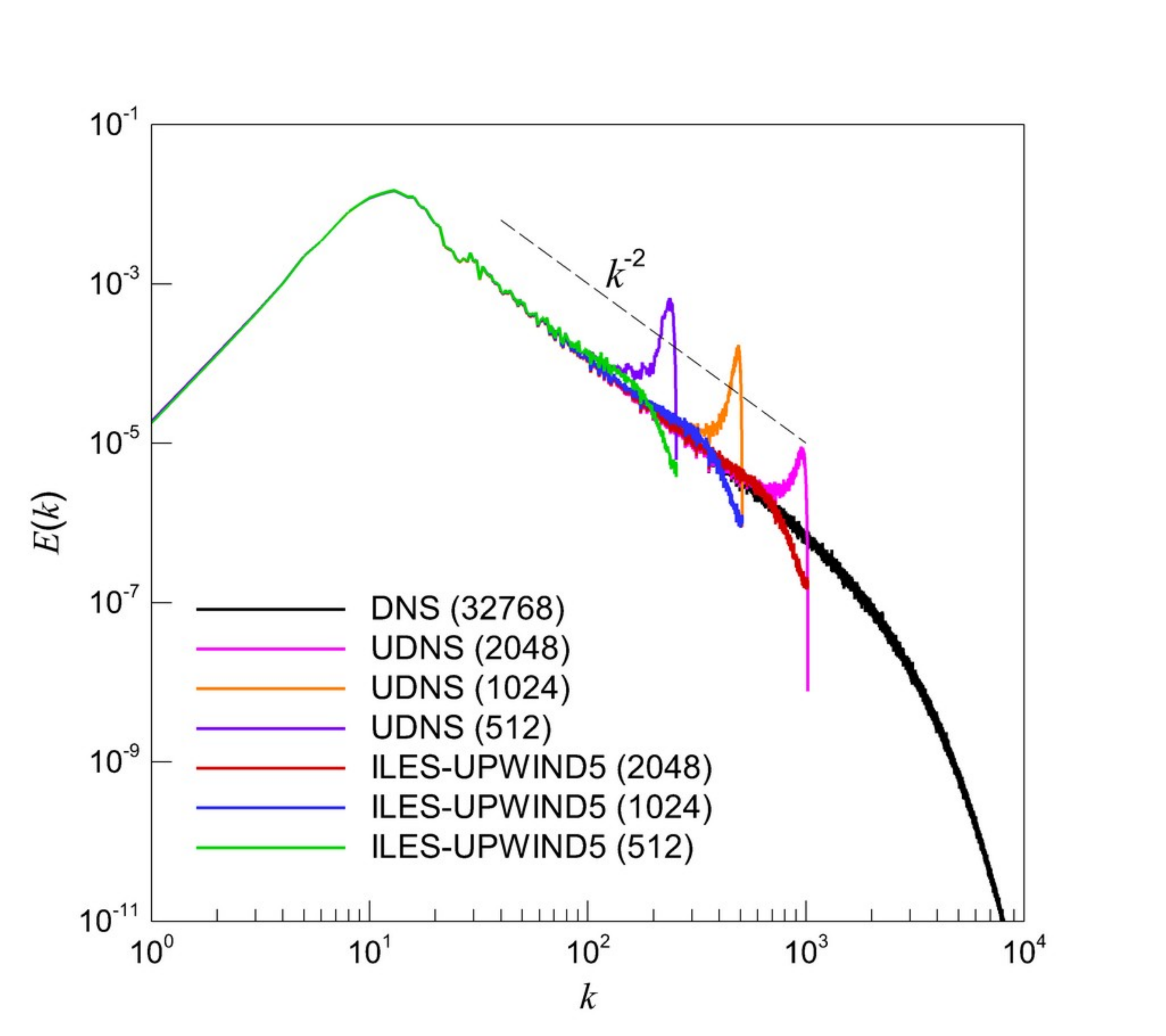}}
\subfigure[CU5]{\includegraphics[width=0.45\textwidth]{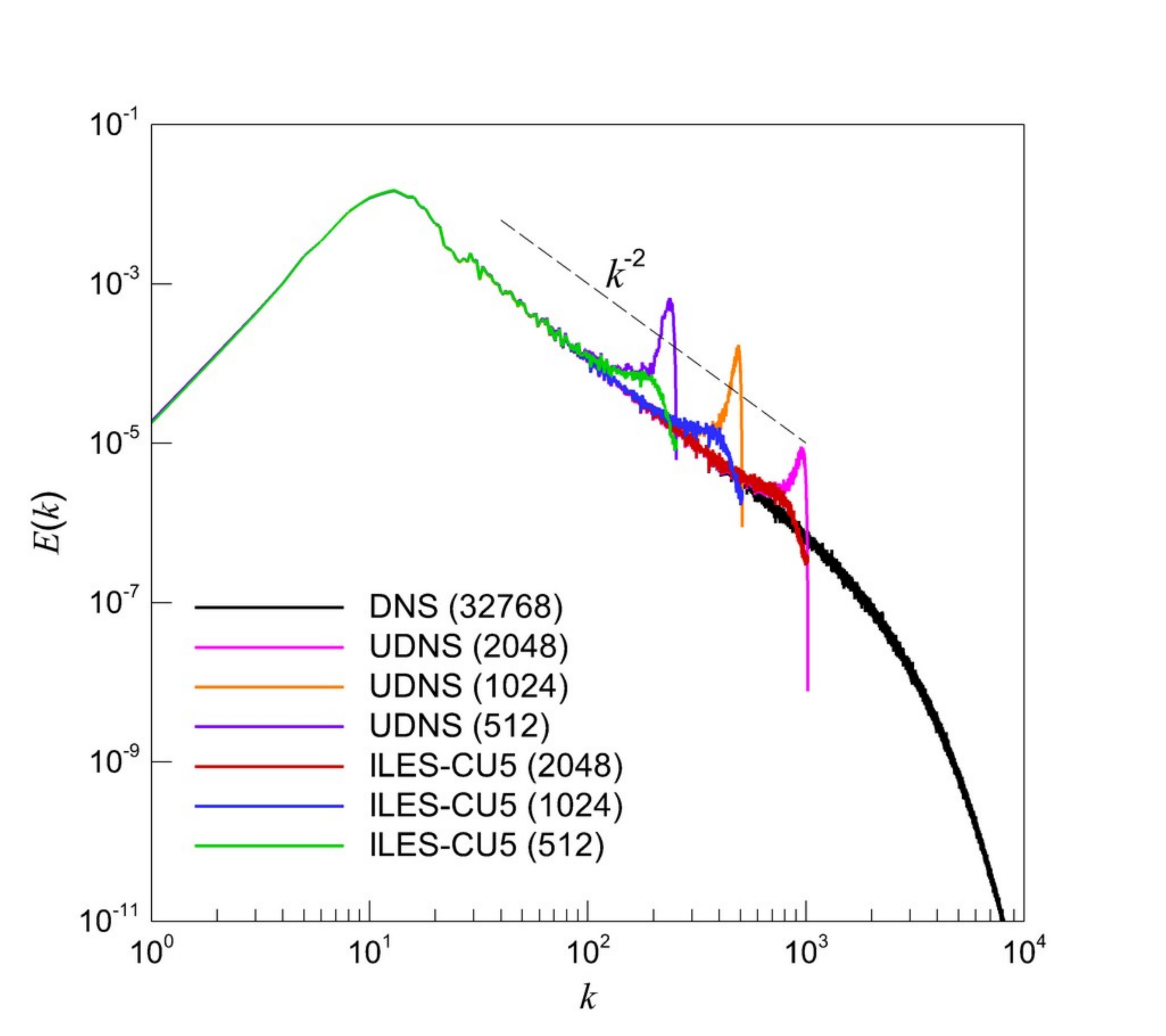}}
}
\caption{Energy spectra obtained by ILES models using local pointwise flux splitting given by Equation (\ref{eq:ls}).}
\label{fig:a10}
\end{figure}

\begin{figure}[!t]
\centering
\mbox{
\subfigure[WENO5]{\includegraphics[width=0.45\textwidth]{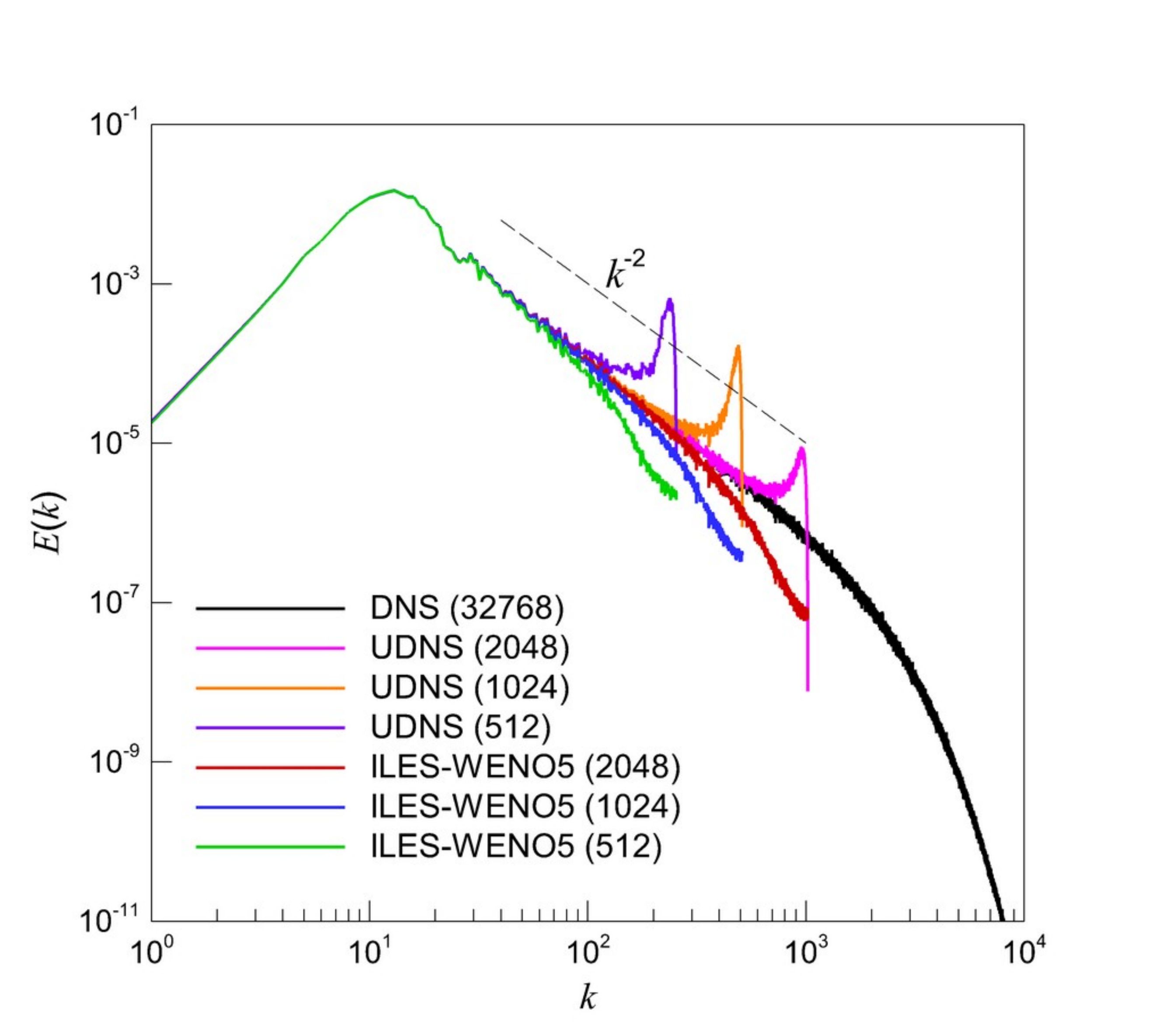}}
\subfigure[CRWENO5]{\includegraphics[width=0.45\textwidth]{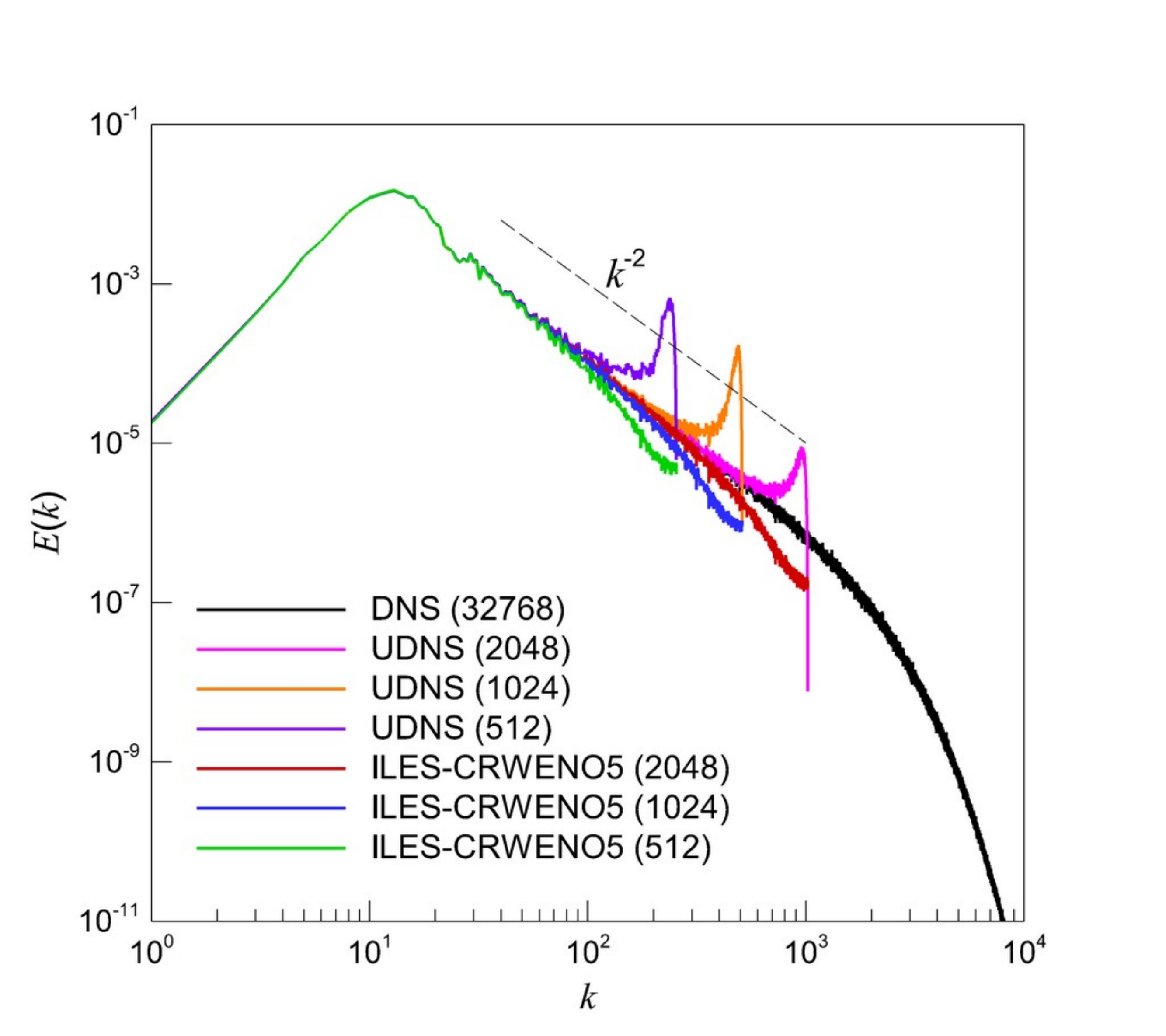}}
}\\
\mbox{
\subfigure[UPWIND5]{\includegraphics[width=0.45\textwidth]{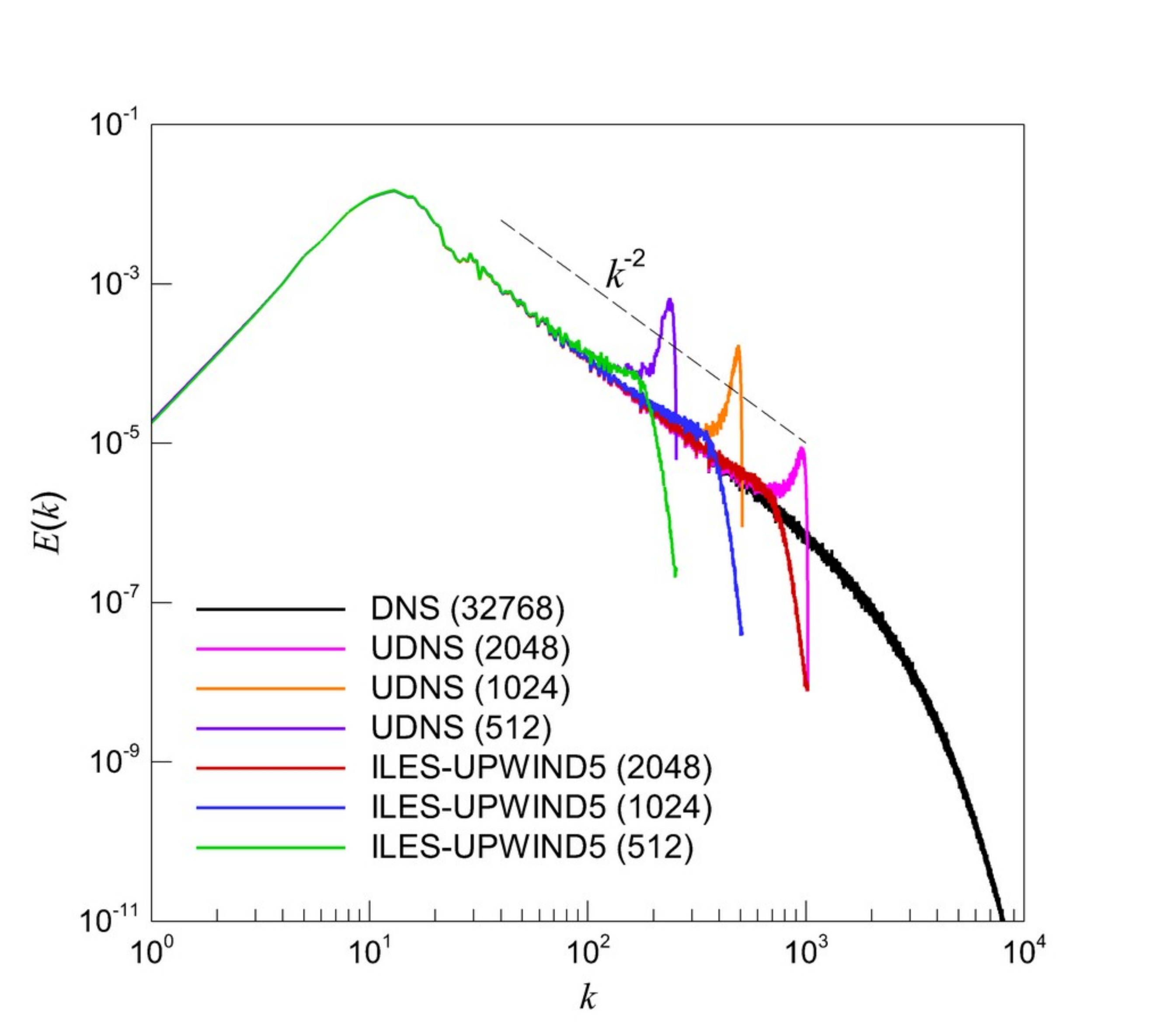}}
\subfigure[CU5]{\includegraphics[width=0.45\textwidth]{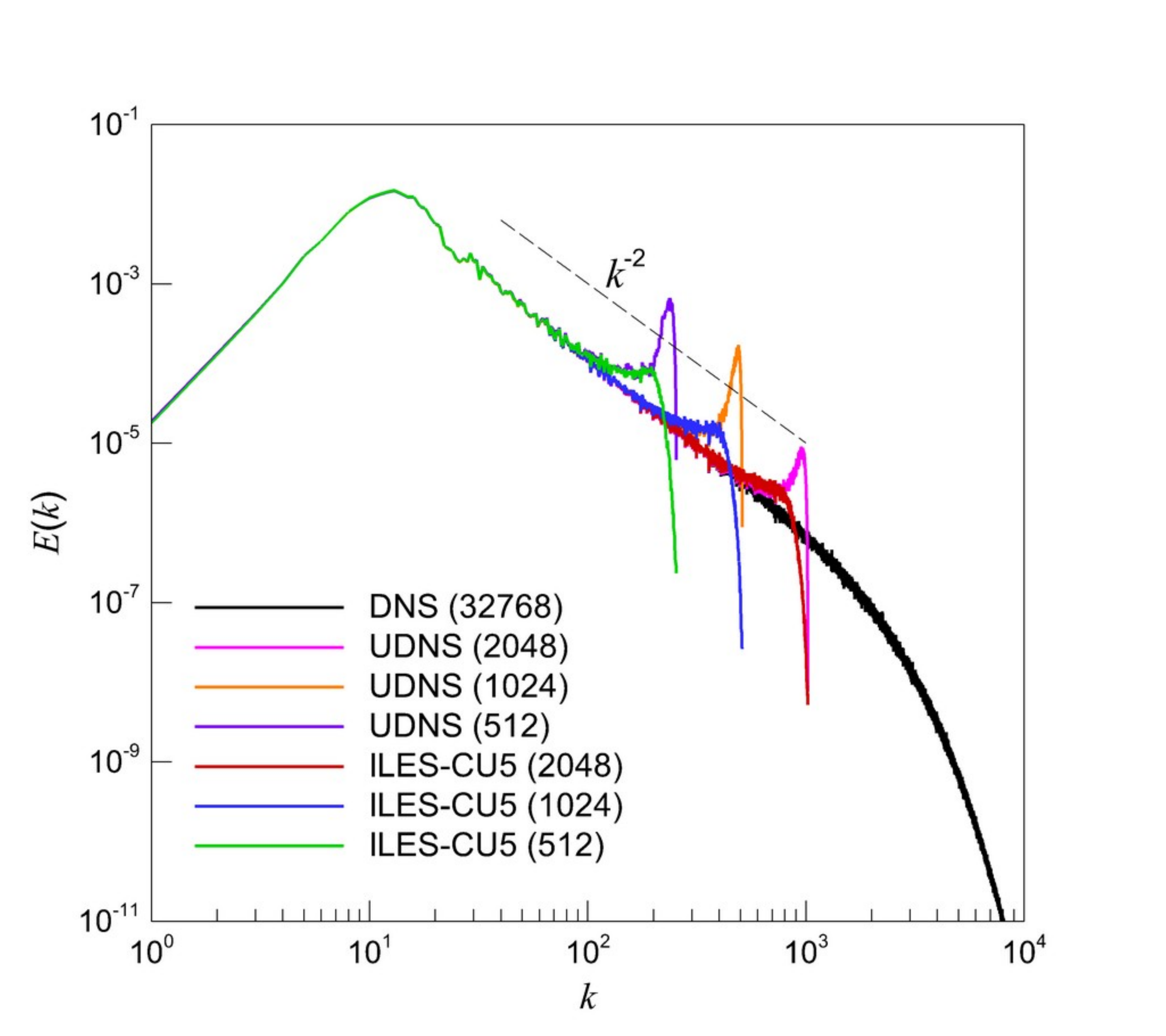}}
}
\caption{Energy spectra obtained by ILES models using local stencil flux splitting given by Equation (\ref{eq:ss}).}
\label{fig:a11}
\end{figure}

\begin{figure}[!t]
\centering
\mbox{
\subfigure[WENO5]{\includegraphics[width=0.45\textwidth]{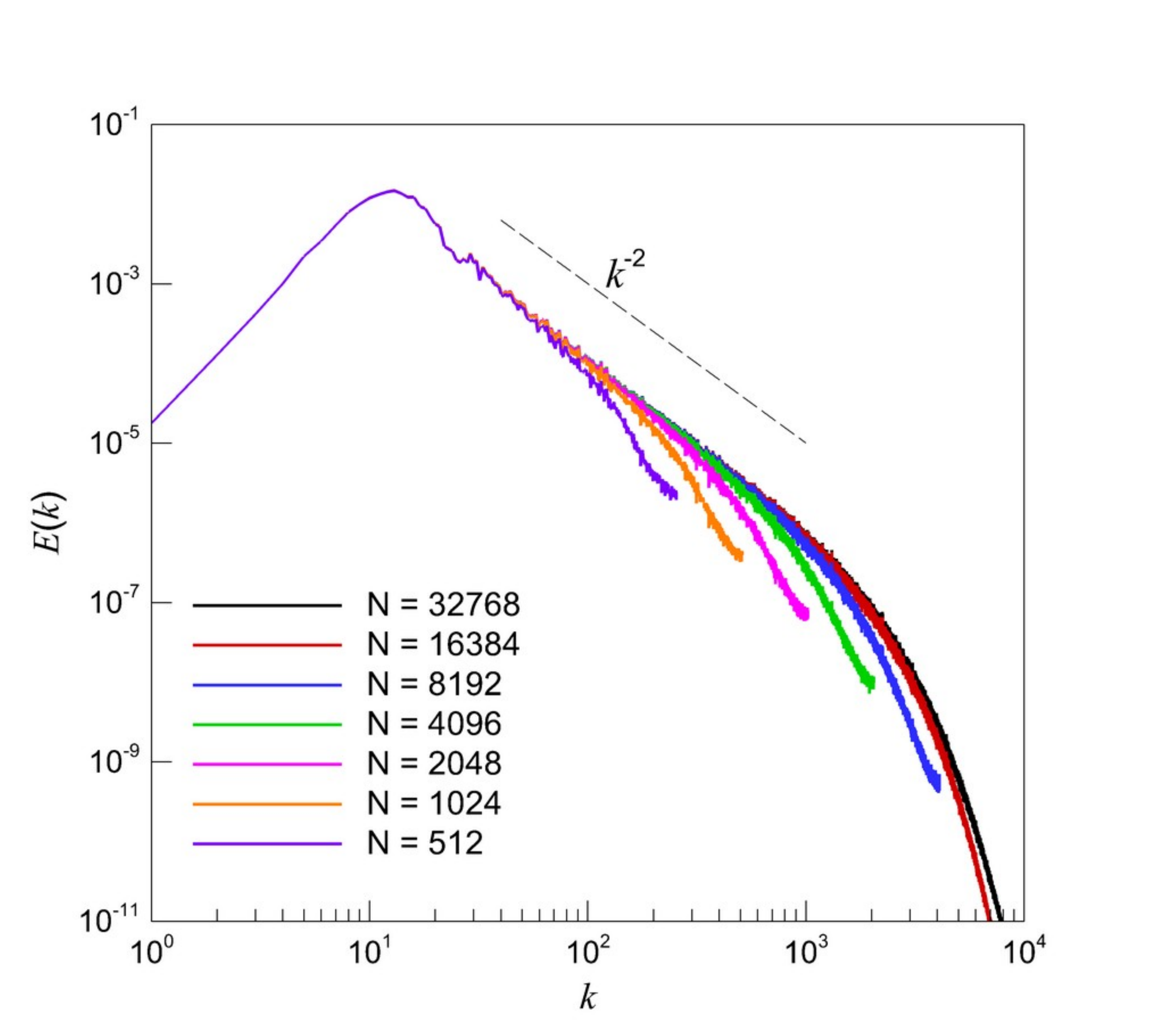}}
\subfigure[CRWENO5]{\includegraphics[width=0.45\textwidth]{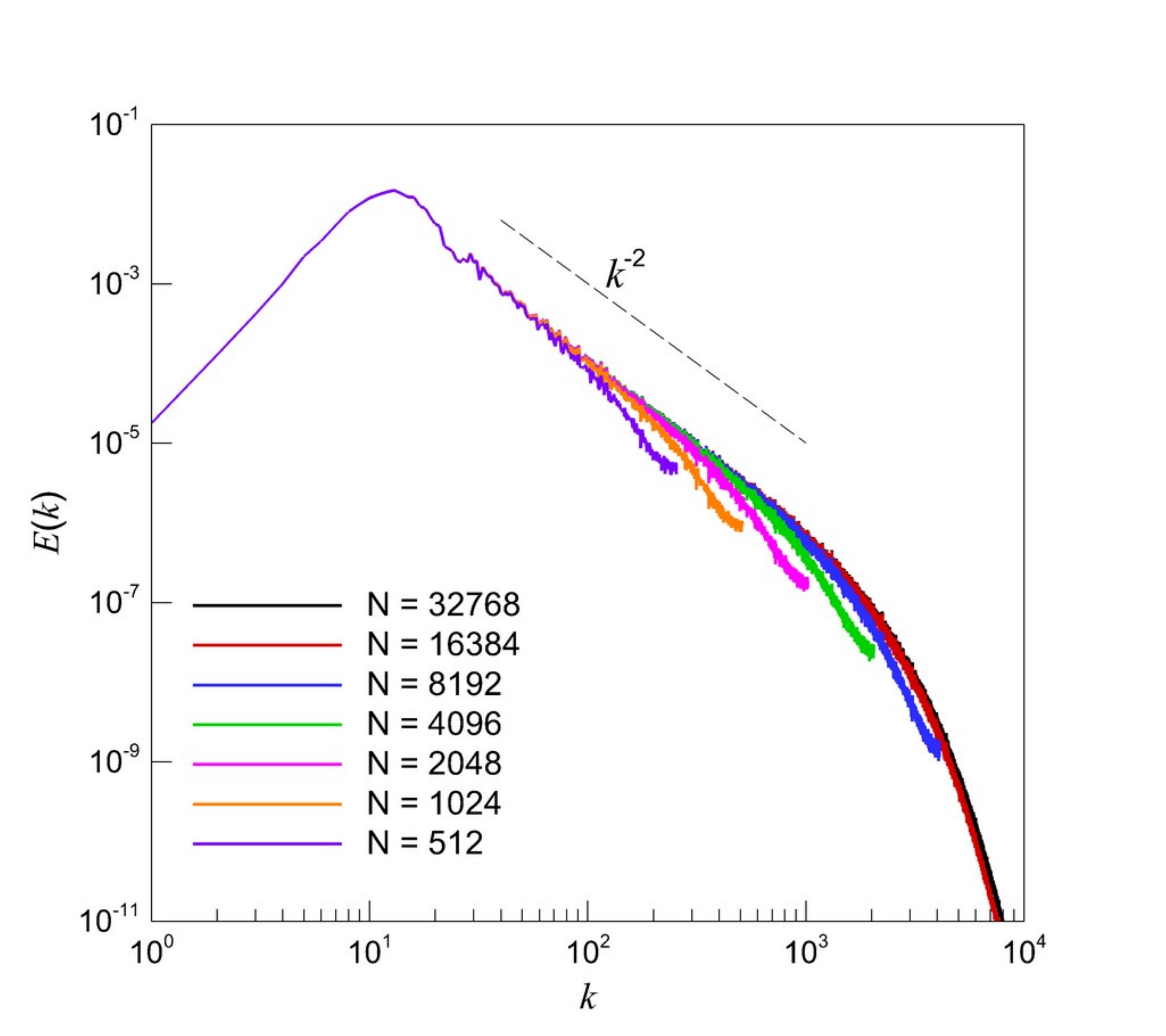}}
}\\
\mbox{
\subfigure[UPWIND5]{\includegraphics[width=0.45\textwidth]{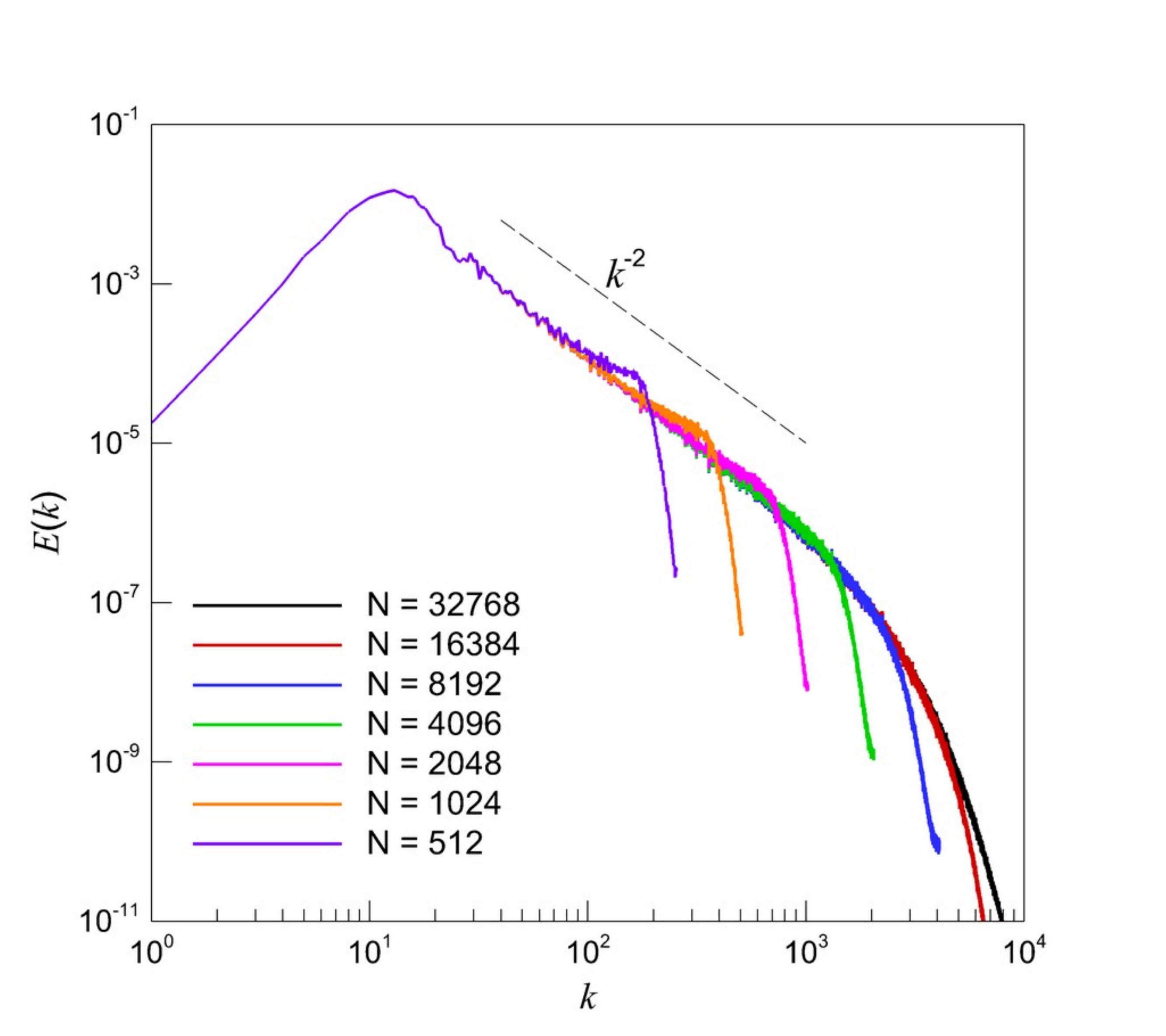}}
\subfigure[CU5]{\includegraphics[width=0.45\textwidth]{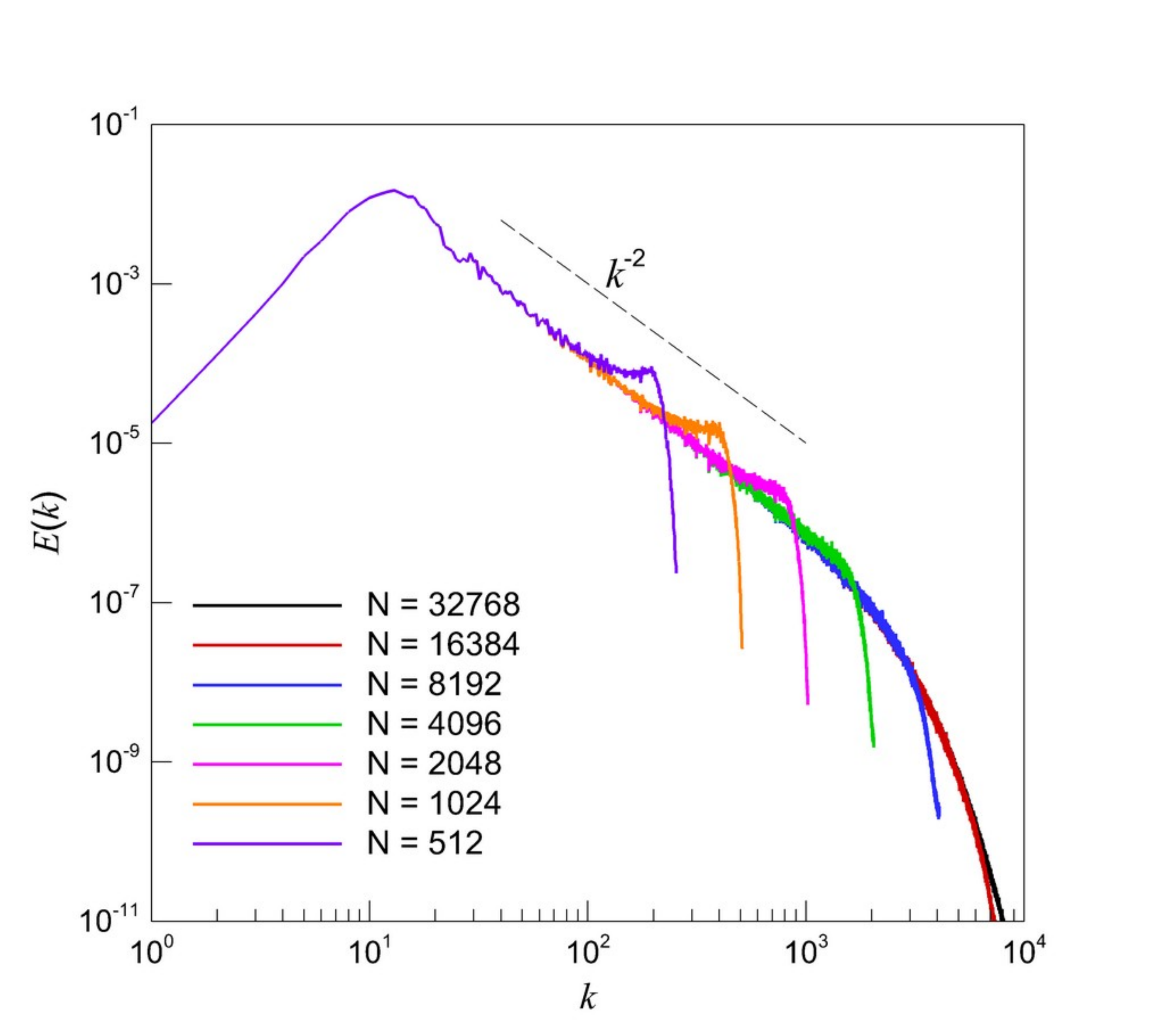}}
}
\caption{Grid converge study for ILES schemes.}
\label{fig:a12}
\end{figure}

The performance of the implicit filtering methods using the UPWIND5, CU5 WENO5, and CRWENO5 schemes was tested as shown in Figures (\ref{fig:a10}) and (\ref{fig:a11}) for both the local pointweise and stencil based flux splitting versions of the formulation. It can be easily seen from these figures that the pointwise flux splitting provides less amount of dissipation compared to the stencil based flux splitting procedure. In terms of reconstruction methods, it was seen that the WENO reconstructions (WENO5 and CRWENO5) were too dissipative and predicts slightly steeper scaling for the wide range of the inertial range. This issue was marginally remedied in the CRWENO5 formulation using the local point flux splitting method but the WENO5 method continued to remain too dissipative. However, it was noticed in both WENO schemes that the problem of pile-up was avoided since these schemes originally designed to eliminate grid-to-grid oscillations for sharp discontinuities and shocks. The developed shocks in Burgers turbulence were captured for the price of adding numerical dissipation through a nonlinear weighting process described in Section \ref{sec:ILES3} and Section \ref{sec:ILES4}. For the upwind type schemes it was seen that non-compact version of the formulation (UPWIND5) was fairly accurate in capturing the energy spectrum. The UPWIND5 scheme is less dissipative than the WENO schemes and it shows less degree of pile-up phenomenon compared to the its compact version (CU5). A pronounced pile-up phenomenon was seen in the CU5 for both local pointwise and stencil based flux splitting approaches. Finally, we present in Figure (\ref{fig:a12}) a grid convergence study for ILES methods using Equation (\ref{eq:ss}). It is important to note that all ILES methods converges to the DNS for increasing resolutions. Based on our study we can conclude that CRWENO scheme performs better in terms of dissipation mechanism and free of pile-up phenomenon at all resolutions.

\section{Conclusion}
\label{sec:sum}

This study presents an aposteriori analysis of both implicit and explicit closures for LES of the 1D viscous Burgers equation. Numerical experiments are carried out using DNS which serve as our benchmark for comparison. Variants of the eddy viscosity type functional closures have been included in our analysis. The explicit closures used include the AD-LES framework which attempts to solve the closure problem without the use of phenomenological arguments by utilizing the Van Cittert iterations to approximately recover the true solution at each time step from the spatial smoothed field variable. AD-LES is applied with the use of the binomial, the binomial smoothing and Pad\'{e} low pass filters with a wide range of filter parameters to establish a link between the transfer functions of the these filters and the accuracy in the approximation of the energy spectrum. The binomial filter is seen to add excessive dissipation and is ill-suited to the role of a low-pass filter for AD-LES, especially when we increase the order of the filter. However, the binomial smoothing filters yield more accurate results providing less dissipation.  It is seen that the Pad\'{e} filter performs better than the binomial smoothing filter with respect to the capturing of the ideal scaling of energy spectrum in the highest resolutions before cut-off although both filters may be used for the AD-LES purpose with an intelligent choice of parameters.

Other explicit approaches include the use of a secondary filter to smooth the field variable at the end of each time step. This secondary filter may be used explicitly without iterative AD-LES procedure or may be used in conjunction with the AD-LES method. In any case, it is seen to dominate the dissipative effect when used together with AD-LES. Our studies show us that the RF process due to explicit filtering adds too much dissipation and causes a substantial loss of information when a dissipative low-pass filter is used. We observed that the RF approach reduces the effective resolution of the simulation compared to the dynamic inertial range supported by the underlying computational mesh. The level of attenuation depends on the free filtering parameter and we found that $\alpha=0.49$ would be ideal for the purpose of relaxation filtering. It also avoids the computational expense of deconvolution at each time step associated with AD-LES procedure by effectively removing the high-frequency contents of the solutions. Therefore, following \citep{mathew2003explicit} we can interpret RF model as a compact and nearly equivalent procedure of AD-LES between time integration steps.

A sensitivity analysis of the Van Cittert iterative process is also carried out here which shows that the value of over-relaxation parameter $\beta$ controls the speed of convergence of the iterative process in AD-LES. The larger values of $\beta$ shows faster convergence and the number of iterations becomes optimal at $Q = 2$. Although the theoretical maximum for $\beta$ is two for most of the practical low-pass filters, our numerical experiments also showed that the value of $\beta \approx 2.5$ yields numerically stable results. Therefore, we can safely use the upper-bound $\beta=2$ in practice to accelerate Van Cittert iterative process with smaller number of $Q$ in repeated filtering process of AD procedure.

A hybrid structural-functional numerical method which uses a combination of AD-LES and a Smagorinsky type eddy viscosity to add dissipation is also detailed here. This regularized AD-LES has been found to yield significantly improved results when compared with standard AD-LES in terms of pile-up phenomenon for very coarse grids. The performances of the Smagorinsky model and variants of its dynamic version are also detailed to serve as a comparison to the other models investigated here. We found that explicit filtering models perform better than the eddy viscosity models studied here. Instantaneous dynamic model (DEV-LES) shows pile-up for coarse grid and its averaged variant (ADEV-LES) yield pile-up free solutions for all grids but add excessive dissipation.

Implicit schemes investigated in this work include the WENO reconstruction schemes (both non-compact and compact) which are seen to add too much dissipation for both the local stencil based flux splitting and pointwise based flux splitting versions of the formulation although it is marginally less dissipative for the latter. WENO and CRWENO schemes found to be effective for eliminating grid-to-grid oscillations and no pile-up phenomenon is observed even for very coarse grid. However, they are more dissipative than the linear upwind biased schemes (i.e., UPWIND5 and CU5) because of the presence of shock waves in the Burgers turbulence. Due to dynamic stencil formulation of WENO and CRWENO schemes, the amount of dissipation has been increased by decreasing the formal order of accuracy in shock regions. On the other hand, linear upwind schemes are also prone to adding considerable amount dissipation, although less than nonlinear ones, with the compact upwind version being unable to prevent pile-up at the higher resolutions near grid cut-off. In practice, the most important factor of an LES closure is its ability to capture the inertial range scaling of the spectrum as well as ensure total attenuation at the grid cut-off wavenumber without showing any pile-up phenomena. With this in mind, we can conclude that CRWENO, regularized AD and RF (with carefully selected relaxation filtering parameter) models would be accurate in capturing long range of inertial range without showing a pile-up phenomenon near the grid cut-off scale.





\bibliographystyle{gCFD}
\bibliography{references}

\end{document}